\documentclass[10pt, aps, prc, superscriptaddress, nofootinbib, 
               amsmath, %amsfonts, amsmath, amssymb, 
               twocolumn, %preprintnumbers, 
               showpacs, % Not for RMP
               raggedbottom, 
               floatfix, 
               longbibliography, %noeprint, 
               colorlinks=true, linkcolor=blue, citecolor=blue, urlcolor=blue, 
              ]{revtex4-1}
\pdfoutput=1
\usepackage[T1]{fontenc}
\usepackage[utf8]{inputenc}
\usepackage[secnoindent, 
            ]{my_paper} % Use this for APS
\usepackage{my_acronyms}        % acronyms for this paper.
\usepackage[english]{babel}     % Not exactly sure why this is
\usepackage{adjustbox}          % Used to adjust figure frames
\usepackage{colortbl}
\usepackage{xparse}             % Improved \DeclareDocumentCommand
\usepackage[caption=false]{subfig}
\usepackage[percent]{overpic} % https://tex.stackexchange.com/a/20802/6903

\usepackage[%active
            tightpage]{preview}

\renewcommand{\vect}[1]{\boldsymbol{#1}}
\newcommand{\grad}{\vect{\nabla}}
\newcommand{\laplacian}{{\vect\nabla}^2}
\newcommand{\bGGA}{\kappa}
\newcommand{\Ekin}{\mathcal{E}_{\text{kin}}}
\newcommand{\Etau}{\mathcal{E}_{\tau}}
\newcommand{\Ehomo}{\mathcal{E}_{\text{homo}}}
\newcommand{\Eint}{\mathcal{E}_{\text{int}}}
\newcommand{\Ec}{\mathcal{E}_{C}}
\newcommand{\Egrad}{\mathcal{E}_{{\vect \nabla}\n}}
\newcommand{\Eso}{\mathcal{E}_{\text{SO}}}
\newcommand{\Epair}{\mathcal{E}_{\Delta}}
\newcommand{\Eentrain}{\mathcal{E}_{\text{entrain}}}
\newcommand{\n}{n}  % Use this for densities so we can switch to \rho if needed

\let\norm\abs
\let\Norm\Abs

\begin{document}

\newlength{\mybaselineskip}
\setlength{\mybaselineskip}{\baselineskip}

% 10 word limit
\title{A Minimal Nuclear Energy Density Functional}

\author{Aurel Bulgac}%
\email{bulgac@uw.edu}%
\affiliation{Department of Physics, %
  University of Washington, Seattle, Washington 98195--1560, USA}

\author{Michael McNeil Forbes}%
\email{mforbes@alum.mit.edu}%
\affiliation{Department of Physics \& Astronomy, %
  Washington State University, Pullman, Washington 99164--2814, USA}%
\affiliation{Department of Physics, %
  University of Washington, Seattle, Washington 98195--1560, USA}

\author{Shi Jin}%
\email{js1421@uw.edu}%
\affiliation{Department of Physics, %
  University of Washington, Seattle, 
  Washington 98195--1560, USA}
  
\author{Rodrigo Navarro Perez}%
\email{navarrop@ohio.edu}
\affiliation{Nuclear and Chemical Science Division, 
  Lawrence Livermore National Laboratory, Livermore, 
  CA 94551, USA}
%\affiliation{Department of Physics and Astronomy and
%  Institute of Nuclear and Particle Physics, Ohio University,
%  Athens, Ohio 45701, USA}

\author{Nicolas Schunck}%
\email{schunck1@llnl.gov}
\affiliation{Nuclear and Chemical Science Division, 
  Lawrence Livermore National Laboratory, Livermore, 
  CA 94551, USA}

\date{\today}
\preprint{NT@UW-17-12}
%\pacs{21.10.Dr, 21.65.Cd, 21.65.Ef, 21.65.Mn}

% 21.10.Dr 	Binding energies and masses
% 21.65.Cd 	Asymmetric matter, neutron matter
% 21.65.Ef 	Symmetry energy
% 26.60.Kp 	Equations of state of neutron-star matter
% 21.65.Mn 	Equations of state of nuclear matter

\begin{abstract}
  We present a minimal \gls{NEDF} called ``SeaLL1'' that has the smallest number
  of possible phenomenological parameters to date. SeaLL1 is defined by 7
  significant phenomenological parameters, each related to a specific nuclear
  property. It describes the nuclear masses of even-even nuclei with a mean
  energy error of \SI{0.97}{MeV} and a standard deviation \SI{1.46}{MeV},
  two-neutron and two-proton separation energies with \gls{rms} errors of
  \SI{0.69}{MeV} and \SI{0.59}{MeV} respectively, and the charge radii of
  \num{345} even-even nuclei with a mean error $\epsilon_r=\SI{0.022}{fm}$ and a
  standard deviation $\sigma_r=\SI{0.025}{fm}$.  SeaLL1 incorporates constraints
  on the \gls{EoS} of pure neutron matter from quantum Monte Carlo calculations
  with chiral effective field theory two-body (NN) interactions at \gls{N3LO}
  level and three-body (NNN) interactions at the \gls{N2LO} level. Two of the
  seven parameters are related to the saturation density and the energy per
  particle of the homogeneous symmetric nuclear matter, one is related to the
  nuclear surface tension, two are related to the symmetry energy and its
  density dependence, one is related to the strength of the spin-orbit
  interaction, and one is the coupling constant of the pairing interaction. We
  identify additional phenomenological parameters that have little effect on
  ground-state properties, but can be used to fine-tune features such as the
  Thomas-Reiche-Kuhn sum rule, the excitation energy of the giant dipole and
  Gamow-Teller resonances, the static dipole electric polarizability, and the
  neutron skin thickness.
\end{abstract}

\glsreset{NEDF}
\glsreset{GDR}
\glsreset{DFT}
\glsreset{UFG}

\maketitle

%\tableofcontents

\glsreset{NEDF}
\glsreset{GDR}
\glsreset{DFT}
\glsreset{UFG}

\section{Introduction}

The accurate and precise calculation of ground-state nuclear properties and 
nuclear dynamics represent a formidable challenge for quantum many-body theory. 
While there exist a variety of techniques for directly solving the many-body 
Schr\"odinger equation, most of them are often limited to static properties, 
and do not scale well with the number of particles in the system. In contrast, 
\gls{DFT} provides a unified framework for computing both static and dynamic
properties. Although in principle exact, at least for atomic 
systems~\cite{HK:1964,Dreizler:1990lr}, the theory does not provide the form of 
the energy functional. A successful implementation of \gls{DFT} thus requires a
physically-motivated functional form, together with carefully fitted 
phenomenological parameters, or alternatively, a first-principle derivation. 
Most \gls{NEDF} in the literature are typically constructed by building the 
functional from the expectation value of effective nuclear forces on Slater 
determinants, such as the Skyrme and Gogny parameters, or by considering the 
average values of effective Lagrangians as in the relativistic mean-field
theory \cite{Bender:2003}.  Despite a significant
research investment~\cite{UNEDF, NUCLEI, Reinhard:2016, Goriely:2015,
  Goriely:2016, Goriely:2016a}, improvements to these functionals have
been incremental.

In this paper, we present a different approach, revisiting the motivation behind
the form of current \glspl{DFT}.  We systematically construct a truly minimal
\gls{NEDF}, which we call SeaLL1, that cleanly separates the phenomenological
parameters into hierarchies.  Unlike typical \glspl{NEDF}, which are built
directly from the approach of \textcite{Kohn:1965fk}, we start with a minimal
orbital-free formulation functional of neutron and proton densities in the
spirit of \textcite{HK:1964}, along the lines delineated by
\textcite{Weizsacker:1935}.  Built on a core of four dominant parameters, this
orbital-free \gls{NEDF} obtains a global mass fit better than the four-parameter
Bethe-Weizsäcker mass formula~\cite{Bethe:1936}, but in addition provides quite
accurate charge radii.  The orbital based SeaLL1 functional then minimally
extends this four-parameter \gls{NEDF} by adding three parameters to describe shell
effects, pairing correlations, and the density dependence of the symmetry
energy, the latter which governs the neutron skin thickness of \ce{^{208}Pb} and
\ce{^{48}Ca}.  In this form, the seven parameter SeaLL1 functional displays extremely
reasonably single-particle spectra, globally fitting masses, charge radii, and
two-nucleon separation energies.  As the nucleon effective mass in SeaLL1 is the
bare nucleon mass, we expect the total energy level densities to be in much
better agreement with experiment than for typical Skyrme-like \glspl{NEDF}.

Since we advocate a new paradigm for constructing and improving a \gls{NEDF}, 
we begin in~\cref{sec:backgr-motiv} with a somewhat lengthy historical 
background to motivate our approach in~\cref{sec:functional-form-nedf}. The 
form of the SeaLL1 functional is presented in detail 
\cref{sec:functional-form-nedf} along with its orbital-free formulation.
\Cref{sec:physical-properties} discusses a number of nuclear properties that 
have been used to validate the predictive power of our \gls{NEDF}. 
\Cref{sec:perspectives} identifies how the \gls{NEDF} could be systematically 
improved for applications either to static or dynamical properties. Finally, we
summarize our results in \cref{sec:conclusions}. The hurried reader can 
just read \cref{sec:functional-form-nedf}  
and \cref{sec:physical-properties}, which contain all the results. 

For the interested reader, we 
provide additional material in \cref{app:orbit-free-funct}, where we discuss in more 
details the orbital free formulation and illustrate  the dominance or 
sub-dominance of various parameters.  Numerical values for the functional 
parameters, as well as tables of quantities used in our fits, are provided as 
Supplementary Material~\cite{EPAPS:apsSupp}.

\section{Historical Background and Motivations}\label{sec:backgr-motiv}

Almost a century ago, \textcite{Aston:1920} realized that a nucleus is
not quite the sum of its parts.  This led \textcite{Eddington:1920}
to correctly conjecture a link between nuclear masses, the conversion
of hydrogen into heavier elements, and the energy radiated by the
stars.  An accurate theoretical model of nuclear masses, particularly
close to the neutron drip line and with an uncertainty of better than
100 keV (an accuracy which has not been achieved yet even for known
stable nuclei) will have a great impact on predicting the origin and
the abundances of elements in the Universe~\cite{Mumpower:2016,
  Mumpower:2016a}.

When quantum mechanics was first applied to many-body systems,
\textcite{Weizsacker:1935} proposed that an energy density approach
could be an effective tool for calculating nuclear binding energies.
This was the first instance of an energy density functional being
applied in nuclear physics, several decades before the
foundation of \gls{DFT}~\cite{HK:1964, Kohn:1965fk, Dreizler:1990lr}
was formulated.  \textcite{Bethe:1936} further developed Weizsäcker's
ideas and introduced the nuclear mass formula (the Bethe-Weizsäcker
formula) for the ground-state energies of nuclei with $A=N+Z$ nucleons
($N$ neutrons and $Z$ protons):
\begin{equation}\label{eq:Bethe}
  E(N, Z) = a_v A + a_s A^{2/3} + a_C\frac{Z^2}{A^{1/3}}
  + a_I\frac{(N-Z)^2}{A}.
\end{equation}
Unlike electrons in atoms, nuclei are saturating systems with a nearly
constant interior density.  This yields the terms in \cref{eq:Bethe}
referred to as volume energy, surface tension, non-extensive
Coulomb energy, and symmetry energy that favors similar numbers of
protons and neutrons.  (Because of the long-range Coulomb
interaction, the terms ``volume'' and ``surface'' do not have a strict
thermodynamic meaning.)  As shown in the first row of
\cref{table:liquid_drop}, these four terms alone fit the AME2012
evaluated nuclear masses~\cite{Audi:2012, Wang:2012} with a \gls{rms}
error of $\chi_E =\SI{3.30}{MeV}$ per nucleus.  This is a remarkable
result: the nuclear binding energy of heavy nuclei can reach
\SI{2000}{MeV}, hence the errors are at the sub-percent level.

\renewcommand{\th}[1]{\multicolumn{1}{c}{#1}}  % Use for the table headers
\newcolumntype{3}{D{.}{.}{3}}
\newcolumntype{2}{D{.}{.}{2}}
\newcolumntype{1}{D{.}{.}{1}}
\newcolumntype{i}{D{.}{.}{0}}
\begin{table}[t]
  \begin{ruledtabular}
    \begin{tabular}{22223322}
      \th{$a_v$} & \th{$a_s$} & \th{$a_I$} & \th{$a'_I$} & \th{$a_C$} &
      \th{$a'_C$} & \th{$\delta$} & \th{$\chi_E$} \\
      \hline
      -15.47 & 16.73 & 22.87 & 0 & 0.699 & 0 & 0 & 3.30 \\
      -15.49 & 16.78 & 22.91 & 0 & 0.700 & 0 & 12.29 & 3.18 \\[0.5\mybaselineskip]
      -15.32 & 17.76 & 24.96 & -22.60 & 0.767 & -0.675 & 0 & 2.64 \\
      -15.34 & 17.80 & 25.01 & -22.43 & 0.767 & -0.661 & 11.46 & 2.50 \\[0.5\mybaselineskip]
      -15.77 & 17.50 & 23.65 &   0    & 0.723 &  0  & 0  & 1.87 \\
      -15.46 & 18.29 & 25.72 & -26.00 & 0.792& -0.773 & 0    & 1.53 \\
    \end{tabular}
  \end{ruledtabular}
  \caption{Parameters and the energy \gls{rms} of the mass formulas
    \cref{eq:Bethe} or \cref{eq:masses}, with or without the even-odd
    staggering correction \cref{eq:odd-even}.  Here
    $\chi_E^2 =\sum\abs{E_{N, Z} - E(N, Z)}^2/N_E$ and we fit the
    $N_E = 2375$ measured (not extrapolated) nuclear masses of nuclei
    with $A\geq 16$ from~\textcite{Audi:2012, Wang:2012} and an
    evaluated uncertainty less than \SI{1}{MeV} with the electronic
    correction.  (All quantities expressed in \si{MeV}.)  The last two
    rows show how the mass formulas \cref{eq:Bethe} or
    \cref{eq:masses} fit the theoretical nuclear masses computed using
    the SeaLL1 functional.}
  \label{table:liquid_drop}
\end{table}

\begin{figure}[t]
  \includegraphics[width=\columnwidth]{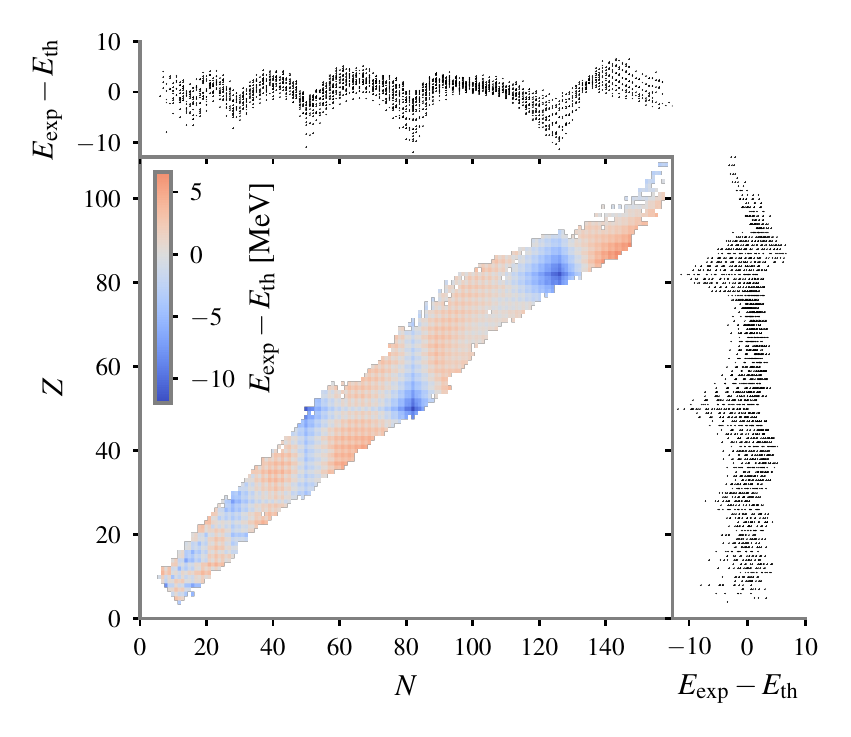}
  \caption{(Color online) The differences $E_{\text{exp}} - E_{\mathrm{th}}$
    in \si{MeV}s between the evaluated ground state energies
    $E_{\text{exp}}(N,Z)$~\cite{Audi:2012, Wang:2012} of 2375 nuclei
    with $A\ge 16$ and fitted with the six-parameter mass formula
    $E_{\mathrm{th}}=E(N,Z)$ \cref{eq:masses} with $\Delta \equiv 0$.
    One can easily identify the location of closed shells (the blue
    regions) for protons and neutrons.}
  \label{fig:masses}
\end{figure}

\begin{figure}[t]
  \includegraphics[width=\columnwidth]{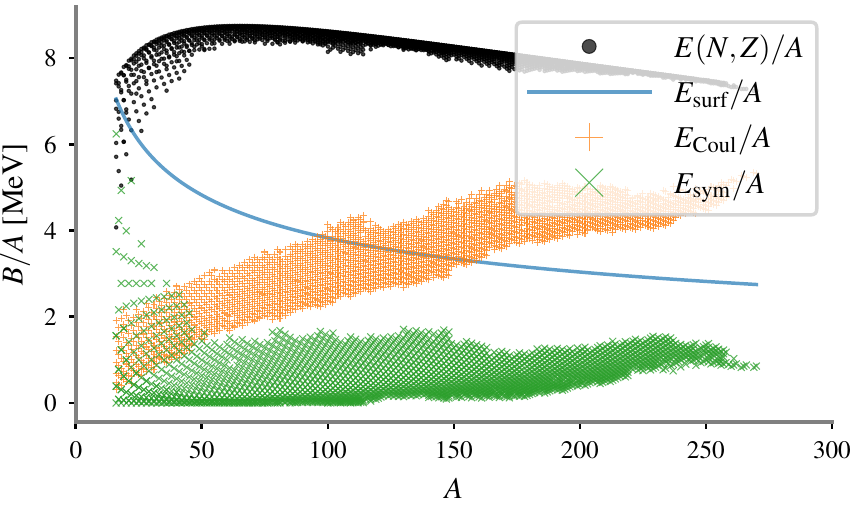}
  \caption{(Color online) The binding energy per nucleon
    $B/A = \abs{E(N, Z)}/A$ and the Coulomb, surface and symmetry
    energy per nucleon in \cref{eq:masses} for the measured 2375
    nuclei with $A \ge 16$~\cite{Audi:2012, Wang:2012}.}
  \label{fig:mass_contributions}
\end{figure}

A slightly better fit is obtained using a mass formula with surface
corrections terms to the symmetry and Coulomb energies, as well as
odd-even staggering correction due to pairing:
\begin{subequations}
 \label{eq:masses}
 \begin{multline}
    E(N, Z) = a_v A + a_s A^{2/3} + a_C\frac{Z^2}{A^{1/3}}
    + a'_C\frac{Z^2}{A^{2/3}} \\
    + a_I\frac{(N - Z)^2}{A} + a'_I\frac{(N - Z)^2}{A^{4/3}}
    + \Delta.
  \end{multline}
  \begin{gather}
    \Delta = \begin{cases}
      -\delta A^{-1/2} & \text{even-even nuclei, } \\
      \hfil 0 & \text{odd nuclei, } \\
      \hfil\hphantom{-}\delta A^{-1/2} & \text{odd-odd nuclei.}
    \end{cases} \label{eq:odd-even}
  \end{gather}
\end{subequations}
This pairing contribution is significantly smaller than the others,
with an amplitude $\approx \SI{12}{MeV}/A^{1/2}$.  It is also smaller
than contributions arising from shell-correction energies (discussed
below), changing the \gls{rms} error $\chi_E$ by about at most
\SI{150}{keV}.  This fit is shown in \cref{table:liquid_drop} and the
residuals are displayed in \cref{fig:masses}.  The magnitudes of the
various terms are compared in \cref{fig:mass_contributions}, which
shows that the volume, surface, and Coulomb contributions are
dominant, while the symmetry energy contribution is roughly at the
level of 10\%.

There are several possible ways to determine the volume, surface,
symmetry, etc.\@ coefficients of \cref{eq:Bethe} or \cref{eq:masses}.
For example, one may turn off the Coulomb interaction, and extract
volume, surface, and symmetry energy from the asymptotic behavior of
the energy of nuclei with very large numbers of protons and
neutrons~\cite{Reinhard:2006}.  This corresponds to considering the
thermodynamic limit, which is not realized in real nuclei due to the
presence of the long-range Coulomb interaction among the protons.  We
prefer instead a unified approach, determining the parameters by
directly fitting almost all nuclear binding energies, whether
experimental or computed.  (See last two rows of
\cref{table:liquid_drop}.)

In a parallel development, properties of many-fermion systems were
understood in mathematical physics by tying together the roles of the
geometry and of the periodic trajectories in cavities.  As early as
1911, \textcite{Weyl:1911, Weyl:1912, Weyl:1912a, Weyl:1912b,
  Weyl:1913, Weyl:1915, Weyl:1950} and others related the wave
eigenstate density in boxes of various shapes and boundary conditions
to the geometrical shape of the box~\cite{Kac:1966, Waechter:1972,
  Baltes:1976, Brack:1997}.  In a manner similar to the nuclear mass
formula \cref{eq:Bethe}, this approach can be applied to saturating
systems, relating the ground state energy to the volume ($V$), surface
area ($A$), and mean curvature radius $R$ of the many-particle
system~\cite{Reinhard:2006}:
\begin{gather}\label{eq:Weyl}
  E = a_V V + a_S S + a_R R + \dots.
\end{gather}

The similarity between \cref{eq:Weyl} and the nuclear mass formula
\cref{eq:Bethe} becomes apparent after relating the volume to the
particle number $\n = A/V\approx \text{const}$.
The ground state energy can thus be rewritten in terms of particle number
$A$ (here for only one kind of particles)
\begin{gather}\label{eq:EA}
  E = b_V A + b_S A^{2/3} + b_R A^{1/3} + \dots.
\end{gather}
The coefficient $b_V$ is the energy per particle in infinite matter
and $a_S$ represents the surface tension.  These types of expansion
are classical in character: Planck's constant plays no explicit role.
Their accuracy for many-fermion systems is thus limited by the lack of
quantum effects (often referred to as shell effects).  It appears that
for nuclei, the mass formula \cref{eq:masses} is about as good as one
can achieve without introducing the quantum effects.

There is a long debate in literature, fueled mainly by studies of
quantum chaos, about whether an expansion in powers of $A$ can be
extended beyond the terms present in \cref{eq:EA}.  Naïvely, one might
expect the next terms to be proportional to $A^0$, $A^{-1/3}$, and so
forth, but a a more careful analysis shows that that is not correct.
(See for example \textcite{Brack:1997}.)  The next term is instead
proportional to $A^{1/6}$~\cite{Balian:1970, Balian:1971, Balian:1972,
  Strutinsky:1976}, arising from the contribution of periodic orbits.
Subsequent terms appear to be stochastic, due to the inherent chaotic
character of the interacting many-body systems~\cite{Bohigas:1993}.
It is well established by now that ideas originating from quantum
chaos and random matrices provide extremely useful tools to study
properties of neutron resonances, for example, in the region of
nuclear spectra where the level density is quite high.  Subsequent
works have shown~\cite{Johnson:1998} that even the properties of
ground states in many fermion systems are amenable to study using
similar ideas.  Thus it should not be surprising that small
contributions to the nuclear binding energies might be interpreted
using similar ideas.

\textcite{Gutzwiller:1971}, \textcite{Balian:1970, Balian:1971,
  Balian:1972}, and \textcite{Berry:1976, Berry:1977} observed that
quantum states in a finite system can be quite accurately reproduced
by quantizing the periodic classical trajectories.  (See also
\textcite{Brack:1997}.)  Combining the idea of geometric quantization,
with the Thomas-Fermi model, the Pauli principle, and copious
empirical evidence that strongly interacting fermionic systems share
many similarities with non-interacting systems~\cite{Haxel:1949,
  Mayer:1949, Mayer:1950, Mayer:1950a, Landau:1956, Landau:1957,
  Migdal:1967}, one can quite accurately construct the single-particle
density of states and binding energies as a function of the particle
number, eventually correcting this by the shape of the system.

The single-particle density of states $\n(\varepsilon)$ in a given
potential has a smooth and an oscillating components:
\begin{subequations} \label{eq:rho_eo}
  \begin{gather}
    \n(\varepsilon) = \n_{\text{TF}}(\varepsilon) + \n_{\text{osc}}(\varepsilon), 
    \label{eq:rho_e}\\
    \n_{\text{osc}}(\varepsilon)= \sum_{\text{PO}}a_{\text{PO}}(\varepsilon)
    \sin\left(
      \frac{S_{\text{PO}}(\varepsilon)}{\hbar}+\phi_{\text{PO}}\frac{\pi}{2}
    \right) +\dots, \label{eq:rho_osc}
  \end{gather}
\end{subequations}
where the sum is performed over classical periodic orbits (PO)
(diameter, triangles, squares, etc.).  Here,
$a_{\text{PO}}(\varepsilon)$ is the stability amplitude,
$S_{\text{PO}}(\varepsilon)$ is the action, and $\phi_{\text{PO}}$ is the
Maslov index of each orbit at the energy
$\varepsilon$~\cite{Balian:1970, Balian:1971, Balian:1972,
  Nishioka:1990, Brack:1997}.  The single-particle density of states
in the Thomas-Fermi approximation $\n_{\text{TF}}$~\cite{Weyl:1911,
  Weyl:1912, Weyl:1912a, Weyl:1912b, Weyl:1913, Weyl:1915, Weyl:1950,
  Baltes:1976, Kac:1966, Waechter:1972, Brack:1997} has a clear
dependence on the size and shape of the system, and leads to
\cref{eq:Weyl,eq:EA} for a square-well potential.  At the same time,
the nature of the periodic orbits also depends on the size and shape
of the single-particle potential.  Knowing $\n(\varepsilon)$, one can
calculate the particle number $A$ and shell-corrections (SC)
$E_{\text{SC}} = E - E_{\text{TF}}$ for a many-fermion system by
integrating up to the chemical potential $\mu$:
\begin{align}
  A &=\int_{-\infty}^\mu\n(\varepsilon) \d\varepsilon, 
  & E_{\text{SC}}= \int_{-\infty}^\mu
    \varepsilon \n_{\text{osc}}(\varepsilon)\d\varepsilon.
    \label{eq:shell}
\end{align}

The theory of periodic orbits and structure of these shell corrections
has been studied extensively.  For example, in a three-dimensional
spherical cavity, quantum effects can be reproduced by including only
triangular and square orbits~\cite{Balian:1970, Balian:1971,
  Balian:1972, Nishioka:1990, Brack:1997}.  The emergence of magic
numbers, and the role of the shapes of many-fermion systems have been
tested in theory and validated against experimental results in fermion
systems with up to \num{3000} electrons~\cite{Heer:1993, Brack:1993,
  Pedersen:1991}.  In particular, in atomic clusters, the emergence of
super-shells has been predicted theoretically~\cite{Nishioka:1990,
  Brack:1993, Bulgac:1993} and confirmed
experimentally~\cite{Heer:1993, Pedersen:1991}.  (Nuclei are too small
to exhibit of super-shells.)

In nuclear physics, a similar line of inquiry is encapsulated in the
method of shell-corrections, developed by
Strutinsky~\cite{Strutinsky:1966, Strutinsky:1967, Strutinsky:1968}
and many others~\cite{Strutinsky:1976, Brack:1972, Brack:1985x,
  Myers:1966, Myers:1969, Myers:1974, Myers:1990, Myers:1991,
  Myers:1996, Ring:2004, Bohr:1998, Moller:1995, Moller:2012,
  Moller:2016}.  This method shows that $\n(\varepsilon)$ has a well
defined dependence on the particle number.  The smooth part of the
density of states is quite well described by the Thomas-Fermi
approximation (and by the smoothing procedure introduced by
Strutinsky).  The leading terms are the volume ($\sim A$), surface
($\sim A^{2/3}$), Coulomb ($\sim Z^2/A^{1/3}$), and symmetry energy
[$\sim (N-Z)^2/A$] contributions encoded in the Bethe-Weizsäcker mass
formula~\eqref{eq:Bethe}.  The oscillating part is dominated by the
nuclear shape and the shell effects from the periodic orbits, where
the amplitude depends on the particle number as
$A^{1/6}$~\cite{Strutinsky:1976}.

The separation of $\n(\varepsilon)$ into the smooth and oscillating
parts~\eqref{eq:rho_e} is a general characteristic of the many fermion
systems.  Both the macroscopic-microscopic
method~\cite{Strutinsky:1966, Strutinsky:1967, Strutinsky:1968,
  Strutinsky:1976, Brack:1972, Brack:1985x, Myers:1966, Myers:1969,
  Myers:1974, Myers:1990, Myers:1991, Myers:1996, Ring:2004,
  Bohr:1998, Moller:1995, Moller:2012, Moller:2016} and
self-consistent approaches~\cite{Bender:2003, Goriely:2009,
  Delaroche:2010, Goriely:2013, Goriely:2014, Kortelainen:2010,
  Kortelainen:2012, Kortelainen:2014_2} lead to the same conclusions
about the various contributions described above, and agree with
experimental data~\cite{Lunney:2003}.  In all previous considerations
of mass tables, either in self-consistent approaches or in
microscopic-macroscopic models, the single-particle spectroscopic
factors are modified only by pairing correlations.  It is well known,
however, that the coupling between collective degrees of freedom and
single-particle degrees of freedom lead to a significant fragmentation
of the single-particle occupation probabilities, which are measured in
pick-up and knock-out reactions~\cite{Bohr:1969, Ring:2004}.  This
fragmentation of the single-particle occupation probabilities is not
taken into account in the single-particle density of states
Eqs.~\eqref{eq:rho_eo} or in the definition of the single-particle
densities Eqs.~\eqref{eq:KS_densities}, and is likely to affect the
exact magnitude of the shell-effects.  The order of magnitude of these
effects is perhaps a (small) fraction of the \gls{rms} error
$\chi_E=\SI{3.3}{MeV}$ of the Bethe-Weizs\"{a}cker mass formula
\eqref{eq:Bethe}.  All of this begs the question: To what order
can one expand the density of states in powers of the particle numbers
and periodic orbits?

There is a reasonable consensus that, beyond the leading contributions
from the periodic orbits and shell-corrections, any such expansion
fails due to the effects of quantum chaos -- i.e\@., contributions from
classically chaotic trajectories through the many-body phase
space~\cite{Bohigas:1993}.  Stable periodic orbits provide the
strongest shell effects in quantum systems, evidenced by the
appearance of magic numbers (see e.g., \cref{fig:masses}).  Unstable periodic orbits
also produce shell effects, but with smaller weights.  In
contrast, chaotic orbits appear to produce irregular oscillations in
the single-particle density of states with a rather small amplitude.
Various estimates suggest that chaotic fluctuations appear at the
level of \SI{0.5}{MeV} per nucleus~\cite{Bohigas:2002, *Bohigas:2002E,
  Aberg:2002, Olofsson:2006, Olofsson:2008, Molinari:2004,
  Molinari:2006, Hirsch:2004, Hirsch:2005, Barea:2005}, noticeably
smaller than shell effects contributions due to periodic orbits and
deformations, which are of the order of several \si{MeV}s.

The effect of periodic orbits is not limited to finite systems: the
Casimir energy in quantum field theory~\cite{Casimir:1948,
  Klimchitskaya:2009}, critical phenomena~\cite{Fisher:1978,
  Hanke:1998}, and strongly interacting infinite inhomogeneous
systems, e.g\@., nuclear pasta phase in neutron
stars~\cite{Bulgac:2001y, *Bulgac:2002x, Magierski:2003,
  Magierski:2004x, Magierski:2002a, Bulgac:2005x, Yu:2000,
  Bulgac:2006x}, can also be explained and calculated to high
precision by evaluating the contributions from periodic orbits.  This
method has become the standard approach for evaluating the Casimir
energy in a variety of fields~\cite{Bordag:2010, Rahi:2009,
  Graham:2014, Canaguier-Durand:2010, Schaden:2010}.

It is somewhat surprising that shell effects from periodic orbits
appear at the same level as deformation effects in the energy of
nuclear systems.  Naïvely one might expect the deformation energy to
be controlled by the surface area of a saturating system, and thus to
contribute as a correction to the surface term in nuclear mass
formulas like Eqs.~\eqref{eq:Bethe} and~\eqref{eq:masses}.  However,
the deformation energy in nuclei has a quantum nature, and is
determined by a delicate interplay between the change in surface area
and the shell effects.  A similar behavior has been observed in the
case of atomic clusters with up to 3000 electrons~\cite{Bulgac:1993}.
This leads to a leveling of the peaks, which one would otherwise
expect in the absence of deformation, leaving in place only the large
negative shell-corrections for the magic spherical systems, as seen in
\cref{fig:masses} for the case of nuclei.

The shape stability of a many-fermion system is controlled by the
single-particle level density at the Fermi level.  In an open-shell
system this level density is high; the system can thus deform quite
easily and single-particle levels can rearrange until the level
density is low enough to render the system stable.  The stabilization
process of the nuclear deformation in the ground state is analogous to
the Jahn-Teller effect in polyatomic molecules~\cite{Jahn:1937}, where
the high degeneracy of the ground state is lifted by the deformation
of the system.  This mechanism leads to new ``magic numbers'' in
deformed systems as Strutinsky discussed in his seminal
papers~\cite{Strutinsky:1966, Strutinsky:1967, Strutinsky:1968}.  The
increase in surface area and the energy penalty incurred (deformation
energy) is canceled to a large extent by the shell-corrections (due to
periodic orbits in the deformed potential), unless the system is
``magic'' or ``semi magic''.  The cancellation between deformation
energy and shell effects suggests that open-shell systems should be
easier to deform than magic systems.  This is consistent with the
character of the residuals remaining after fitting the nuclear binding
energies with Bethe-Weizsäcker formulas like Eqs.~\eqref{eq:Bethe} and
\eqref{eq:masses} as shown in \cref{fig:masses} and \cref{fig:be_zn}.
The largest residuals appear as large (negative) spikes at the shell
closures for spherical nuclei with magic numbers of either protons
and/or neutrons, while the expected (positive) peaks in between magic
numbers are flattened.  From the nature of the residuals
$E_{\text{exp}} - E_{\text{th}}$ in \cref{fig:masses} -- sharp
negative spikes at the magic numbers, but roughly constant
fluctuations in between -- one can conclude that mass formulas of the
type in \cref{eq:masses} do encode the role of the nuclear deformation.
For open shell nuclei it thus appears that the deformation energy is
roughly compensated by the shell-correction energy, and shell effects
only survive near magic and semi-magic nuclei.

A number of corrective terms might be considered to improve the
accuracy of the nuclear mass formulas Eqs.~\eqref{eq:Bethe}
and~\eqref{eq:masses}.  For example, in the Coulomb term, one might
replace $Z^2$ with $Z(Z-1)$ to correctly count the number of proton
pairs, and one might add an additional term proportional to $Z$ to
account for the Coulomb exchange interaction and
screening~\cite{Bulgac:1996}.  Motivated by \cref{eq:EA}, one might
also consider including terms proportional to $A^{1/3}$ and $A^0$.
The symmetry energy terms might also be ``corrected'' by replacing
$(N-Z)^2/4$ with $T(T+1)$ where $T=\abs{N-Z}/2$.  Finally, one might
introduce an additional correction to account for the Wigner energy
$\propto \abs{N-Z}$, which appears as a cusp in the nuclear binding
energies as a function of $N-Z$ (basically only for nuclei with small values of
$\abs{N-Z}$)~\cite{Moeller:1997}.  However, including these
corrections lead to very small improvements in the energy \gls{rms} $\chi_E$
below the value \SI{2.64}{MeV} obtained with the main terms of
\cref{eq:masses}.  All these corrections are eclipsed by the shell effects
as seen in \cref{fig:masses}.

There are a variety of many-body approaches based on the
Schr\"{o}dinger equation: the \gls{QMC} method~\cite{Carlson:2015,
  Pederiva:2017}, the self-consistent Green's function
method~\cite{Barbieri:2017}, the coupled-cluster
method~\cite{Lietz:2017}, and the in-medium renormalization
method~\cite{Gebrerufael:2017}.  In all these approaches one has to
specify the two-body (NN), three-body (NNN), etc.\@, interactions
between nucleons, the form of which is ambiguous and depends on how the
theory is regularized.  Chiral \gls{EFT}~\cite{Epelbaum:2011,
  Machleidt:2011} provides a framework for organizing these
interactions using the symmetries of the underlying theory \gls{QCD}
of quarks and gluons with the hope that physical results are
independent of the energy cutoff.  In general, there is still no guarantee,
however, that this many-body expansion converges quickly enough using 
a naïve sum of diagrams~\cite{Kozik:2010, Rossi:2017}.

The \gls{DFT} approach differs from approaches based on the
Schr\"{o}dinger equation.  For many-electron systems, it has been
established that there is a mathematical one-to-one correspondence
between the number density and the wavefunction of a many-body
system~\cite{HK:1964, Dreizler:1990lr}, and this one-to-one
correspondence leads to the existence of an exact energy density
functional.  In practice, however, this functional is extremely
complicated and establishing a useful form is more of an art than a
science.  One particularly successful example is the \gls{UFG}, which
shares many properties with dilute neutron matter, and is also a
superfluid with a large pairing gap~\cite{CCPS:2003}.  In this case, the form of a
local energy density functional follows using only dimensional
arguments, renormalizability of the theory, Galilean invariance, and
symmetries.  The functional and the corresponding framework needed to
treat fermionic superfluids is called the \gls{SLDA} (extending the
\gls{LDA} acronym of Kohn and Sham~\cite{Kohn:1965fk}), and has been
verified and validated against both \gls{QMC} calculations and
experiments at the few percent level for a wide range of
systems~\cite{Bulgac:2013b, Bulgac:2016}.  Our approach here is
motivated by similar considerations, leading to a simple and compact
functional in which time-dependent phenomena can be treated easily as well.  Thus,
unlike approaches based on the Schr\"{o}dinger equation, which are
primarily limited to static properties, the \gls{DFT} can be applied
to reactions, fission, time-dependent non-equilibrium phenomena, and
for very heavy systems with remarkable accuracy.
 
\glsreset{NEDF}
\glsreset{UFG}

\section{Form of the Functional}\label{sec:functional-form-nedf}

The lesson from our brief historical review is that, since nuclei are
saturating systems with a rather well defined saturation density, the
bulk of the nuclear binding energy should be fixed by the geometry of
the nuclei (volume, surface area, curvature radius) to sub-percent
accuracy.  As demonstrated in \cref{table:liquid_drop}, the accuracy
of the mass formulas Eqs.~\eqref{eq:Bethe} and~\eqref{eq:masses} --
which both lack shell effects, deformation, spin-orbit effects,
pairing, etc\@. -- suggests that such a \gls{NEDF} should be capable
of describing at a similar level of accuracy both the nuclear binding
energies, and the proton and neutron matter density distribution.
Therefore, we might reasonably expect that a \gls{NEDF} will also
describe the nuclear charge radii, for which there is a large amount
of accumulated data~\cite{Angeli:2013}.  Shell effects, pairing correlations, 
and beyond mean-field corrections, enter at the
level of a few \si{MeV}s per nucleus, reducing the \gls{rms} energy error
$\chi_E$ from around \SI{3}{MeV} to about
\SI{0.5}{MeV}~\cite{Moller:1995, Moller:2012, Moller:2016}, and are
most pronounced for magic or semi-magic nuclei, see \cref{fig:masses}.
 
\begin{subequations}
  We will describe a \gls{NEDF} that depends on the smallest number of
  phenomenological parameters needed to account for all the
  contributions in the nuclear mass formulas Eqs.~\eqref{eq:Bethe} and
  \eqref{eq:masses}.  First we relate these parameters to various
  physical quantities relevant for nuclear physics.  For a large
  nucleus, the Coulomb energy can be used to estimate the saturation
  density $\n_0$ by approximating the nucleus as a uniformly charged
  sphere with $E_C = 3Z^2e^2/5R =a_CZ^2/A^{1/3}$, where $R=r_0A^{1/3}$
  and $r_0\approx \SI{1.2}{fm}$ is a nuclear length scale:
  \begin{gather}
    \n_0 =\frac{3}{4\pi r_0^3}, \quad \text{where} \quad r_0 = \frac{3e^2}{5a_C}.
  \end{gather}
  One can further estimate the ground-state energy of infinite nuclear
  matter per nucleon $\varepsilon_0$, the nuclear surface tension
  $\sigma$, and their dependence on the isospin $(N-Z)/2$:
  \begin{align}
    \varepsilon_0 &= \frac{E(N, Z)}{A} = a_v + a_I\frac{(N-Z)^2}{A^2}, \\
    \sigma &= a_s + a'_I\frac{(N-Z)^2}{A^2}.
  \end{align}
  Finally, one can relate the value of the coefficient $a'_C$ (or of
  the alternative coefficient of the contribution $a''_CZ^2/A$ to the
  mass formula~\cite{Myers:1969}) with the nuclear surface
  diffuseness.
\end{subequations}

For a \gls{NEDF} to be as accurate as the mass formula, one expects no more than
five or six significant parameters.  As we shall see, such a functional does
exist, requiring as few as four parameters, and demonstrating better accuracy
than the original Bethe-Weizs\"{a}cker mass formula, with the additional
property of predicting charge radii.  That a functional depending on such a
small number of phenomenological parameters can go beyond the capabilities of
the empirical mass formula and also describe density distributions is truly
remarkable.

We postulate a \gls{NEDF} with three main contributions, which
significantly improves on the Weizsäcker's original
idea~\cite{Weizsacker:1935}:
\newcommand\overcomment[2]{\overbrace{#1}^{\clap{\small\text{#2}}}}
\newcommand\undercomment[2]{\underbrace{#1}_{\clap{\small\text{#2}}}}
\begin{gather}
  \label{eq:NEDF}
  \mathcal{E}[\n_n, \n_p] =
  \overcomment{\Ekin}{kinetic}
  + \undercomment{\Ec}{Coulomb}
  + \overcomment{\Eint}{interactions}.
\end{gather}
The first two terms -- the kinetic energy and Coulomb energy -- are
well motivated and have no free parameters.  All phenomenological
parameters of the model appear in the interaction term
$\mathcal{E}_{\text{int}}$:
\begin{gather}
  \label{eq:NEDF_int}
    \Eint =
    \overcomment{\Ehomo}{homogeneous}
    + \undercomment{\Egrad}{gradients}
    + \overcomment{\Eso}{spin-orbit}
    + \undercomment{\Epair}{pairing}
    + \overcomment{\Eentrain}{entrainment}, 
\end{gather}
The Kohn-Sham formulation of the functional is specified in terms of
the single-particle orbitals
$v_{k\sigma}(\vect{r}), v_{k\sigma}(\vect{r})$ through the time-even number,
anomalous, kinetic, and spin-current densites (for both neutrons and protons), 
\begin{subequations}
  \label{eq:KS_densities}
  \begin{align}
    \n(\vect{r})
    &= \sum_{k, \sigma}v^*_{k\sigma}(\vect{r})v_{k\sigma}(\vect{r}), 
      %\tag{number}
      \\
    \nu(\vect{r})
    &= \sum_{k}v^*_{k\uparrow}(\vect{r})u_{k\downarrow}(\vect{r}), 
      %\tag{anomalous}
      \\
    \tau(\vect{r})
    &= \sum_{k, \sigma}\grad{v}^*_{k\sigma}(\vect{r})\cdot \grad{v}_{k\sigma}(\vect{r}), 
      %\tag{kinetic}
       \\
    \vect{J}(\vect{r})
    &= \left.\frac{\grad - \grad'}{2\I}
      \times
 \sum_{k, \sigma, \sigma'}  v^*_{k\sigma}(\vect{r}) \vect{\sigma}_{\sigma, \sigma'} v_{k\sigma'}(\vect{r}')\right |_{ \vect{r}=\vect{r}' }. ,  
      \label{eq:spin_curr}
      %\tag{spin current}
  \end{align}
  as well as the time-odd spin-density and current (which are non-vanishing 
  if time-reversal symmetry is broken)
  \begin{align}
    \vect{s}(\vect{r})
     &=  \sum_{k, \sigma, \sigma'} v^*_{k\sigma}(\vect{r}) \vect{\sigma}_{\sigma, \sigma'} v_{k\sigma'}(\vect{r}),      \label{eq:spin_dens}
        %\tag{spin-density}
      \\       
    \vect{j}(\vect{r})
    &= \sum_{k, \sigma}\left.\frac{\grad-\grad'}{2\I}v^*_{k\sigma}(\vect{r}')v_{k\sigma}(\vect{r})\right |_{ \vect{r}=\vect{r}' },
    \label{eq:curr}
      %\tag{current}
       \end{align}
\end{subequations}
see Refs.~\cite{Bender:2003, Jin:2017} and references therein for details.
[Note: In nuclear physics literature proton and neutron number densities are
typically denoted with the symbols $\rho_{n, p}(\vect{r})$.  In accordance with
the wider physics literature, we reserve $\rho$ for mass densities, which are
related to number densities by $\rho_{n, p}(\vect{r}) = m \n_{n, p}(\vect{r})$.]

Developing an orbital-free version of \eqref{eq:NEDF_int} would
require expressing all the various terms exclusively in terms of the
number density $\n(\vect{r})$.  Whether such a \gls{NEDF} exists and
how it should be implemented remains an open question.  In this work,
we will implement an orbital-free functional by approximating all the
auxiliary densities \eqref{eq:KS_densities} as 
functions of the number
density; see \cref{sec:orbit-free-functional} for details.

\subsection{Kinetic Terms}\label{sec:kinetic-terms}

The kinetic energy density derives from the energy density of a
non-interacting system of protons and neutrons and contains no free
parameters:
\begin{gather}
  \label{eq:NEDF_kin}
  \mathcal{E}_{\text{kin}}
  = \frac{\hbar^2}{2m} (\tau_n + \tau_p) - \frac{\delta m}{2m} \frac{\hbar^2}{2m} (\tau_n - \tau_p)
  +\order\left(\frac{\delta m}{2m}\right)^2, 
  \end{gather}
  where $\tau_{n, p}$ are the kinetic densities in the \gls{HFB}
  formulation with neutron and proton $m_{n, p} = m \pm \delta m/2$.
  In principle, one should include an explicit isospin splitting due
  to the different proton and neutron masses, but we follow here
  common practice in nuclear theory to use a common average mass
  $m = (m_n+m_p)/2$ and neglect $\delta m = m_n-m_p$.  Note that since
  we are using the bare masses here, the theory is covariant under
  Galilean boosts.  The consideration of terms with a more complex
  dependence on the kinetic energy densities requires adding
  current terms to restore the Galilean covariance of the theory (see
  e.g\@., Refs.~\cite{Engel:1975, Dobaczewski:1995, Nesterenko:2008,
    Bender:2003, Bulgac:2011}.)

\subsection{Coulomb Terms}\label{sec:coulomb-terms}

\begin{subequations}
  \label{eq:NEDF_C}
  The direct Coulomb energy and exchange contribution in the Slater
  approximation are:
  \begin{align}
    \mathcal{E}_{C}(\vect{r})
    &=
      \frac{1}{2} V_C(\vect{r})\n_{\text{ch}}(\vect{r}) - \frac{e^2\pi}{4} \left(
      \frac{3\n_p(\vect{r})}{\pi}\right)^{4/3}, 
      \label{eq:Slater}\\
    V_C(\vect{r})
    &=e^2 \int \d^3\vect{r}'\; \frac{\n_{\text{ch}}(\vect{r}')}
      {\norm{\vect{r} - \vect{r}'}}, 
  \end{align}
  where $e$ is the proton charge and $\n_{\text{ch}}$ is the charge
  density, which is obtained from the proton and neutron densities by
  convolution (here noted with an asterisk, ``$*$'') with the appropriate charge form
  factors (see \cref{app:charge-form-factors} for details):
  \begin{gather}
    \label{eq:FF}
    \n_{\text{ch}} = G^n_E * \n_n + G^p_E * \n_p.
  \end{gather}
  Including the form factors does not significantly improve the mass
  fits, but improves somewhat the fit of the charge radii.  In
  principle, one might allow the coefficient of the Coulomb exchange
  term to vary; this is done, for example, in atomic physics in order to obtain
  better estimates of the Coulomb exchange energy.  We find, however,
  that fitting the nuclear binding energies leads with high accuracy
  to the same coefficient presented in \cref{eq:Slater}, so we leave
  it fixed and do not include this as a parameter in our model.
\end{subequations}

We require our energy density functional to be an isoscalar and
include no isospin breaking terms other than those due to the
neutron-proton mass difference (which we neglect here) and the Coulomb
interaction.  Additional isospin violation due to up and down quark
mass differences and electromagnetic effects~\cite{Miller:1990iz,
  Miller:2006tv, Muther:1999, Machleidt:2001, Meisner:2008} beyond
these two contributions are much smaller and are partly responsible
for the Nolen-Schiffer anomaly~\cite{Nolen:1969}, to which the
screening of the Coulomb exchange also contributes at a comparable
level~\cite{Bulgac:1996, Bulgac:1999}.

\subsection{Homogeneous Terms: Infinite Nuclear and Neutron Matter}\label{sec:homogeneous-terms}

We parameterize the nuclear \gls{EoS} as:
\begin{subequations}
  \label{eq:NEDF_homo}
  \begin{gather}
    \Ehomo = \sum_{j=0}^2\mathcal{E}_j(\n)\beta^{2j}\\
    \mathcal{E}_j(\n) = \varepsilon_j(\n)\n= a_j \n^{5/3} + b_j \n^2 + c_j \n^{7/3}, 
  \end{gather}
  where $\n$ is the total density, and $\beta$ is the asymmetry:
  \begin{gather}
    \n = \n_n + \n_p, \qquad
    \beta = \frac{\n_n - \n_p}{\n_n + \n_p}.
  \end{gather}
\end{subequations}
We have considered terms with powers of the density
$\n^{8/3}\sim \n\tau$ and higher, but in all our fits of the nuclear
masses, we found such terms to be unconstrained in magnitude, barely
improving the quality of the fits.

In infinite homogeneous nuclear matter, as might be found in a neutron
star for example, the gradient, spin-orbit, entrainment, and Coulomb
terms vanish (charge neutrality is maintained by a background of
electrons).  The semiclassical expansion of the kinetic energy density
$\Ekin$ becomes exact in the leading Thomas Fermi term
$\tau = \tau_{TF}$.  Thus, neglecting the small neutron-proton mass
difference $m_n \approx m_p \approx m$, the functional acquires the
simple form:
\begin{figure}[tb]
  \includegraphics[width=\columnwidth]{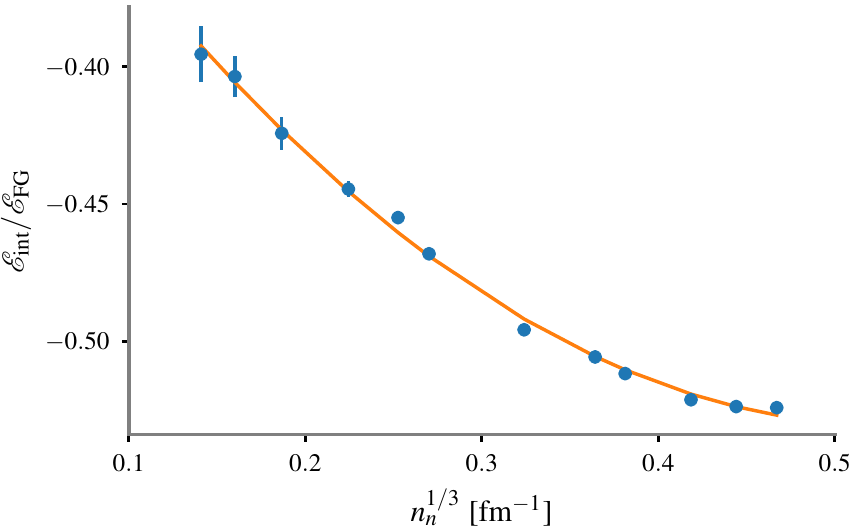}
  \caption{(Color online) The \gls{QMC} results of
    \textcite{Wlazlowski:2014a} for the interaction energy per neutron
    displayed as the ratio $\mathcal{E}_{\text{int}}/\mathcal{E}_{\text{FG}}$
    defined in \cref{eq:nm} (with $\beta = 1$), where
    $\mathcal{E}_{\text{FG}} = 3 \hbar^2 (3\pi^2\n_n)^{2/3}\n_n/(10m_n)$.  If
    $a_n=0$ in \cref{eq:nm}, the ratio
    $\mathcal{E}_{\text{int}}/\mathcal{E}_{\text{FG}}$ would tend to 0 for
    $\n_n\rightarrow 0$.  For densities $\n_n^{1/3}\abs{a_{nn}} < 1$
    (where $a_{nn}=-\SI{18.9}{fm}$ is the $s$-wave neutron-neutron
    scattering length), the leading order correction to the kinetic
    energy density per particle contribution would be instead linear
    in density $4\pi\hbar^2 a_{nn}\n_n/m_n$.}
  \label{fig:QMC}
\end{figure}
\begin{multline}
  \label{eq:inm}
  \mathcal{E}(\n_n, \n_p) =
  \frac{3\hbar^2(3\pi^2)^{2/3}}{10m}(\n_n^{5/3} + \n_p^{5/3})\\
  +
  \sum_{j=0}^{2} \left(a_j\n^{5/3} + b_j\n^2 + c_j
    \n^{7/3}\right)\beta^{2j}, 
\end{multline}
This portion of the functional is essentially an expansion in powers
of the Fermi momenta $k_F$: $k_{n, p} = (3\pi^2 \n_{n, p})^{1/3}$ with
only three terms $k_F^{5}$, $k_F^6$, and $k_F^{7}$.  This type of
expansion is ubiquitous in many-body perturbation theory, and also
applies to fitting the neutron matter \gls{EoS} with high accuracy
($\n_p=0$, $\beta=1$):
\begin{subequations}
  \begin{align}
    \mathcal{E}_n(\n_n)
    &=
      \frac{3\hbar^2}{10m_n} (3\pi^2\n_n)^{2/3}\n_n+\mathcal{E}_\text{int}(\n_n), \label{eq:NM}\\
    \mathcal{E}_\text{int}(\n_n)
    &= a_n\n_n^{5/3}
      + b_n\n_n^2 + c_n \n_n^{7/3}, \label{eq:nm}
  \end{align}
\end{subequations}
The coefficients $a_n$, $b_n$, and $c_n$ are fixed by fitting the
neutron matter \gls{EoS} as calculated with \gls{QMC} including up to
N$^3$LO two-body and up to N$^2$LO three-body interactions from chiral
perturbation theory~\cite{Wlazlowski:2014a}:
\begin{align}
  \label{eq:neut_mat}
  \begin{split}
    a_n &= a_0 + a_1 + a_2 =  \SI{-32.6}{MeV.fm^2}, \\
    b_n &= b_0 + b_1 + b_2 = \SI{-115.4}{MeV.fm^3}, \\
    c_n &= c_0 + c_1 + c_2 =  \SI{109.1}{MeV.fm^4}.
  \end{split}
\end{align}
As seen from \cref{fig:QMC}, all three terms (but no more) are needed
in \cref{eq:nm} for an accurate reproduction of the neutron \gls{EoS}
(see also \cref{sec:satur-symm-prop}).  When we include the $j=2$
quartic terms in \cref{eq:inm} the values of $a_2$, $b_2$, and $c_2$
are determined from the values of $a_n$, $b_n$, and $c_n$ describing
the \gls{QMC} results \eqref{eq:neut_mat}, without adding additional
free parameters to the \gls{NEDF}.\footnote{%
  We have also performed a fully self-consistent mass fit with
  additional powers of densities
  $\sum_{j=0, 1} (a_j\n^{5/3} + b_j\n^2 +
  c_j\n^{7/3}+d_j\n^{8/3})\beta^{2j}$.  While this kind of fit leads
  to a lower energy \gls{rms} $\chi_E\approx \SI{1.2}{MeV}$, the charge
  radii \gls{rms} increases to $\chi_r\approx \SI{0.1}{fm}$ and the value of
  the incompressibility $K_0\approx \SI{170}{MeV}$ is very low.
  Typically in these cases the parameter $a_0$ becomes significant and
  acquires relatively large negative values, similar to the behavior
  seen in \cref{fig:hfbfit_a0}.  See also the discussion in
  \cref{subsec:seall1-param}.  }

\begin{figure}[tb]
  \includegraphics[width=\columnwidth]{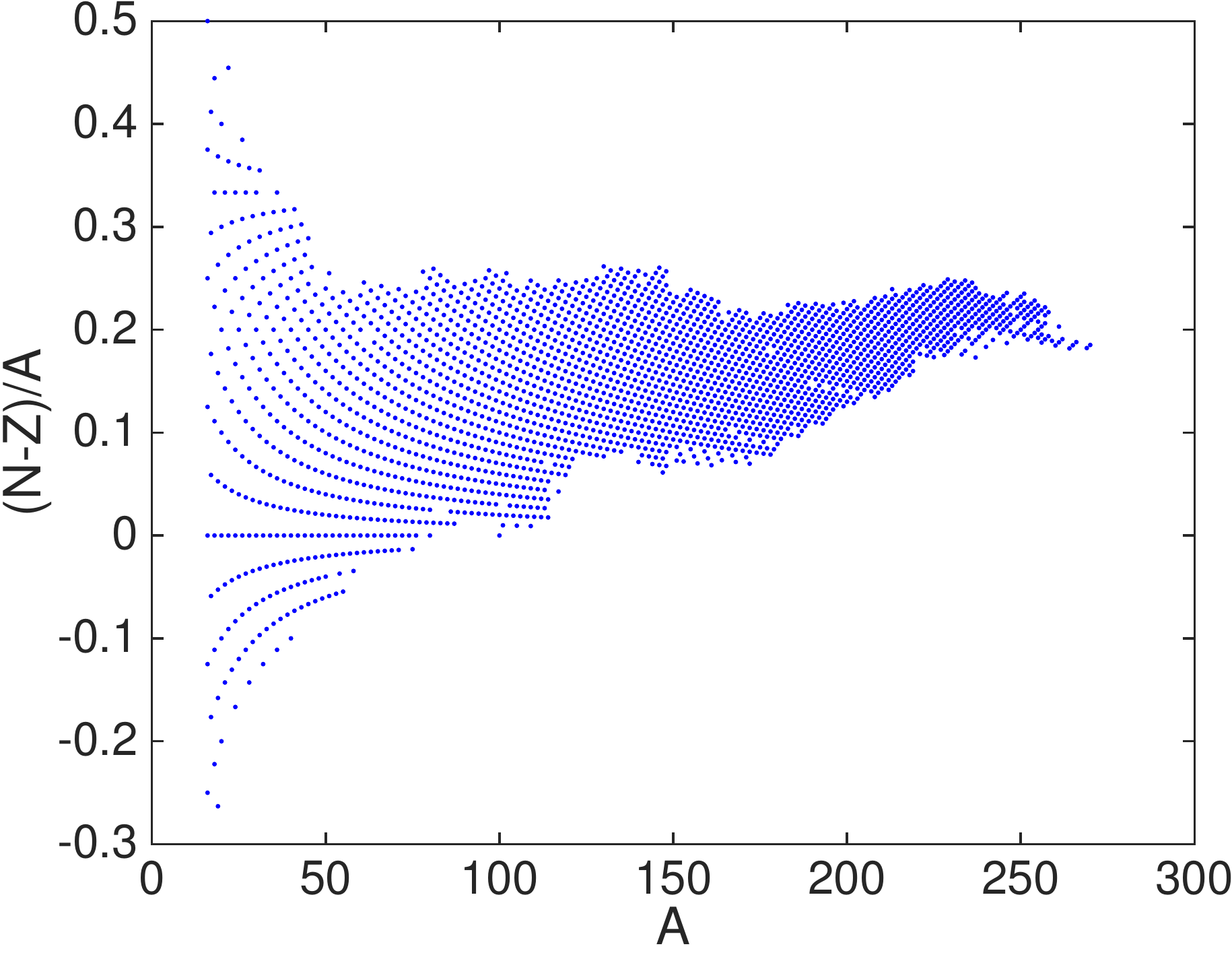}
  \caption{(Color online) The contribution to the ground state energies of
    the terms quartic in isospin density
    $\delta E_{I4}=\int \d^3\vect{r}\; \mathcal{E}_2(\n)\beta^4$,
    evaluated perturbatively with NEDF-1, see
    \cref{table:NEDF_A_supp}.  In the lower panel we display the ratio
    $(N-Z)/A$ for the nuclei we have considered.  Among the 2375
    nuclei we have considered, there are 33 nuclei with $N=Z$, 78
    nuclei with $Z>N$, and 70 nuclei with $\abs{N-Z}/A > 1/4$.}
  \label{fig:I4}
\end{figure}

The contribution of quartic terms to nuclear masses is small (typically less 
than \SI{1}{MeV}) since in most nuclei $\beta < 0.25$, see \cref{fig:I4} and
\cref{subsec:seall1-param}. However, the best fit functional with only 
quadratic $\beta^2$ $(j=1)$ terms, does not reproduce the neutron matter 
\gls{EoS}, especially near $\n \approx \SI{0.1}{fm^{-3}}$ (see \cref{fig:INM}). 
Quartic terms are thus needed to reproduce the  neutron matter \gls{EoS}, but 
are not constrained by nuclear binding energies. Therefore, they provide a 
direct (and independent) way to incorporate the \gls{EoS} of neutron matter 
into the \gls{NEDF}.

At this time we do not have an equally accurate \gls{QMC} calculation
of nuclear matter with varying isospin composition, so we must rely
instead on a phenomenological approach.  Our main assumption is that
we can describe both the isoscalar ($j=0$, $\beta^0$) and isovector
($j=1$, $\beta^2$) parts of the nuclear \gls{EoS} using the same three
powers of Fermi momenta Eqs.~\eqref{eq:nm} and \eqref{eq:neut_mat} as
required to fit the \gls{EoS} of pure neutron matter.  This approach
differs from typical Skyrme-like parameterizations, which include
terms with higher powers of densities, e.g\@. $\n^{8/3}$ arising from
$\tau\n$ type of terms, where $\tau$ is kinetic energy density.

One could in principle consider additional terms of the type
$\tau \n^{1/3}\propto \n^{2}$, $\tau \n^{2/3}\propto \n^{7/3}$, and
$\tau \n\propto \n^{8/3}$, but the contribution to the bulk energy of
such terms would be practically indistinguishable from terms $\n^2$,
$\n^{7/3}$, and $\n^{8/3}$.  Their contribution might become important
only in the surface region, and since
\begin{subequations}
  \begin{align} \label{eq:tau_n}
    \tau\n^{1/3}-\frac{3}{5}(3\pi^2)^{2/3}\n^2     &\propto \frac{\norm{\grad\n}^2 }{\n^{2/3}}, \\
    \tau\n^{2/3}-\frac{3}{5}(3\pi^2)^{2/3}\n^{7/3} &\propto  \frac{\norm{\grad\n}^2 }{\n^{1/3}}, \\
    \tau\n      -\frac{3}{5}(3\pi^2)^{2/3}\n^{8/3} &\propto  \norm{\grad\n}^2, 
  \end{align}
\end{subequations}
most of these terms could be incorporated effectively in gradient
corrections (see \cref{sec:gradients,sec:orbit-free-functional}).

The terms $a_j\n^{5/3}$ are somewhat unexpected and are not included
in Skyrme-like parameterizations.  \textcite{Tondeur:1978} introduced
only a term $a_1$ (without theoretical justification), but it makes
sense to include the other $a_j$ for several reasons.  First, the
\gls{QMC} calculations of \textcite{Gezerlis:2010a, Wlazlowski:2014a,
  Gandolfi:2015} (see \cref{fig:QMC}) are consistent with the
existence of a non-vanishing parameter $a_n$ in the neutron \gls{EoS},
which implies that $a_n=\sum_{j=0}^2 a_j\neq 0$.  Then, these terms
also appear naturally in the case of the \gls{UFG}~\cite{Zwerger:2011}, 
which has been confirmed to high precision in
many experiments.  The \gls{UFG} is a system of two species of
fermions, interacting with an $s$-wave interaction with zero range and
infinite scattering length.  In response to the Many-Body X challenge
posed by Bertsch in 1999, \textcite{Baker:1999} showed that the system
was stable.  The energy density of the \gls{UFG} scales exactly like
the kinetic energy density of a free Fermi gas
$\mathcal{E} \propto\n^{5/3}$.  Since both neutron and protons have
similar $s$-wave interaction properties, one expects the nuclear
energy density to behave somewhat like the unitary Fermi
gas at low densities.\footnote{%
  Subsequent to our introduction of terms $\propto \n^{5/3}$ in
  Ref.~\cite{Bulgac:2015}, Reinhard~\cite{Reinhard:2016} also
  considered these, but with a strength corresponding to a pure
  \gls{UFG}, which is quantitatively very different from neutron
  matter.  His conclusions, that the properties of the low-density
  neutron matter cannot be incorporated into the \gls{NEDF}, differ
  from ours.}

Although the energy density of the \gls{UFG} scales as the kinetic
energy, this is not necessarily due to a mass renormalization as one
might naïvely suspect.  \Gls{QMC} calculations of the single
quasi-particle dispersion~\cite{Carlson:2005kg} and spectral weight
function~\cite{Magierski:2009, Magierski:2011} both arive at the
conclusion that the effective mass in the \gls{UFG} is close to the
bare mass $\approx m$.  However, this does not preclude the
interpretation that some part of the energy arises from the kinetic
energy density $\tau$ (if $m_{\text{eff}}\neq m$) as is the case in the
\gls{UFG}~\cite{Bulgac:2007a, Bulgac:2011, Bulgac:2013b}.  The
\gls{QMC} calculations are simply not yet of sufficient accuracy to
confirm or exclude an effective mass different from unity.

\subsection{Gradient terms}\label{sec:gradients}

We include a gradient term of the following form, similar to terms
considered in Skyrme \glspl{NEDF}~\cite{Chabanat:1998}:
\begin{gather}
  \label{eq:NEDF_grad}
  \Egrad = \eta_s \sum_{q=n, p} \frac{\hbar^2}{2m}\norm{\grad\n_{q}}^2.
\end{gather}
One might consider a more general term of the form
\begin{gather}
  \label{eq:NEDF_grad1}
  \Egrad = \eta_0 \frac{\hbar^2}{2m} \norm{\grad\n_n+\grad\n_p}^2
              + \eta_1 \frac{\hbar^2}{2m} \norm{\grad\n_n-\grad\n_p}^2 .
\end{gather}
Note that this form of gradient term alone in an orbital-free theory
leads to unphysical density profiles with a discontinuity in $\grad\n$
at a finite radius, beyond which the density vanishes exactly.
However, in the presence of $\mathcal{E}_{\text{kin}}$ in an
orbital-based approach the density is well behaved.  We have found
that the nuclear mass fits are basically insensitive to the linear
combination $\eta_m=\eta_0-\eta_1$, and we use
$\eta_s=(\eta_0+\eta_1)/2$ and $\eta_m=(\eta_0-\eta_1)/2=0$.  The
linear combination $\eta_m=(\eta_0-\eta_1)/2$ can instead be used to
independently fit the static isovector dipole polarizability of
nuclei, as it favors a small separation between the neutron and proton
surfaces if $\eta_1>0$.

\subsection{Spin-Orbit Coupling}\label{sec:spin-orbit-coupling}

Related to the gradient term is the spin-orbit coupling, which we
include in the same form as in the Skyrme
\gls{NEDF}~\cite{Chabanat:1998}:
\begin{gather}
  \label{eq:NEDF_so}
  \Eso = W_0\vect{J}\cdot\grad\n
  %\Eso = \frac{W_0}{2}[\vect{J}\cdot\grad\n + \vect{J}_n\cdot\grad\n_n + \vect{J}_p\cdot\grad\n_p]
\end{gather}
where $\vect{J} = \vect{J}_{n} + \vect{J}_{p}$ is the total spin
current.  Following~\textcite{Fayans:1998}, we only include the isoscalar
portion here as the isovector contribution is small; see
\cref{sec:stat-prop-corr} for possible extensions).

\subsection{Pairing interaction}\label{sec:pairing}

The pairing energy depends on the anomalous density
\begin{align}\label{eq:NEDF_pairing}
  \Epair = \sum_{q=n, p} \int \d^3 \vect{r}\; g_{\mathrm{eff}}(\vect{r}) \lvert \nu_q(\vect{r}) \rvert^2
\end{align}
and the effective pairing coupling strength
$g_{\mathrm{eff}}(\vect{r})$ is obtained via a
renormalization~\cite{BY:2002fk, Yu:2003, Borycki:2006} of the bare
pairing strength, which may depend on neutron and proton densities.

In the case of pairing one can consider volume, surface, or mixed
pairing coupling constants, but previous studies of large sets of
nuclei have shown~\cite{Yu:2003, Bertsch:2009} that there is little
evidence preferring one form to another.  Phenomenological
studies~\cite{Bertsch:2009} also show that the proton pairing coupling
is stronger than the neutron pairing coupling, a result at odds with
the naïve expectation that the proton pairing coupling should be
weaker due to the Coulomb interaction~\cite{Lesinski:2009,
  Hebeler:2009, Yamagami:2012}.
It would also be peculiar to find that isospin invariance is broken by
the pairing interaction in this manner, when no other more
important terms of the \gls{NEDF} break isospin symmetry.  For
now, we will also not account for the role of the Coulomb interaction
on the pairing of the protons.

In an orbital-free approach the role of pairing is revealed only by
the presence of the odd-even staggering of the energy term.  As shown
in \cref{table:liquid_drop}, it has a small effect on the overall
quality of global mass fits and it may be omitted as a variational
parameter.

\subsection{SeaLL1 NEDF}\label{subsec:seall1-param}

We characterize the parameters of the theory according to their
significance for mass fits and dynamics.  We define a parameter as
\emph{dominant} if varying this parameter by less than 5\% or so
reduces the $\chi_E$ of the best fit by $\SI{0.1}{MeV}$ per nucleon.
We define a parameter as \emph{subdominant} if it can be varied by
10\% or more with a similar decrease in the quality of the fit.  We
define a parameter as \emph{unconstrained} if it can be set to zero at
this level of accuracy.

Our analysis shows that a minimal orbital-free \gls{NEDF} has 4
dominant parameters, and 2 subdominant parameters, consistent with the
analysis presented above.

\begin{description}
\item[Kinetic (none)] The kinetic energy density $\Ekin$
  \cref{eq:NEDF_kin} contains no free parameters -- just $\hbar$ and
  the bare nucleon masses $m_n$ and $m_p$ and the kinetic densities
  $\tau_{n, p}$.  However, since the orbital-free approach depends on
  densities alone, an approximation of the kinetic energy densities in
  terms of densities introduces a single parameter $\bGGA$.  This is
  discussed in \cref{sec:kinetic-terms,sec:orbit-free-functional}.

\item[Coulomb (none)] The Coulomb interactions $\Ec$ \cref{eq:NEDF_C}
  also contains no free parameter in either formulation.  In
  principle, the proton and neutron form-factors can be included, but
  these have only a small effect.  This is discussed in
  \cref{sec:coulomb-terms}.

\item[Homogeneous (3 dominant, 1 subdominant)] The homogeneous portion
  of the functional $\Ehomo$ \cref{eq:NEDF_homo} adds only three
  significant parameters.  In principle, up to nine parameters $a_j$,
  $b_j$, and $c_j$ for $j\in\{0, 1, 2\}$ describe the \gls{EoS} for
  homogeneous nuclear matter.  However, three of these nine (for
  $j=2$) are fixed by the \gls{EoS} of neutron matter as determined
  {\it in ab initio} calculations.  Two of the remaining six
  parameters ($a_0$, and the combination of $a_1-b_1\n_0^{1/3}$, where
  $\n_0$ is symmetric matter saturation density) are found to be
  unconstrained at the level of changing the energy \gls{rms} by
  $\delta \chi_E < \SI{0.1}{MeV}$ and are thus set to 0.  In our full
  \gls{SeaLL1}, we keep $c_1$ as a fitting parameter, although it is
  significantly less dominant than the others.  We fix $c_1$ sometimes
  in the orbital-free theory to provide a reasonable description of
  the neutron skins, see \cref{sec:hydrodynamic-model}.  Either $c_1$
  or the linear combination $a_1 - b_1\n^{1/3}$ can be used to tune
  the density dependence of the symmetry energy.
  
  This counting echoes the dominant and subdominant roles of the
  various nuclear saturation and symmetry properties in fitting
  masses.  In particular, the dominant parameters fix the saturation
  density $\n_0$, saturation energy $\varepsilon_0$, and quadratic
  symmetry energy $S_2$.  The slope of the quadratic symmetry energy
  $L_2$ is subdominant as far as mass fits are concerned, but
  important for properties such as the neutron skin thickness, which
  is why we keep an additional parameter in the \gls{SeaLL1}
  functional.

\item[Gradients (1 dominant)] The gradient corrections $\Egrad$
  \cref{eq:NEDF_grad} add a single new parameter $\eta_s$.

\item[Spin-orbit (1 subdominant)] The spin-orbit coupling term $\Eso$
  \cref{eq:NEDF_so} add a single new parameter $W_0$.  This parameter
  is subdominant for the mass fits, but is crucial for producing the
  shell structure of nuclei.  In the orbital-free approach this term
  is practically incorporated in the gradient contribution.

\item[Pairing (1 parameter)] The pairing interaction $\Epair$
  \cref{eq:NEDF_pairing} adds an additional parameter $g_0$ in the
  orbital-based approach.  Its contribution is practically
  incorporated in the homogeneous isoscalar terms in the orbital-free
  approach.  A different parameter $\delta$ measuring the odd-even
  staggering is required for the orbital-free formulation.  However,
  as is seen for the liquid drop models in \cref{table:liquid_drop},
  this additional parameter is quite unconstrained.

\end{description}

The orbital-based approach is specified by seven parameters: $b_0$, $c_0$,
characterizing isoscalar nuclear properties; $b_1$, $c_1$, defining
the isovector nuclear properties; $\eta_s$, defining the surface
tension; $W_0$. the strength of the isoscalar spin-orbit interaction;
and the bare (unrenormalized) pairing coupling constant $g$.  In the
orbital-free approach, we are left with only 4 significant
phenomenological parameters: $\eta_s$, $b_0$, $c_0$, and a linear
combination $a_1=b_1n^{1/3}$, since $c_1$ is unconstrained.  The
orbital-free approach has the additional parameter $\bGGA$ controlling
the Pad\'e gradient approximation of the kinetic energy density.

The full form of the functional \gls{SeaLL1} is:

\begin{widetext}
  \begin{multline}\label{eq:nedf_formula}
    \mathcal{E}[n_n, n_p] =
    \overbrace{\frac{\hbar^2}{2m} (\tau_n + \tau_p)}^{\text{kinetic}}
    + \overbrace{
      \sum_{j=0}^2 \left(
        a_j n^{5/3} + b_j n^{2} + c_j n^{7/3}\right)\beta^{2j}
    }^{\text{homogeneous}}
    + \overbrace{\eta_s \sum_{q=n, p} \frac{\hbar^2}{2m} \norm{\grad n_q}^2}^{\text{gradient}}\\
    + \underbrace{W_0 \vect{J} \cdot \grad n}_{\text{spin-orbit}}
    + \underbrace{\sum_{q=n, p} g_{\mathrm{eff}}(\vect{r}) \abs{\nu_q(\vect{r})}^2}_{\text{pairing}}
    + \underbrace{\frac{e^2}{2} \int \d^3\vect{r}' \frac{ n_p(\vect{r}) n_p(\vect{r}')}{\norm{\vect{r} - \vect{r}'}}
      - \frac{e^2\pi}{4} \left(\frac{3n_p(\vect{r})}{\pi} \right)^{4/3}}_{\text{Coulomb}}.
  \end{multline}
\end{widetext}

\begin{table}[htbp]
  \sisetup{
    detect-weight = true,
    detect-all = true,
    %round-mode = places,
    %round-precision = 3,
    table-number-alignment = center,
    table-format = 4.6,
    table-unit-alignment = left,
    % table-space-text-pre = xx,
    % table-space-text-post = xx,
  } %

  \robustify\bfseries
  \robustify\itshape
  % \begin{ruledtabular}  % Breaks the S column formatting
  \newlength{\paramsep}
  \setlength{\paramsep}{2px}
  \addtolength{\paramsep}{-\baselineskip}
  \begin{tabular}{lSSl}
    \toprule
    & {\textbf{\gls{SeaLL1}}} & {\textit{hydro}} & Comments\\
    \midrule

\rowcolor[gray]{0.85}[\the\tabcolsep][\the\tabcolsep]    $\n_0$ & \bfseries 0.154 & \itshape 0.154 & Adjusted (see Fig.~\ref{fig:hfbfit_rho})\cellcolor{white}\\ \\[\paramsep]
    $a_0$ & \bfseries    0 & \itshape {{\itshape same}} & Insignificant\\ \\[\paramsep]
\rowcolor[gray]{0.85}[\the\tabcolsep][\the\tabcolsep]    $b_0$ & \bfseries -684.5+-1.0 & \itshape -685.6+-0.2 & \cellcolor{white}\\ \\[\paramsep]
    $c_0$ & \bfseries  827.26 & \itshape  828.76 & $2c_0n_0^{\frac{2}{3}} = -\frac{3\hbar^2}{10m}\left(\frac{3\pi^2}{2}\right)^{\frac{2}{3}} \!\! - \frac{3}{2}b_0n_0^{\frac{1}{3}}$\\ \\[\paramsep]
    $a_1$ & \bfseries   64.3 & \itshape   50.9 & $a_1 = n_0^{1/3}b_1$\\ \\[\paramsep]
\rowcolor[gray]{0.85}[\the\tabcolsep][\the\tabcolsep]    $b_1$ & \bfseries  119.9+-6.1 & \itshape  94.9+-1.4 & \cellcolor{white}\\ \\[\paramsep]
\rowcolor[gray]{0.85}[\the\tabcolsep][\the\tabcolsep]    $c_1$ & \bfseries -256+-25 & \itshape -160.0\cellcolor{white} & Fixed in orbital-free theory\cellcolor{white}\\ \\[\paramsep]
    $a_2$ & \bfseries  -96.8 & \itshape  -83.5 & $a_{2} = a_{n}-a_{0}-a_{1}$\\ \\[\paramsep]
    $b_2$ & \bfseries  449.2 & \itshape  475.2 & $b_{2} = b_{n}-b_{0}-b_{1}$\\ \\[\paramsep]
    $c_2$ & \bfseries -461.7 & \itshape  559.6 & $c_{2} = c_{n}-c_{0}-c_{1}$\\ \\[\paramsep]
    $a_n$ & \bfseries  -32.6 & \itshape {{\itshape same}} & from neutron matter \gls{EoS} \eqref{eq:neut_mat}\\ \\[\paramsep]
    $b_n$ & \bfseries -115.4 & \itshape {{\itshape same}} & from neutron matter \gls{EoS} \eqref{eq:neut_mat}\\ \\[\paramsep]
    $c_n$ & \bfseries  109.1 & \itshape {{\itshape same}} & from neutron matter \gls{EoS} \eqref{eq:neut_mat}\\ \\[\paramsep]
\rowcolor[gray]{0.85}[\the\tabcolsep][\the\tabcolsep]    $\eta_s$ & \bfseries    3.93+-0.15 & \itshape    3.370+-0.050 & \cellcolor{white}\\ \\[\paramsep]
\rowcolor[gray]{0.85}[\the\tabcolsep][\the\tabcolsep]    $W_0$ & \bfseries   73.5+-5.2 & \itshape    0.0\cellcolor{white} & Fixed in orbital-free theory\cellcolor{white}\\ \\[\paramsep]
\rowcolor[gray]{0.85}[\the\tabcolsep][\the\tabcolsep]    $g_0$ & \bfseries -200 & \itshape {N/A}\cellcolor{white} & $g_0$ fit in Ref.~\cite{Yu:2003}\cellcolor{white}\\ \\[\paramsep]
    $\kappa$ & \bfseries {N/A} & \itshape 0.2 & Semi-classical (see section~\ref{sec:orbit-free-functional})\\ \\[\paramsep]
\midrule
    $\frac{\hbar^2}{2m}$ & \bfseries 20.7355 & \itshape {{\itshape same}} & units ($\si{MeV}=\si{fm}=1$)\\ \\[\paramsep]
    $e^2$ & \bfseries 1.43996 & \itshape {{\itshape same}} & cgs units ($4\pi\epsilon_0=1$)\\ \\[\paramsep]
\midrule
    $\chi_E$ & \bfseries 1.74 & \itshape 3.04 & 606 even-even nuclei\\ \\[\paramsep]
    $$ & \bfseries  & \itshape 2.86 & 2375 nuclei\\ \\[\paramsep]
    $\chi_r$ & \bfseries 0.034 & \itshape 0.038 & 345 charge radii\\ \\[\paramsep]
    $$ & \bfseries  & \itshape 0.041 & 883 charge radii\\ \\[\paramsep]

  \bottomrule
  \end{tabular}
  \caption{\label{tab:SeaLL1}
    Best-fit parameters for the \gls{SeaLL1} functional (in bold) and the orbital-free approximation (next column in italic when different).
    The errors quoted for the fit parameters should be interpreted as estimating by how much this parameter can be independently changed while refitting the other and incurring a cost of at most $\delta\chi_E < \SI{0.1}{MeV}$.
    %Extra precision is provided for some parameters to ensure similar accuracy for the most significant principal components. (For the uncertainty of various parameters we follow the \nicolas{convention where the number in parentheses is the numerical value of the standard uncertainty referred to the corresponding last digits of the quoted result.} For example, in case of $b_0 = -684.5(10)$ means $b_0=-684.5\pm 1.0$ and in case of $\eta_s=3.93(15)$ means $\eta_s=3.93\pm0.15$.)
  }
  % \end{ruledtabular}
\end{table}

The parameter values for the \gls{SeaLL1} functional are summarized in
\cref{tab:SeaLL1}.  The seven shaded parameters $b_0$, $c_0$, $b_1$,
$c_1$, $\eta_s$, $W_0$ and $g$ are significant for fitting nuclear
masses and radii.  The other parameters are either fixed independently
(e.g., by the properties of neutron matter) or have been determined
to be unconstrained for mass fits through a principle component
analysis described in \cref{sec:hydrodynamic-model}.

Our fitting strategy is described in detail in
\cref{sec:hydrodynamic-model} and we only recall here its most
important characteristics.  First, we explored the parameter space
with a simplified version of the orbital-free \gls{NEDF}.  This
\gls{NEDF} is characterized by seven parameters ($a_0$, $a_1$, $b_0$,
$b_1$, $c_0$, $c_1$, and $\eta_s$) which we fitted on $N_E = 2375$
experimentally-measured atomic masses (with errors less than
\SI{1}{MeV}) and $N_r = 883$ nuclear charge
radii as listed in \textcite{Audi:2012, Wang:2012}.  From this series
of fits and its statistical analysis, we found that (i) the parameters
$a_0$ and $c_1$ are unconstrained and can be set to zero; (ii) the
mass and radii are sensitive only to a single linear combination of
the parameters $a_1$ and $b_1$.  The parameter $c_1$ can be used
interchangeably with the linearly independent combination
$a_1 - \n_{0}^{1/3}b_1$ to control the slope $L_2$ of the symmetry
energy, which also controls the neutron skin thickness of neutron rich
nuclei; see below \cref{eq:L2} and the related discussion in
\cref{sec:symmetry}.  We will fix here $a_1 = \n_{0}^{1/3}b_{1}$,
where $\n_0 = \SI{0.154}{fm^{-3}}$ is the saturation density (see
discussion below) and $c_1$ to obtain a reasonable neutron
skin-thickness in \ce{^{208}Pb}.  With $c_1=0$ the neutron skin-thickness
of \ce{^{208}Pb} is about $\SI{0.2}{fm}$ and the $\chi_E$ increases by at
most $\SI{0.1}{MeV}$.

The next step consists in minimizing the residuals
$\chi_E^2 =\sum\abs{E_{N, Z} - E(N, Z)}^2/N_E$ over the $N_E = 196$
spherical even-even nuclei with $A\geq 16$ measured (not extrapolated)
from~\textcite{Audi:2012, Wang:2012} with the full orbital-based
functional.  This involves adjusting the five dominant parameters shaded
in \cref{tab:SeaLL1} -- the saturation density having been fixed from
the study of charge radii.  Note that the pairing parameter $g_0$ is
fixed at the value suggested in Ref.~\cite{Yu:2003}: Although this is
in principle a fitting parameter, it plays only a minor role in global
mass fits as discussed in the introduction.  The \gls{SeaLL1}
parameters of the orbital-based \gls{NEDF} (in bold) yield
$\chi_E = \SI{1.51}{MeV}$ over the $N_E = 196$ spherical even-even
nuclei, while the orbital-free \gls{NEDF} yield
$\chi_E = \SI{2.86}{MeV}$ over $N_E = 2375$ nuclei.\footnote{%
  At first sight it is surprising that the value of $\chi_E$ in the
  orbital-free approach over 606 even-even nuclei is larger than the
  value obtained for 2375 nuclei.  The reason is simple: the value
  $\chi_E=\SI{3.04}{MeV}$ was obtained with parameters obtained by
  fine-tuning the masses for spherical nuclei only in the
  orbital-based approach.  This does not minimize the value of
  $\chi_E$ in the orbital-free approach.}  The pairing fields were
treated using the renormalization procedure described in
Refs.~\cite{BY:2002fk, Yu:2003} with a cut-off energy of
\SI{100}{MeV}.

\begin{figure}[tbp]
  \includegraphics[width=\columnwidth]{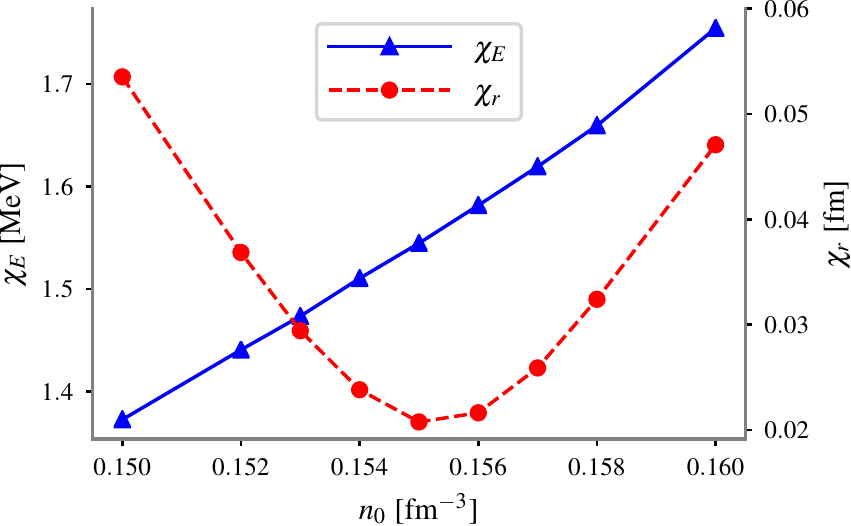}
  \caption{(Color online) Saturation density $\n_0$ dependence of the energy
    residual $\chi_E$ and charge radii residual $\chi_r$ of the
    \gls{SeaLL1} functional.  After holding $\n_0$ fixed (through the
    parameter $c_0$), the remaining five shaded parameters in
    \cref{tab:SeaLL1} were fit by minimizing only
    $\chi_E^2=\sum\abs{E_{N, Z} - E(N, Z)}^2/N_E$ over the $N_E = 196$
    spherical even-even nuclei with $A\geq 16$ measured (not
    extrapolated) from~\textcite{Audi:2012, Wang:2012}.  The value
    $\n_0 = \SI{0.154}{fm^{-3}}$ fixed in the \gls{SeaLL1} functional
    represents a compromise between these residuals here both $\chi_E$
    and $\chi_r$ increase by about 10\%.}
  \label{fig:hfbfit_rho}
\end{figure}

As discussed in \cref{sec:hydrodynamic-model}, we find that fitting
the binding energies alone in the orbital-free approach results in
quite a low saturation density $\n_0 \approx \SI{0.14}{fm^{-3}}$, and
a poorer fit to both charge radii and density profiles.  To explore
the influence of saturation density $\n_0$ on the quality of the fit,
we performed mass-only fits for the remaining five parameters with
various saturation densities $\n_0$ ranging from
\SIrange{0.15}{0.16}{fm^{-3}}.  For each fit, we also calculate the
rms radii residuals $\chi_r^2 = \sum\abs{\delta r}^2/N_r$ for the
$N_r=123$ corresponding nuclei in \cite{Angeli:2013}.  These results
are shown in \cref{fig:hfbfit_rho}, which demonstrates that the charge
radii strongly prefer $\n_0 \approx \SI{0.155}{fm^{-3}}$ in contrast
to the rather weak lower bias from the mass fits.  To incorporate this
preference in our fits, we fix the saturation density
$\n_0 = \SI{0.154}{fm^{-3}}$ by adjusting $c_0$ using the
\cref{eq:satrhoc}.  This represents a compromise between the two
biases where both $\chi_E$ and $\chi_r$ increase by about 10\%.  With
this fixed value of $\n_0$, we fit the remaining five parameters of the
\gls{SeaLL1} functional by minimizing only $\chi_E$ over the $N_E=196$
spherical even-even nuclei as summarized in \cref{tab:SeaLL1}.

\subsection{Orbital-Free Functional}\label{sec:orbit-free-functional}

Although we advocate working with the full orbital-based \gls{SeaLL1}
functional presented above, for tasks such as globally fitting mass
parameters, one can work with a much simpler orbital-free formulation.
The main challenge in formulating an orbital-free theory is to express
terms with the auxiliary densities $\tau_{n, p}$, $\vect{J}_{n, p}$,
and $\vect{j}_{n, p}$ by an appropriate functional of the number
densities $\n_{n, p}$.  Although formally possible, it is still an
open research question as to how best reduce an orbital-based
\gls{DFT} to an orbital-free version.  We discuss in more detail our
approach based on a semiclassical approximation in
\cref{app:orbit-free-funct}.  To summarize here, we suggest using the
following combination for the kinetic and spin-orbit contributions in
an orbital-free theory:
\begin{subequations}
  \begin{multline}
    \label{eq:NEDF_orbital_free}
    \Ekin[\n_n, \n_p] + \Eso[\n_n, \n_p] = \hfill \text{(orbital-free)}
    \\
    = \frac{\hbar^2}{2m}\sum_{q=n, p}\tau_{TF}[\n_q]F(X_q)
    - \frac{W_0^2}{2}\frac{2m}{\hbar^2} \n \abs{\grad\n}^2.
  \end{multline}
  where
  \begin{align}
    \label{eq:F_GGA}
    F(X) &= \frac{1+(1+\bGGA)X + 9\bGGA X^2}{1+\bGGA X}, 
    &
      X &= \frac{\tau_2[\n]}{\tau_{TF}[\n]}, 
    \\
    \tau_{TF}[n] &= \tfrac{3}{5}(3\pi^2)^{2/3} \n^{5/3}, 
    &
      \tau_{2}[n] &= \frac{1}{9} \norm{\grad\sqrt{\n}}^2.
  \end{align}
\end{subequations}
The ratio $X$ characterizes the size of the gradients in the system in
terms of the leading $\tau_{TF}$ and subleading $\tau_2$ terms of the
semiclassical expansion~\cite{Jones:1989, Dreizler:1990lr, Brack:1997}
of the kinetic density $\tau$.  The Padé approximant $F(X)$ suggested
by~\textcite{DePristo:1987} and advocated in~\cite{Dreizler:1990lr}
interpolates between the semiclassical limit $X \ll 1$ valid in the
core of large nuclei, and the approximation
$\tau \approx \tau_{TF} + \norm{\grad\sqrt{\n}}^2$ introduced by
\textcite{Weizsacker:1935} which correctly reproduces the asymptotic
fall off of the density when $X\gg 1$.  When spin-orbit is missing,
$\tau_{TF}[\n_q]F(X_q)$ gives a semi-classical approximation of the
kinetic density $\tau$.  This approximation requires a single
additional parameter $\bGGA$.  The value of $\bGGA$ can be chosen
approximately by comparisons between $\tau$ and
$\tau_{TF}[\n_q]F(X_q)$, and between their resulting kinetic energies
$E_{\mathrm{kin}}$, for the same set of single-particle wavefunctions.
We found $\bGGA \approx 0.2$ will give a reasonable semi-classical
approximation for $\tau$ and $E_{\mathrm{kin}}$.

The semi-classical spin-orbit contribution is suggested by Brack
\textit{et al.} \cite{Brack:1985x}, which brings a parameter $W_0$
corresponding to the one in \cref{eq:NEDF_so}.  Like the full
self-consistent theory, this parameter is also subdominant for the
mass fits and its contribution can be incorporated in the gradient
term.  Furthermore, due to the missing of shell structure in the
orbital-free theory, this parameter is even more unconstrained.

The orbital-free formulation of the \gls{NEDF} requires the additional
parameter $\bGGA$ to approximate the gradient corrections.  As
discussed above we choose $\bGGA = 0.2$.  Following \gls{SeaLL1}, we
fix the saturation density $\n_0 = \SI{0.154}{fm^{-3}}$, and fit the 3
parameters $b_0$, $b_1$ and $\eta_s$ shaded in \cref{tab:SeaLL1}.  The
spin-orbit contribution was absorbed in the gradient term and if
desired the unconstrained parameter $c_1$ can be used to fix the
neutron skin thickness.  The parameter values are determined by
performing the same least squares minimization of the binding energy
residuals as \gls{SeaLL1}, but over all $N_E = 2375$ nuclei (including
the deformed even-even, odd-even, and odd-odd ones) with $A \geq 16$ measured
from~\textcite{Audi:2012, Wang:2012}.

The parameter values and \gls{rms} residuals of orbital-free theory are also
summarized in \cref{tab:SeaLL1}.  As expected, the \gls{rms} residuals
$\chi_E = \SI{2.86}{MeV}$ is larger than the $\chi_E$ of \gls{SeaLL1}
due to the lack of shell corrections in the orbital-free theory, but
are comparable with results from the liquid-drop formula in
\cref{table:liquid_drop}.

\subsection{Principal Component Analysis}\label{sec:PCA}

The parameters listed in \cref{tab:SeaLL1} are highly correlated.  To
analyze these, we consider as significant changes
$\delta\chi_E \approx \SI{0.1}{MeV}$ since this is the typical level
of sensitivity of the mass fits.  We keep the changes relatively small
because otherwise the model is not well approximated by a quadratic
error model if $\delta \chi_E > \SI{0.1}{MeV}$.  Numerically we find
that even \SI{0.1}{MeV} is too large, but yields qualitatively correct
information after a full refitting.  Note that
$\delta(\chi_E^2) = (\chi_E + \delta\chi_E)^2 - \chi_E^2 = 2\,
\chi_E\, \delta\chi_E + (\delta\chi_E)^2$, so we must normalize
$\delta(\chi_E^2)$ by $2\, \chi_E \times \SI{0.1}{MeV}$ in order to
consider changes $\delta\chi_E \approx \SI{0.1}{MeV}$.

To compare the parameters in a meaningful way, we must make them
dimensionless and of order unity.  We do this by scaling them with
appropriate powers of $\n_0 = \SI{0.154}{fm^{-3}}$ and
$\varepsilon_F = \tfrac{\hbar^2}{2m} (3\pi^2\n_0/2)^{2/3} =
\SI{35.29420}{MeV}$, which we take as fixed parameters close to the
saturation values:
\begin{align}
  \tilde{a}_j &= \frac{a_j\n_0^{2/3}}{\varepsilon_F}, 
  & \tilde{b}_j &= \frac{b_j\n_0}{\varepsilon_F}, 
  & \tilde{c}_j &= \frac{a_j\n_0^{4/3}}{\varepsilon_F}.\label{eq:dimensionless}
\end{align}
(It is important to retain a significant number of digits for
isoscalar quantities, as it will be come more clear below.) In
particular, we consider the covariance matrix $\mat{C}$ such that the
residual deviation is
\begin{subequations}
  \begin{gather}
    \frac{\delta(\chi_E^2)}{2\chi_E \times \SI{0.1}{MeV}}
    \approx
    \vect{\delta}^T\cdot\mat{C}^{-1}\cdot\vect{\delta}
    = \sum_{n}\frac{(\delta p_n)^2}{\lambda_n^2}.
  \end{gather}
  where $\vect{\delta}$ is the deviations vector of the dimensionless
  parameters \cref{eq:dimensionless} from their best fit values as
  listed in \cref{tab:SeaLL1}, and we have diagonalized
  ${\mat{C}}{\textbf v}_n = \lambda_n^2{\textbf v}_{n}$ to obtain the
  principal components $p_n$
  \begin{gather}
    \label{eq:pca_p}
    p_n = \vect{v}_n
    \cdot\begin{pmatrix}
      \tilde{a}_0&
      \tilde{b}_0&
      \cdots&
      \tilde{\eta}_s&
      \tilde{W}_0
    \end{pmatrix}.
  \end{gather}
\end{subequations}

Since the parameters are of order unity, we may directly consider the
$\lambda_n$ as a measure of the errors: changing $p_n$ by $\lambda_n$
will affect the fit on the scale of
$\delta\chi_E \approx \SI{0.1}{MeV}$.  Therefore, the smaller the
value of the parameter $\lambda_n$, the more precisely the fit to
nuclear masses constrains the value of the corresponding linear
combination of \gls{NEDF} parameters.  A similar approach was used by
\textcite{Bertsch:2005a} in the analysis of Skyrme \glspl{NEDF}.

\begin{figure}[tbp]
  \subfloat[]{\includegraphics[width=\columnwidth]{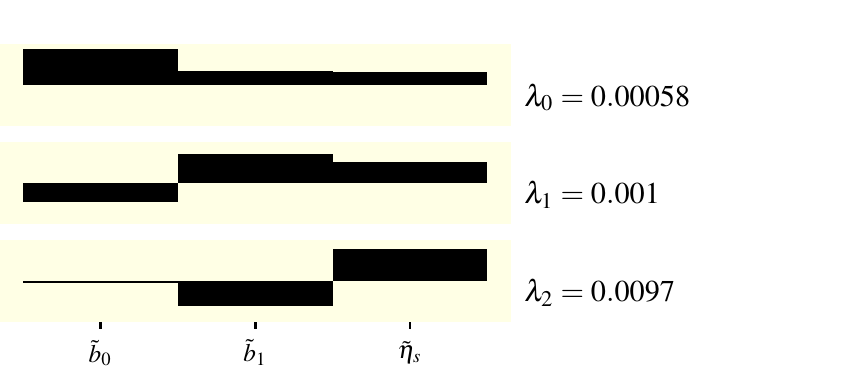}}\\
  \subfloat[]{\includegraphics[width=\columnwidth]{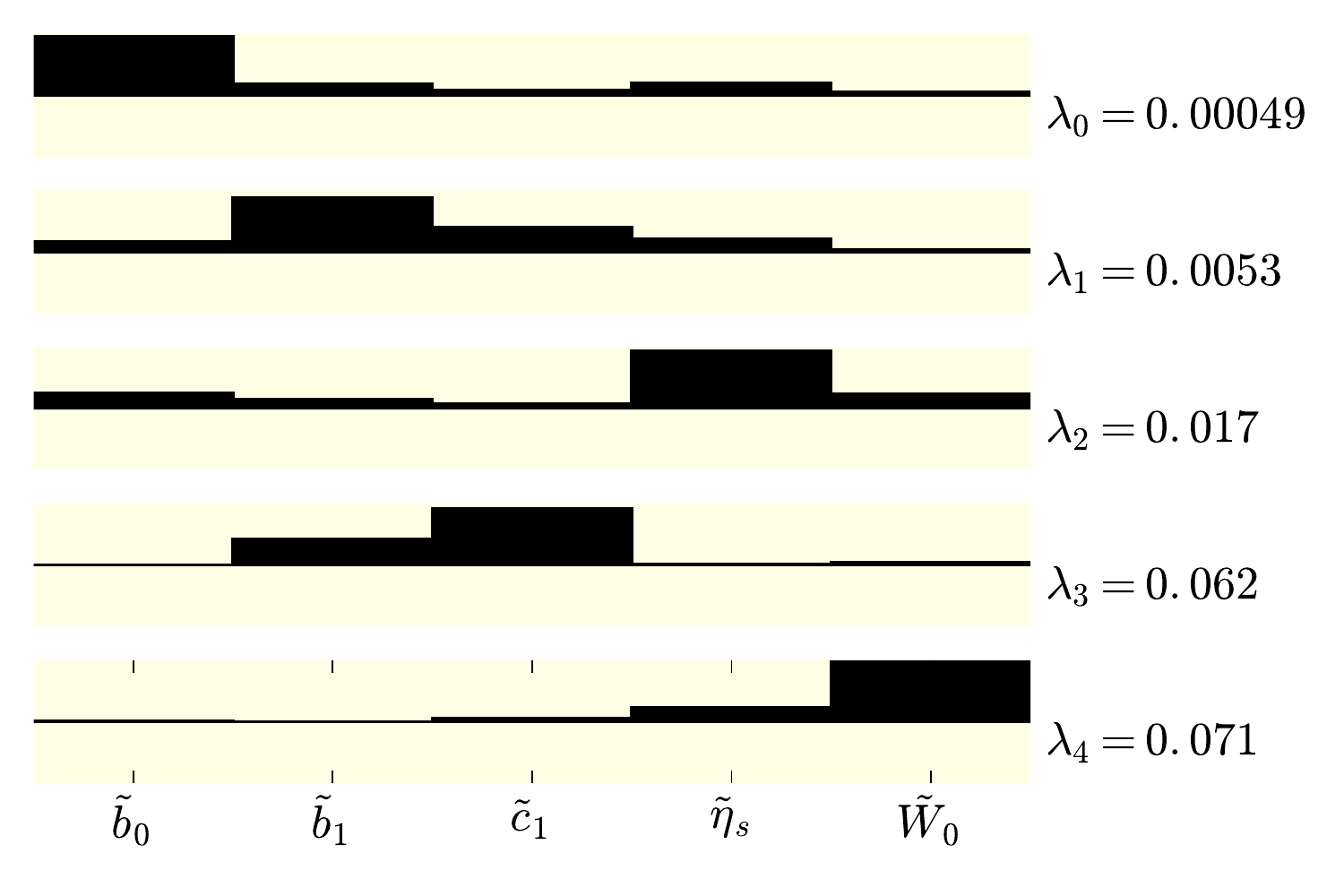}}
  \caption{The principal component analysis of the SeaLL1 \gls{NEDF} in the
    case of the orbital-free \textbf{(a)} and orbital-based
    \textbf{(b)} approach.}
  \label{fig:PCA-SeaLL1}
\end{figure}

\begin{figure}[tbp]
  \includegraphics[width=\columnwidth]{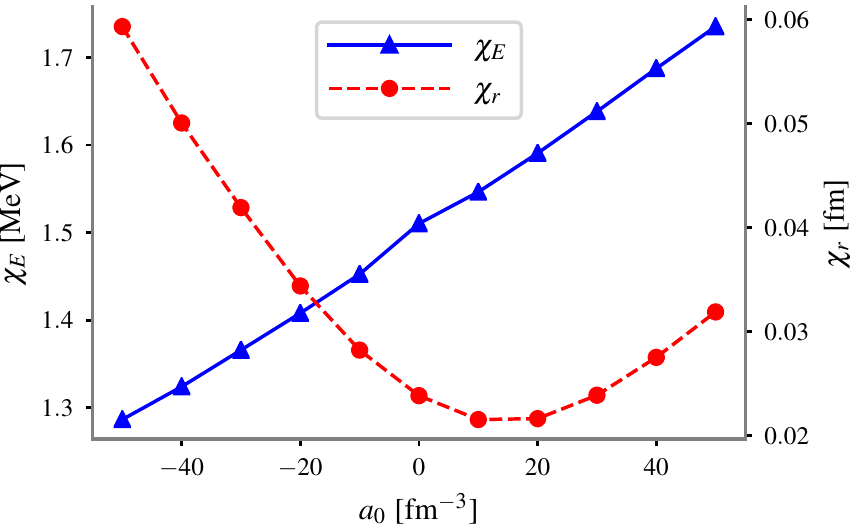}
  \caption{(Color online) The changes in $\chi_E$ and $\chi_r$ for the
    $N_E=196$ even-even spherical nuclei with $ A\geq 16$, similarly
    to \cref{fig:hfbfit_rho} as a function of the fixed parameter
    $a_0$, while the rest of the seven parameters of SeaLL1 specified in
    \cref{tab:SeaLL1} are optimized.}
  \label{fig:hfbfit_a0}
\end{figure}

In \cref{fig:PCA-SeaLL1} we show a principal component analysis of the
SeaLL1 functional.  The orbital-based analysis includes only 196
spherical even-even nuclei used to fine-tune the parameters of the
functional, while the analysis of the orbital-free functional includes
all 2375 nuclei as described in \cref{table:liquid_drop}.  Their
features can be understood in terms of the saturation and symmetry
parameters, see Eqs~\eqref{eq:satsym}.
\begin{subequations}
  \begin{align}
    S &= \frac{\mathcal{E}(\n_0, 0)-\mathcal{E}(\n_0/2, \n_0/2)}{\n_0},
        \label{eq:sym} \\
    L &= \left.3\n \diff{}{\n} \left(
        \frac{\mathcal{E}(\n, 0)}{\n}
        \right) \right |_{\n_0} = 3\n_0 \varepsilon_n'(\n_0)
        \label{eq:LL} \\
      &= \frac{6}{5}\frac{\hbar^2}{2m}(3\pi^2n_0)^{2/3}+ 2a_n \n_0^{2/3} +3b_n \n_0+4c_n\n_0^{4/3}. \nonumber
  \end{align}
\end{subequations}
where $\varepsilon_n(\n)$ is the energy per particle of the neutron
\gls{EoS}~\eqref{eq:NM}.  Since the saturation density $\n_0$
minimizes the energy of symmetric matter, the slope of the full
symmetry energy $L$ at $\n_0$ depends only on the \gls{EoS} of pure
neutron matter.  Thus, the \gls{QMC} neutron \gls{EoS} alone fixes the
global density dependence of the symmetry energy
$L = 3\n_0 \varepsilon_n'(\n_0) \approx \SI{30}{MeV}$.  We may express
these as follows:
\begin{subequations}
  \begin{align}
    \frac{\varepsilon_0}{\varepsilon_F} &= +\frac{3}{5} + \tilde{a}_0 + \tilde{b}_0\; + \tilde{c}_0, 
    \label{eq:e0}\\
    0 &= +\frac{3}{5} + \tilde{a}_0 + \frac{3}{2} \tilde{b}_0\; + 2\tilde{c}_0, 
    \label{eq:rho0}\\
    \frac{K_0}{\varepsilon_F} &= -\frac{6}{5} - 2\tilde{a}_0 + 4\tilde{c}_0,
    \label{eq:K0}\\
    \frac{S}{\varepsilon_F}
    &= \frac{3}{5}(2^{2/3}-1)
      + ( \tilde{a}_1 + \tilde{b}_1 + \tilde{c}_1 )
      + ( \tilde{a}_2 + \tilde{b}_2 + \tilde{c}_2 ), 
      \label{eq:S}\\
    \frac{L}{\varepsilon_F}
    &= \frac{6}{5}2^{2/3}
      + 2\tilde{a}_n   + 3 \tilde{b}_n  +4 \tilde{c}_n, \label{eq:L}
  \end{align}
\end{subequations}
where $K_0$ is the isoscalar incompressibility. The most significant component 
$p_0$ in both fits is the sum of the $j=0$ coefficients 
$\tilde{a}_0 + \tilde{b}_0 + \tilde{c}_0$ which fixes the saturation energy 
$\varepsilon_0$ \cref{eq:e0}, see also \cref{fig:ellipses}.  (Remember that we 
have chosen $a_0=0$ and that $c_0$ is determined from \cref{eq:rho0}.)  Next 
are mixtures of $\eta_s$ and the symmetry energy $S$, \cref{eq:S}, which are
correlated by the finite size of the nuclei; the latter is the sum of
the $j=1$ coefficients $\tilde{a}_1 + \tilde{b}_1 + \tilde{c}_1$.
While we have chosen to keep the value of the parameter $a_0=0$, its
value can be varied without affecting significantly the quality of the
overall mass and charge radii fit, see \cref{fig:hfbfit_a0}.  By
changing the adopted value $a_0=\pm\SI{20}{fm^{-3}}$ and keeping
$\varepsilon_0$ and the saturation density fixed one can change the
incompressibility by
$\delta K_0 = \pm 2\delta \tilde{a}_0\varepsilon_F=\pm 2\delta
a_0\n_0^{2/3} \approx \pm \SI{23}{MeV}$.
The power of this kind of analysis resides in formulating a
``power-counting'' scheme, which organizes the various linear
combinations of parameters in the order of relevance in the mass fit.

\section{Physical Properties}\label{sec:physical-properties}

\subsection{Global mass table} \label{sec:masses}

\begin{figure}[tbp]
  \subfloat[]{\includegraphics[width=\columnwidth]{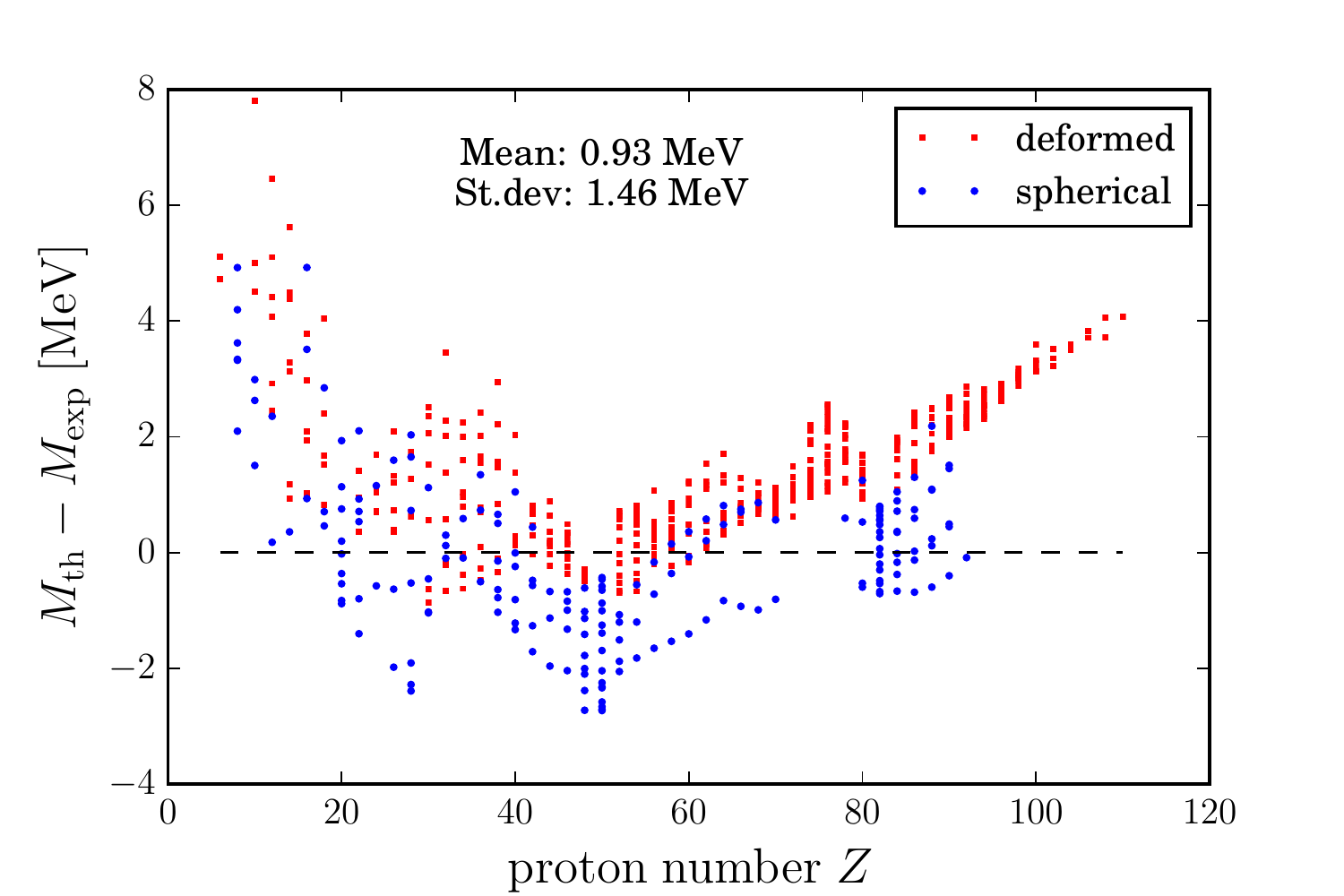}}\\
  \subfloat[]{\includegraphics[width=\columnwidth]{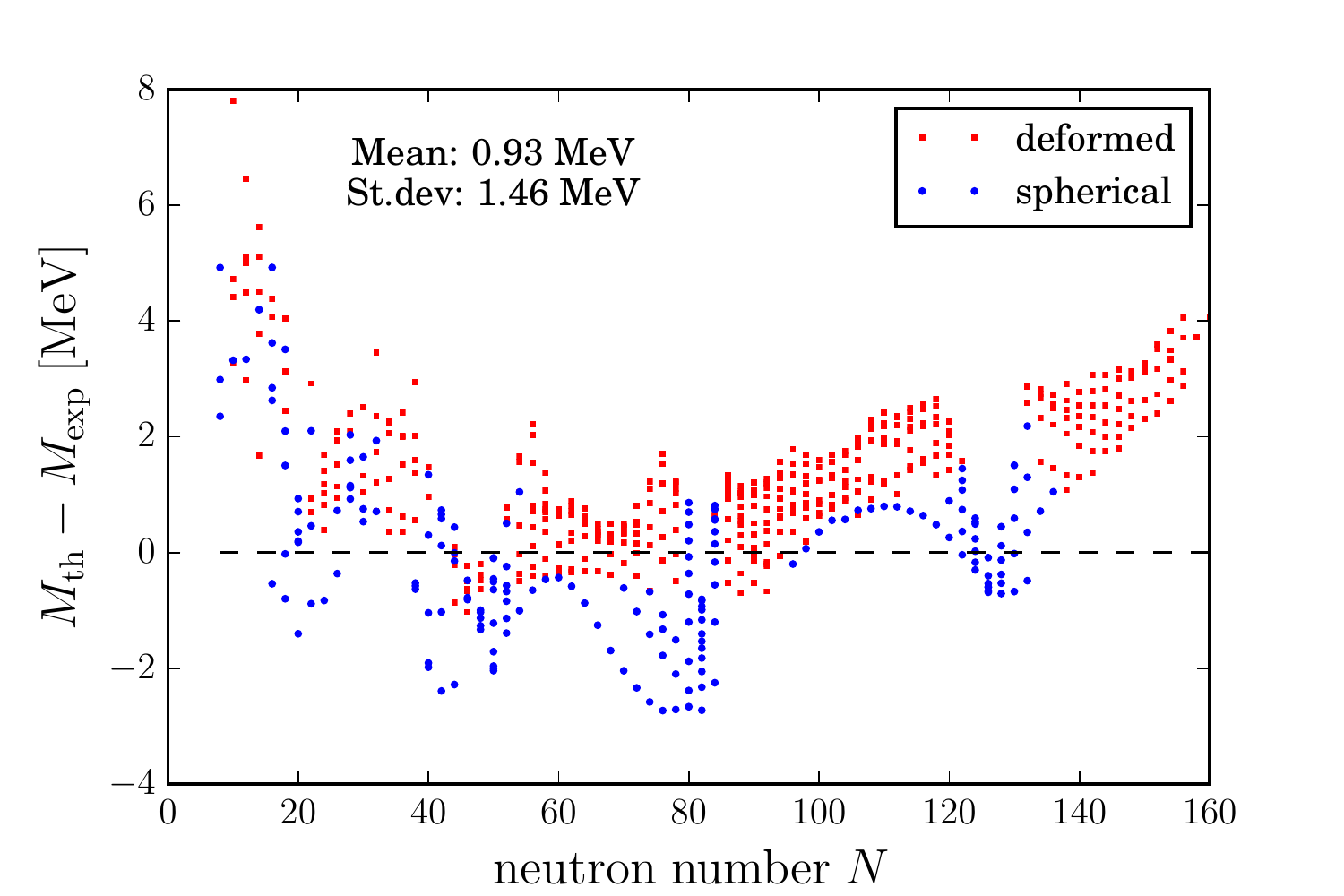}}
  \caption{(Color online) Mass residuals between \gls{SeaLL1} and measured
    masses for 606 even-even nuclei, of which 410 are deformed nuclei and 196
    are spherical nuclei, plotted with red squares and blue bullets respectively
    as a function of proton number $Z$ \textbf{(a)} and neutron number $N$
    \textbf{(b)}.}
  \label{fig:mass_nedf}
\end{figure}

\begin{figure}[tbp]
  \includegraphics[width=\columnwidth]{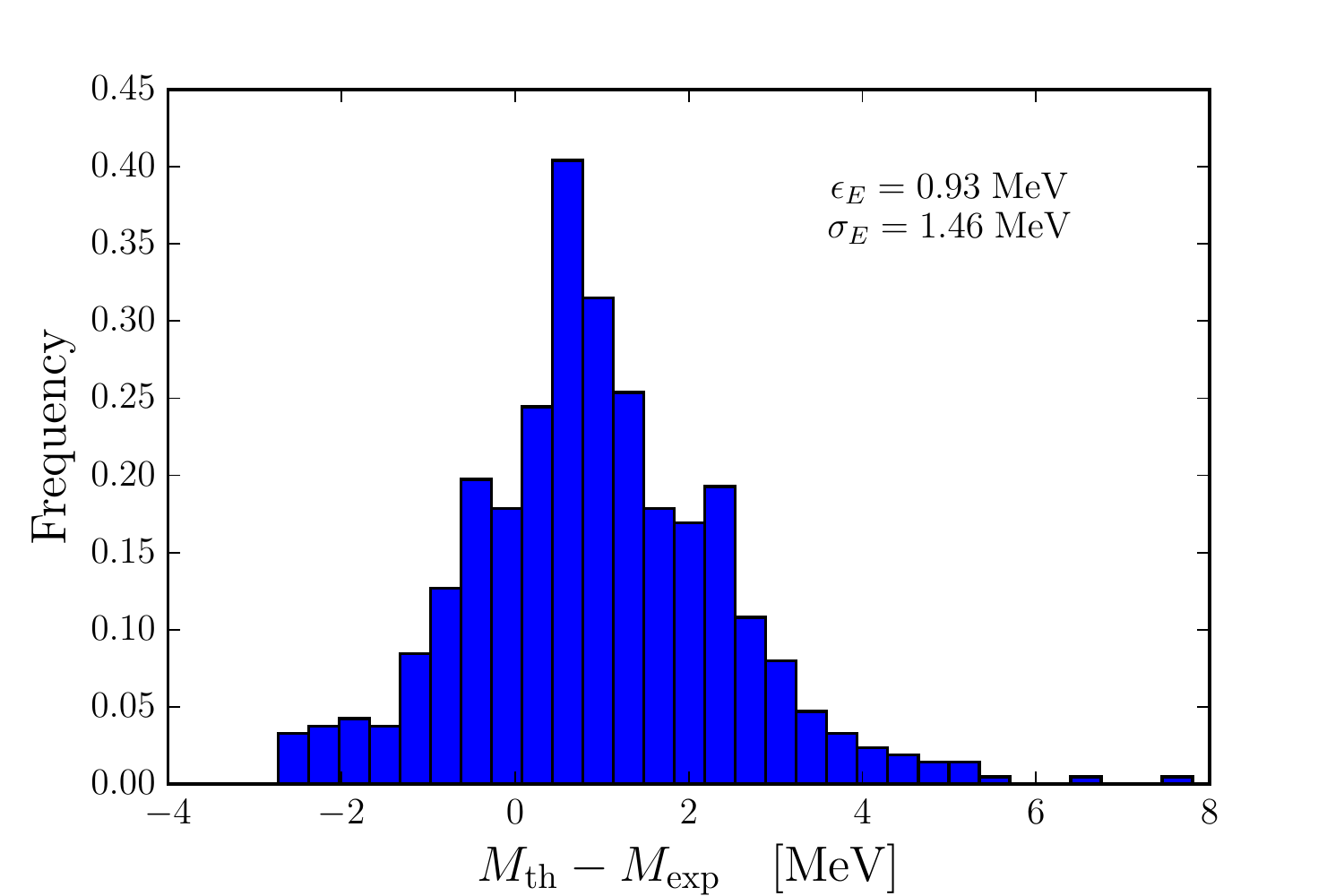}
  \caption{The histogram of the mass residuals between \gls{SeaLL1} and
    experiment for 606 even-even nuclei.}
  \label{fig:mass_hist}
\end{figure}

Since our orbital-based \gls{NEDF} was fit on spherical even-even
nuclei only, we validate its predictive power by performing a fully
microscopic calculation of the nuclear binding energies of 606
even-even nuclei with $A \geq 16$ in \cite{Audi:2012, Wang:2012}.  We
used an extension of the axial \gls{DFT} solver \textsc{hfbtho} code
\cite{Navarro:2017, Stoitsov:2013, Stoitsov:2005} that includes the
\gls{SeaLL1} and the regularization of the pairing
channel~\cite{BY:2002fk}.  Calculations were performed in a deformed
basis of 20 harmonic oscillator shells.  In the pairing channel, a
cut-off of \SI{100}{MeV} was adopted in accordance with
\cite{Yu:2003}.

\Cref{fig:mass_nedf} shows the residuals of the nuclear masses
calculated with \gls{SeaLL1} with respect to the experimental values
of these even-even nuclei. The \gls{rms} of the residuals is 
$\chi_E = \SI{1.74}{MeV}$.  Besides the larger residuals in light
nuclei, we observe the typical arc-like features common
to many \gls{NEDF} calculations, both for isotonic and isotopic
chains.  The poor performance of \gls{SeaLL1} in light nuclei is
likely related to the center-of-mass corrections (not accounted for
here) and is also observed in the UNEDF functionals
\cite{Kortelainen:2010, Kortelainen:2012, Kortelainen:2014_2}.
Since the center-of-mass correction is larger for
light nuclei, our parameter fit limited to spherical nuclei leads to
an underestimate of the masses of heavier spherical nuclei, see
\cref{fig:mass_nedf}. 
Overall the masses have a bias
$\epsilon_E = \braket{\delta E} = \SI{0.93}{MeV}$ and a standard
deviation $\sigma_E = \SI{1.46}{MeV}$, see \cref{fig:mass_hist}.  This
bias enters the \gls{rms} error $\chi_E^2=\sigma_E^2+\epsilon_E^2$ which
leads to a value of $\chi_E = \SI{1.74}{MeV}$.  This $\sigma_E$ is an
upper estimate of the \gls{rms} energy $\chi_E$ we expect if the SeaLL1
parameters would have been instead fitted to all even-even nuclei.

\begin{figure}[tbp]
  \subfloat[]{\includegraphics[width=\columnwidth]{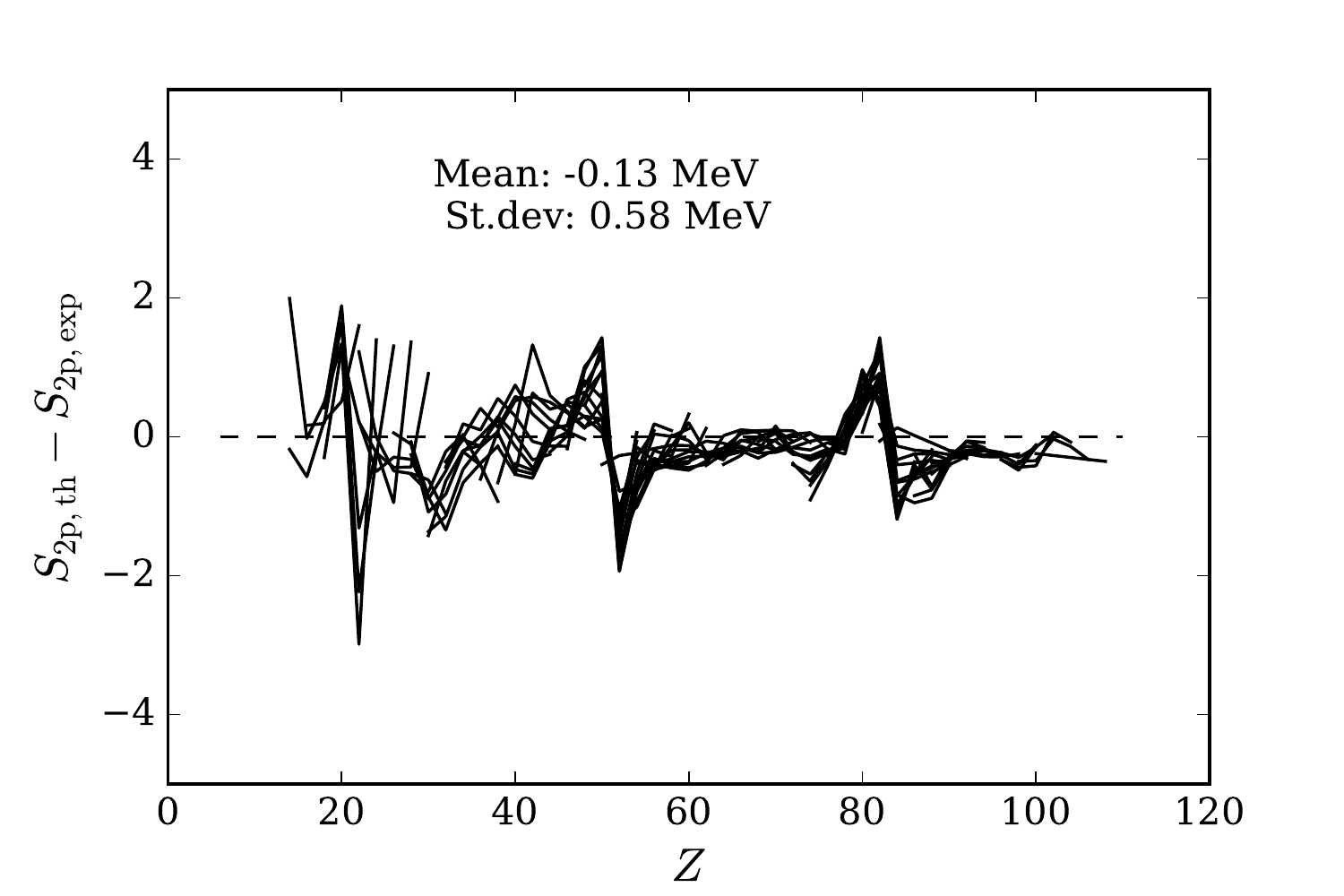}}\\
  \subfloat[]{\includegraphics[width=\columnwidth]{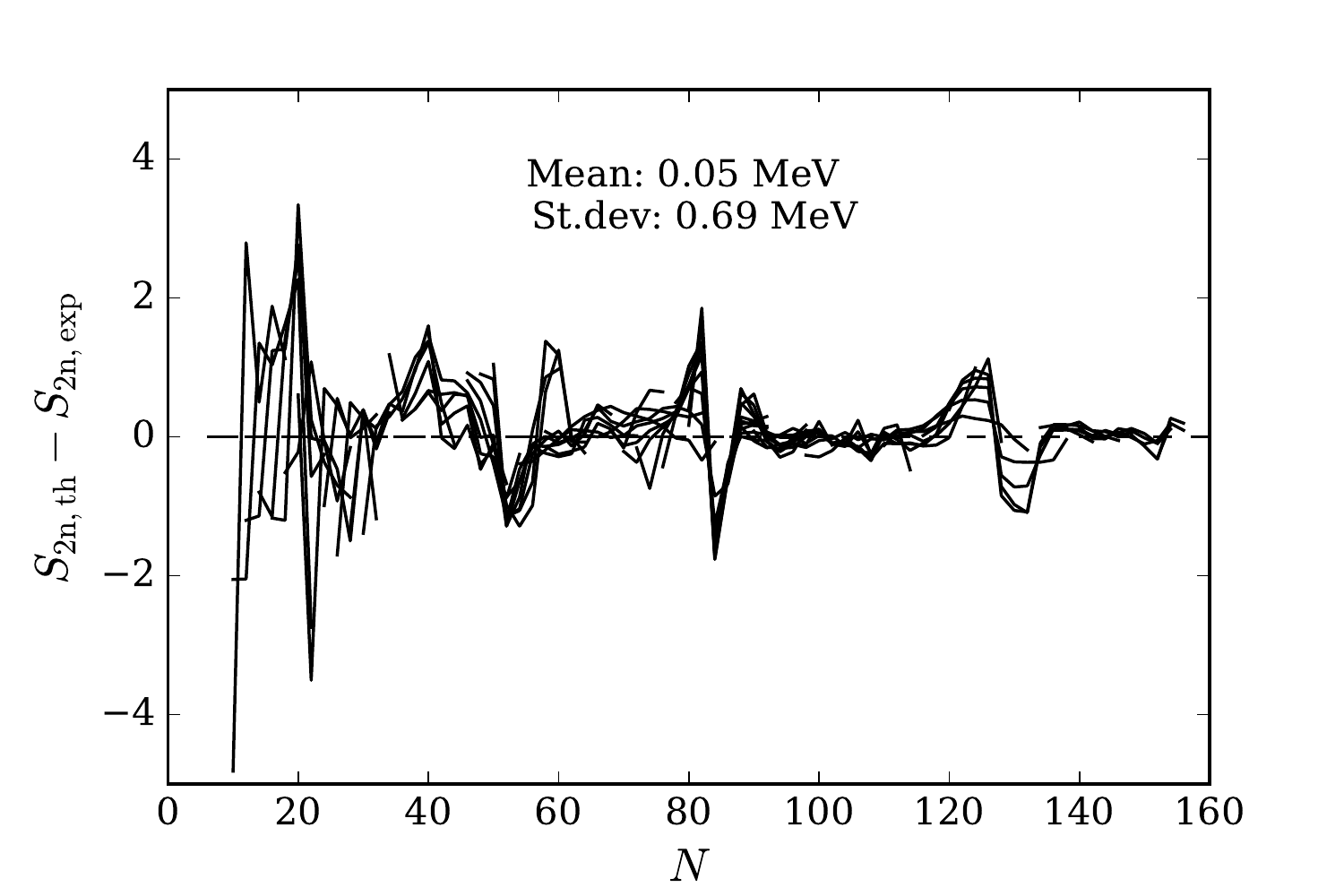}}
  \caption{The residual of the two-nucleon separation energies between
    \gls{SeaLL1} and experiment for 606 even-even nuclei: $S_{2p}(Z)$
    for constant $N$ \textbf{(a)} and $S_{2n}(N)$ for constant $Z$ \textbf{(b)}
    chains connected by lines.}
  \label{fig:S2N}
\end{figure}

The residuals for the two-nucleon separation energies for the same set
of even-even nuclei are shown in \cref{fig:S2N} and they are naturally
less affected by the errors induced by errors on binding energies.

\subsection{Charge radii and density distribution}

\begin{figure}[tbp]
  \includegraphics[width=\columnwidth]{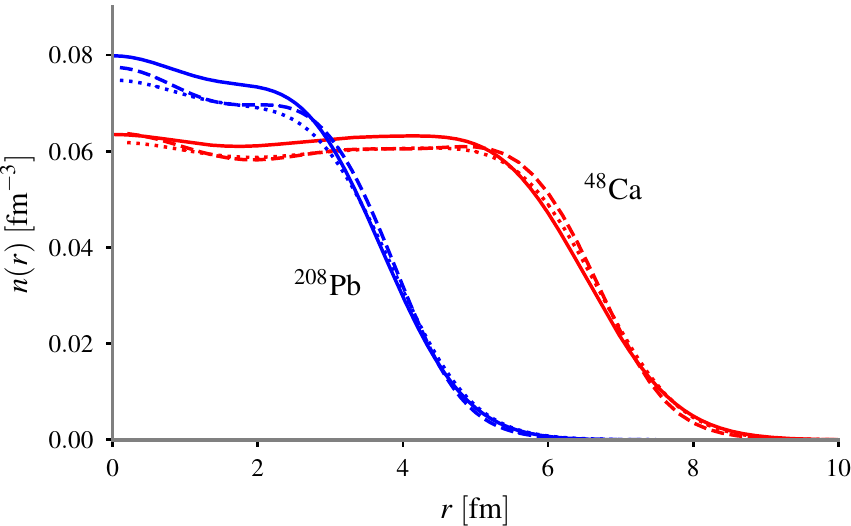}
  \caption{(Color online) The calculated proton $\n_p(r)$ (dashed) and charge
    $\n_{\text{ch}}(r)$ (dotted) densities for $^{48}$Ca (red) and
    $^{208}$Pb (blue), calculated with \gls{SeaLL1} compared to charge
    densities (solid) extracted from electron scattering
    experiments~\cite{Vries:1987}.}
  \label{fig:dens_capb}
\end{figure}

\begin{figure}[tbp]
  \subfloat[]{\includegraphics[width=\columnwidth]{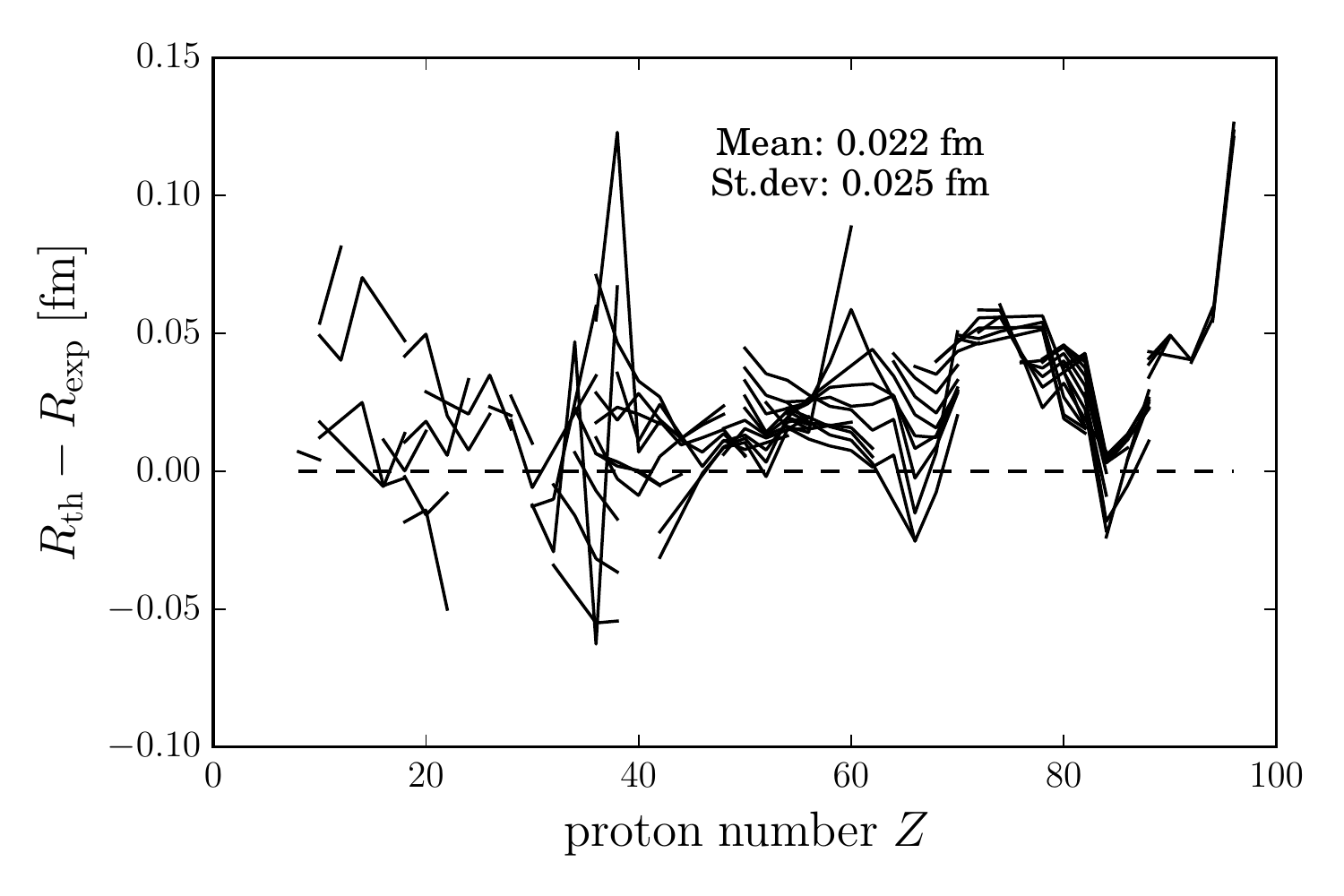}}\\
  \subfloat[]{\includegraphics[width=\columnwidth]{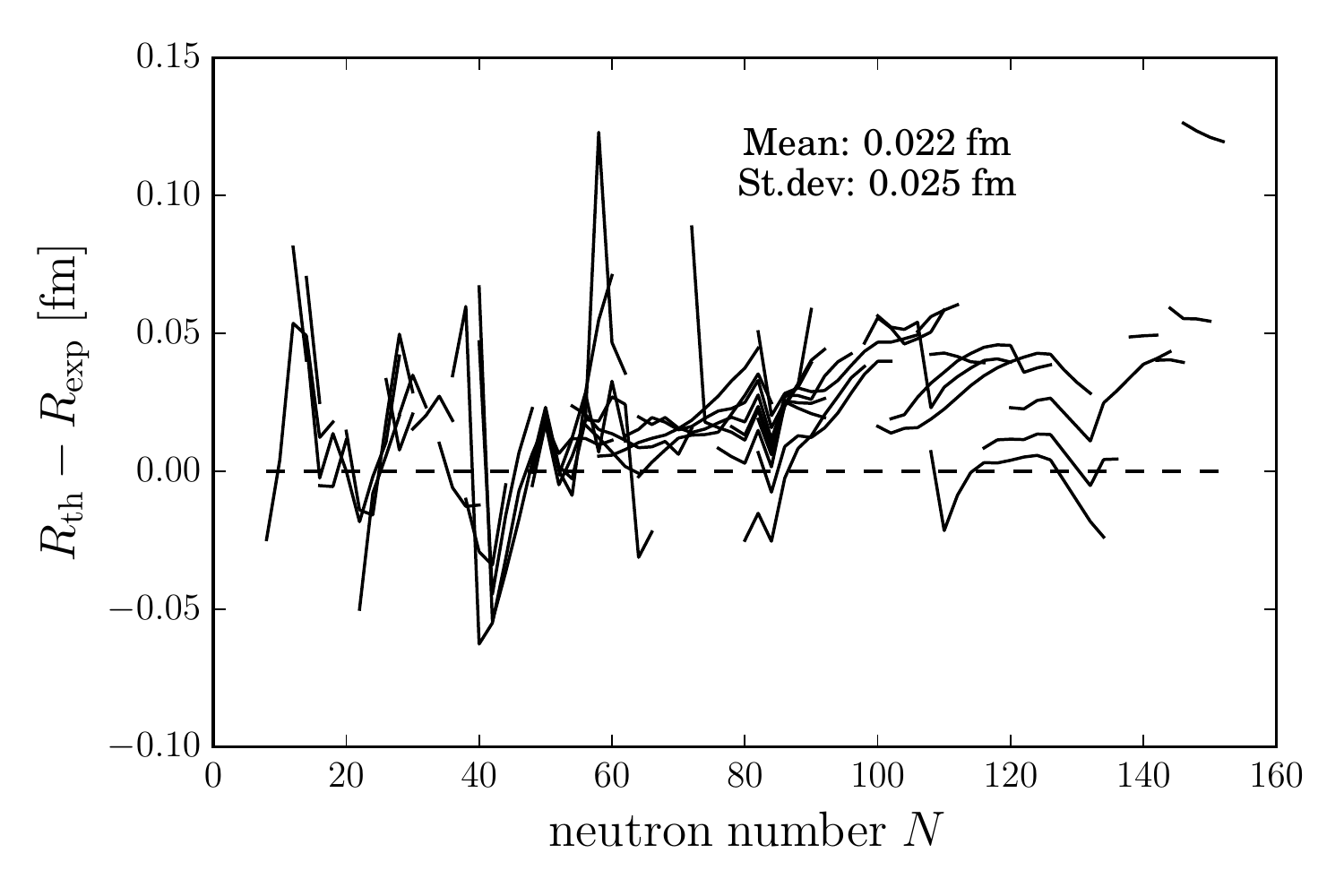}}
  \caption{Radii residuals between \gls{SeaLL1} and experiment for 345
    even-even nuclei.  Isotonic \textbf{(a)} and isotopic \textbf{(b)} chains are
    connected by lines.}
  \label{fig:radii_nedf}
\end{figure}

Using the parameters determined from the mass fits, \gls{SeaLL1} also
models the neutron and proton densities in the nuclei, allowing us to
extract the charge densities for these nuclei using \cref{eq:FF}.  As
a good benchmark, in \cref{fig:dens_capb} we compare the proton and
charge densities of $^{48}$Ca and $^{208}$Pb calculated with
\gls{SeaLL1} with the charge densities extracted from electron
scattering experiments~\cite{Vries:1987}.  The calculated $^{208}$Pb
has a slightly larger radius and slightly smaller diffuseness compared
to those extracted from data, which is consistent with the charge
radii comparison between \gls{SeaLL1} and experiment in
\cref{fig:radii_nedf}.

The residuals of radii for 345 matching even-even nuclei in
\cite{Angeli:2013} are also calculated, with a bias $\epsilon_r = \SI{0.022}{fm}$ and 
a standard deviation $\sigma_r=\SI{0.025}{fm}$, which gives a \gls{rms} residual of
$\chi_r = \SI{0.034}{fm}$, as shown in \cref{fig:radii_nedf}.

\subsection{Symmetry Energy and Neutron Skin Thickness}\label{sec:symmetry}

The isoscalar parameters $j=0$ and quadratic isovector parameters
$j=1$ ($\beta^2$) may be directly related to the saturation and
symmetry properties respectively by expanding the energy per nucleon
of homogeneous nuclear matter \cref{eq:inm} about the symmetric
saturation point $\n_n = \n_p = \n_0/2$:
\begin{gather}
  \label{eq:e024}
  \frac{\mathcal{E}(\n_n, \n_p)}{\n}
  = \epsilon_0(\n) + \epsilon_2(\n)\beta^2 +
  \epsilon_4(\n) \beta^4 + \order(\beta^6).
\end{gather}
The saturation density $\n_0$, energy per nucleon $\varepsilon_0$, and
incompressibility $K_0$ are then defined by the minimum
$\varepsilon_0'(\n_0) = 0$, and depend only on the $j=0$ isoscalar
parameters $a_0$, $b_0$, and $c_0$.  Expanding about $\n_0$ in
$\delta = (\n - \n_0)/3\n_0$ and in powers of
$\beta = (\n_n - \n_p)/\n$, one can define various ``local''
contributions to the symmetry energy $S_{2, 4}$, its density dependent
slope $L_{2, 4}$, etc.:
\begin{gather}
  \label{eq:satsym}
  \setlength{\arraycolsep}{0pt}
  \renewcommand{\arraystretch}{1.2}
  \begin{array}{c >{{}}c<{{}} l >{{}}c<{{}} l >{{}}c<{{}} l >{{}}c<{{}} l}
    \epsilon_0(\n) 
    &=&  
      \frac{6}{5}\varepsilon_F&+&a_0n^{2/3}&+&b_0n&+&c_0n^{4/3} \\
    &=&
        \varepsilon_0 && &+& \frac{1}{2}K_{0} \delta^2 &+& \order(\delta^3), \\
    \epsilon_2(\n)
   &=&
    -\frac{4}{15}\varepsilon_F&+&a_1n^{2/3}&+&b_1n&+&c_1n^{4/3}\\
    &=& 
        S_{2} &+& L_{2}\delta &+& \frac{1}{2}K_{2} \delta^2 &+& \order(\delta^3), \\
    \epsilon_4(\n)
    &=&
        S_{4} &+& L_{4}\delta &+& \frac{1}{2}K_{4} \delta^2 &+& \order(\delta^3)
  \end{array}
\end{gather}

Since we include also quartic terms $\beta^4$, we must differentiate
between these local symmetry parameters $S_{2}$, $L_{2}$, etc\@. and
the full symmetry parameters defined as the difference between
symmetric matter and pure neutron matter (see also the discussion of
\textcite{Lattimer:2014a}).  Using $a_1=b_1n_0^{/3}$, see
\cref{tab:SeaLL1}, we obtain the values for $S_2$ and $L_2$ given by
relations:
\begin{subequations}
  \label{eq:SL2}
  \begin{align}
    S_2 &= \frac{1}{3}\varepsilon_F+ 2a_1n_0^{2/3} + c_1n_0^{4/3}, \label{eq:S2}\\
    L_2 &= \frac{2}{3}\varepsilon_F+ 5a_1n_0^{2/3} + 4c_1n_0^{4/3}.\label{eq:L2}
  \end{align}
\end{subequations}

\begin{table}[htbp]
  \sisetup{round-mode = places,
           round-precision = 3,
           } %
  \begin{ruledtabular}
    \begin{tabular}{ccccccccc}
           &&&&&&& \multicolumn{2}{c}{Neutron skin}\\
      $\rho_0$& $-\epsilon_0$& $K_0$& $S$& $S_2$& $L$& $L_2$& \ce{^{208}Pb}& \ce{^{48}Ca} \\
      {[\si{fm}$^{-3}$]}& & & & & & & {[\si{fm}]}& {[\si{fm}]} \\
      \hline
      %0.154 & 15.6(2) & 230(2) & 31.7(2) & 27.7(6) & 32.4(2) & 32(4) & 0.131 & 0.159 
      0.154 & 15.6 & 230 & 31.7 & 27.7 & 32.4 & 32 & 0.131 & 0.159
    \end{tabular}
  \end{ruledtabular}
  \caption{\label{table:NEDF_Bb} Saturation, symmetry, and neutron skin
    properties for \gls{SeaLL1}.  All values in \si{MeV} unless
    otherwise specified.}
\end{table}

%----
% This figure of s.p. states is here in order to make sure it shows up at the 
% bottom of the proper page
\begin{figure*}[tb]
  \subfloat[]{\includegraphics[width=\textwidth]{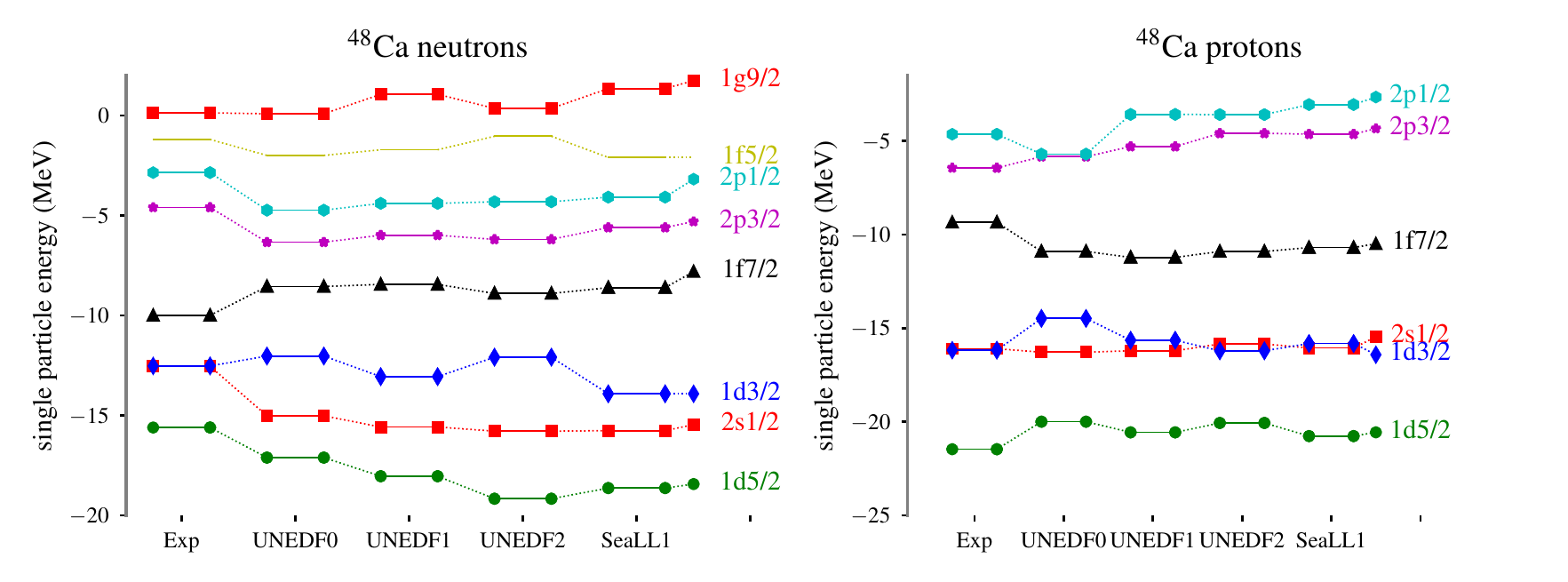}}\\
  \subfloat[]{\includegraphics[width=\textwidth]{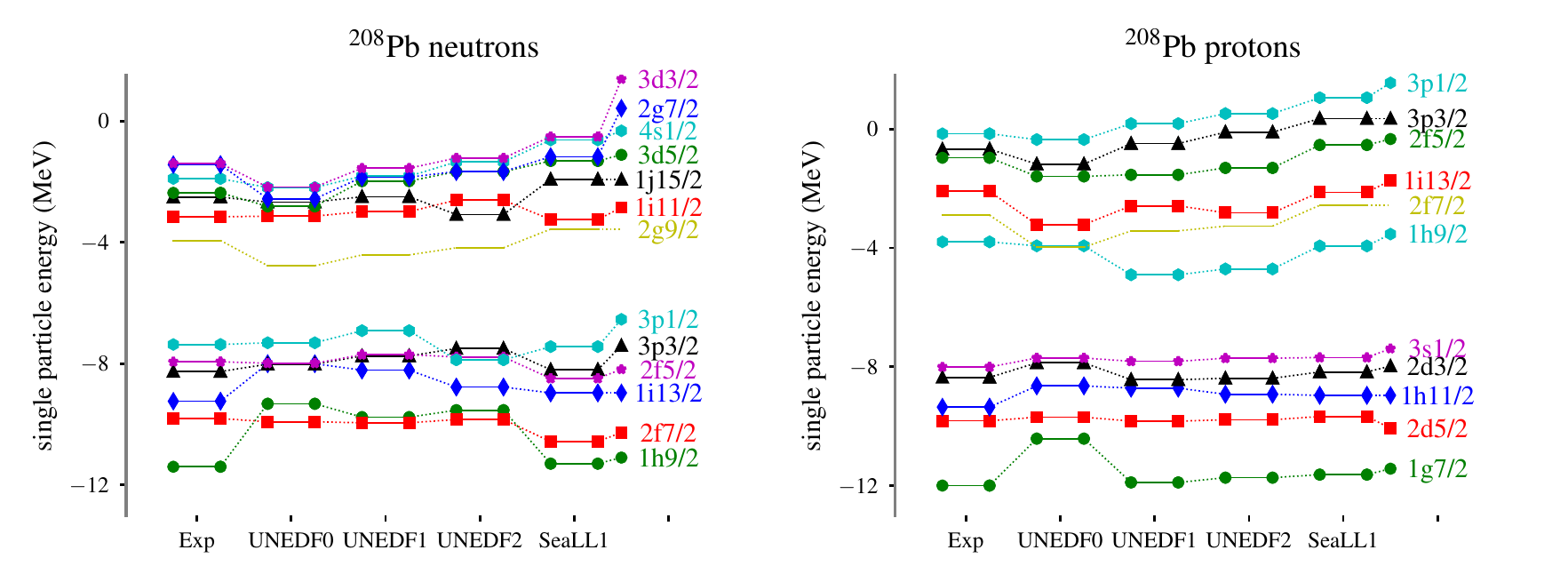}}
  \caption{(Color online) Single particle energies in $^{48}$Ca \textbf{(a)}
    and $^{208}$Pb \textbf{(b)} for a variety of functionals
    UNEDF0-2~\cite{Kortelainen:2010, Kortelainen:2012,
      Kortelainen:2014_2} and \gls{SeaLL1} (calculated using the
    \textsc{hfbtho} \gls{DFT} solver~\cite{Navarro:2017}).}
  \label{fig:spe}
\end{figure*}
%----

As shown in \cref{table:NEDF_Bb}, the binding energy of nuclear matter and the 
symmetry energy predicted by SeaLL1 fit agrees well with the value obtained 
with the mass formula~\eqref{eq:masses}. Our fits generally estimate the slope 
of the symmetry energy $L_2$ from \SIrange{29}{36}{MeV}. However, our fits with 
orbital-free functionals demonstrate that this quantity is not well constrained 
by the masses and can be adjusted independently with the combination 
$a_1 - b_1\n_0^{1/3}$ and/or coefficient $c_1$; see also the discussion in 
\cref{sec:hydrodynamic-model} and \cref{table:NEDF_B}.

We also compute the neutron skin thickness of \ce{^{48}Ca} and \ce{^{208}Pb}, 
for which precision measurements \gls{CREX} and \gls{PREX} are underway; 
see~\cite{Horowitz:2014} for details. The \ce{^{208}Pb} neutron skin is consistent 
with the value 
\SI[parse-numbers=false]{0.156\substack{+0.025\\-0.021}}{fm} of 
\textcite{Tamii:2011} extracted from measurements of the dipole polarizability 
using the method suggested by \textcite{Reinhard:2010} based on observed 
correlations between these two quantities in Skyrme models, and with the recent 
measurement of \SI{0.15(3)}{fm}~\cite{Tarbert:2014x}. Here again, our work with 
orbital-free functionals showed that the neutron skin is controlled by the same 
combination $a_1 - b_1\n_0^{1/3}$ as $L_2$, and hence is unconstrained by the 
masses.

\subsection{Spherical shell structure}

Shell structure is a fundamental property of atomic nuclei.  In an 
independent-particle picture, the shell structure can be associated with the 
single-particle spectra of the mean-field potential. Reproducing the correct 
ordering and distribution of single-particle levels is essential for nuclear 
structure theories, and also important for the application of the \gls{NEDF} in 
nuclear dynamics, such as nuclear fission and collision.  \Cref{fig:spe} display
the single-particle levels for neutrons and protons in $^{48}$Ca and
$^{208}$Pb for the SeaLL1, UNEDF0, UNEDF1, and UNEDF2 \gls{NEDF}. 
Single-particle energies were obtained by blocking calculations in the 
neighboring odd nuclei following the procedure outlined in 
\cite{Kortelainen:2012, Kortelainen:2014_2}.

In $^{48}$Ca, the \gls{rms} deviations for the single-particle energies of UNEDF0, 
UNEDF1, UNEDF2 and SeaLL1 with the empirical values (Exp)~\cite{Schwierz:2007} 
are \SIlist{1.50;1.71;1.92; 1.88}{MeV} and \SIlist{1.22;1.08;1.22;1.17}{MeV} 
for neutrons and protons, respectively.  In $^{208}$Pb, these are 
\SIlist{0.82;0.61;0.69;0.62}{MeV} and \SIlist{0.77;0.49;0.50;0.54}{MeV} for 
neutrons and protons, respectively. 

Compared with the empirical values, the $N=28$ and $Z=20$ gaps in $^{48}$Ca are 
clearly too small with \gls{SeaLL1}.  The single particle proton levels in 
$^{208}$Pb show that the $Z=82$ gap is also smaller in \gls{SeaLL1}.  Such 
patterns are also observed in UNEDF2 functional which, however, included
single-particle spin-orbit splittings in their fit \cite{Kortelainen:2014_2}.  
This might point to the need to consider the contribution from the isovector 
spin-orbit contribution in \cref{eq:NEDF_so1} proportional to $W_1$. Overall, however,
the SeaLL1 single-particle spectra, as quantified in the corresponding 
rms, are of better quality than UNEDF2.

\subsection{Fission pathway of \ce{^{240}Pu}}\label{sec:fiss-pathw-240pu}

\begin{figure}[bp]              % Bottom to fit in section.
  \includegraphics[width=\columnwidth]{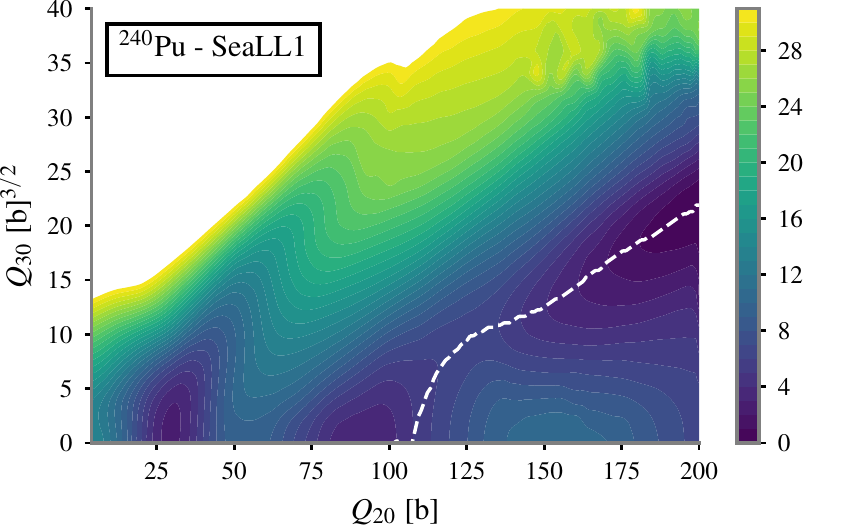}
  \caption{(Color online) Two-dimensional potential energy surface of
    \ce{^{240}Pu} with SeaLL1 for
    $0 \leq Q_{20} \leq \SI{200}{b}, 0 \leq Q_{30} \leq
    \SI{40}{b}^{3/2}$.  The least-energy fission path is marked as
    white dashed line.}
  \label{fig:ensurf}
\end{figure}

One of the important applications of nuclear DFT is the description of
nuclear fission \cite{Schunck:2016}.  In this context, characteristics
of fission pathways such as the excitation energy of fission isomers
or the height of fission barriers are often used to gauge the
predictive power of \glspl{NEDF}.  To this purpose, we computed the
potential energy surface of \ce{^{240}Pu} with SeaLL1 by performing
constrained \gls{HFB} calculations with constraints on the mass
quadrupole $Q_{20}$ and octuple moment $Q_{30}$ in the region
$0 \leq Q_{20} \leq \SI{200}{b}, 0 \leq Q_{30} \leq \SI{40}{b}^{3/2}$.
The definitions and units of $Q_{20}$ and $Q_{30}$ are consistent with
Ref.~\cite{Dobaczewski:2004} and the characteristics of the \gls{HO}
basis used in the calculation are the same as in~\cite{Schunck:2014}.
All calculations were performed with the \textsc{hfbtho} \gls{DFT}
solver \cite{Navarro:2017}.  The results are shown in
\cref{fig:ensurf}.

\begin{figure}[tbp]
  \includegraphics[width=\columnwidth]{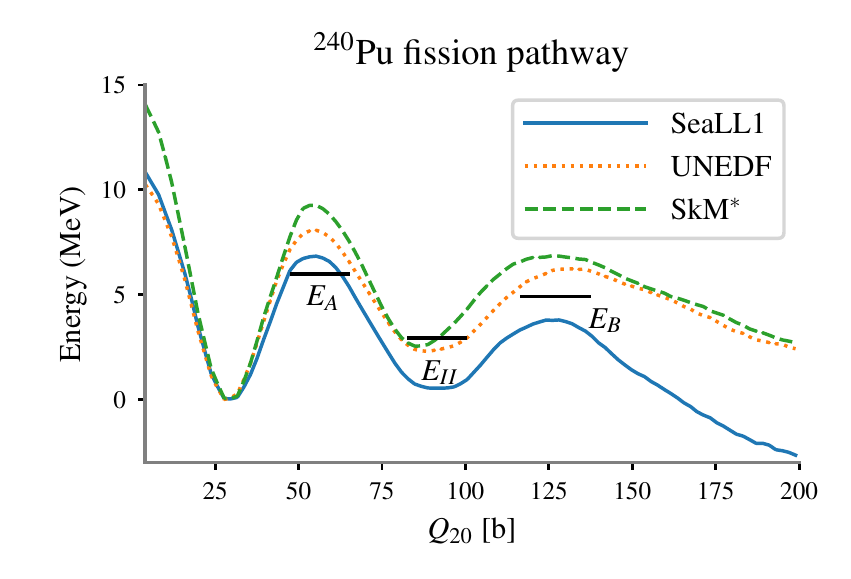}
  \caption{(Color online) Fission pathway for \ce{^{240}Pu} along the mass
    quadrupole moment $Q_{20}$ calculated using \textsc{hfbtho} with
    \gls{SeaLL1}, SkM*, and UNEDF1-HFB.}
  \label{fig:fissionpath}
\end{figure}

From this two-dimensional potential energy surface, we extracted the
least-energy trajectory starting at the ground-state.
\Cref{fig:fissionpath} shows the potential energy curve of
\ce{^{240}Pu} as a function of $Q_{20}$ along this (asymmetric)
fission pathway.  To gain an idea of the quality of \gls{SeaLL1}, we
repeated the calculations with the SkM*~\cite{Bartel:1982}, and
UNEDF1-HFB~\cite{Schunck:2015} energy functionals, both of which
were designed for fission studies.

Since all these calculations were done with the \textsc{hfbtho}
\gls{DFT} solver, triaxiality is not included and the height of the
first fission barrier is typically overestimated for all three functionals
by about \SI{2}{MeV}~\cite{Schunck:2014}.  Compared with SkM* and
UNEDF1-HFB, SeaLL1 underestimates the excitation energy of the fission
isomer ($E_{I} = \SI{0.54}{MeV}$ compared with an experimental value
of 2.8 MeV) and the heights of both fission barriers
($E_{A} = \SI{6.84}{MeV}$ vs.\@ \SI{6.05}{MeV}, and
$E_B = \SI{4.20}{MeV}$ vs.\@ \SI{5.15}{MeV}, respectively, for the
inner and outer barriers) agree within \SI{1}{MeV}.

%--
% This figure is here just so that it can be at the right place in the text
\begin{figure*}[tbp]
  \includegraphics[width=\textwidth]{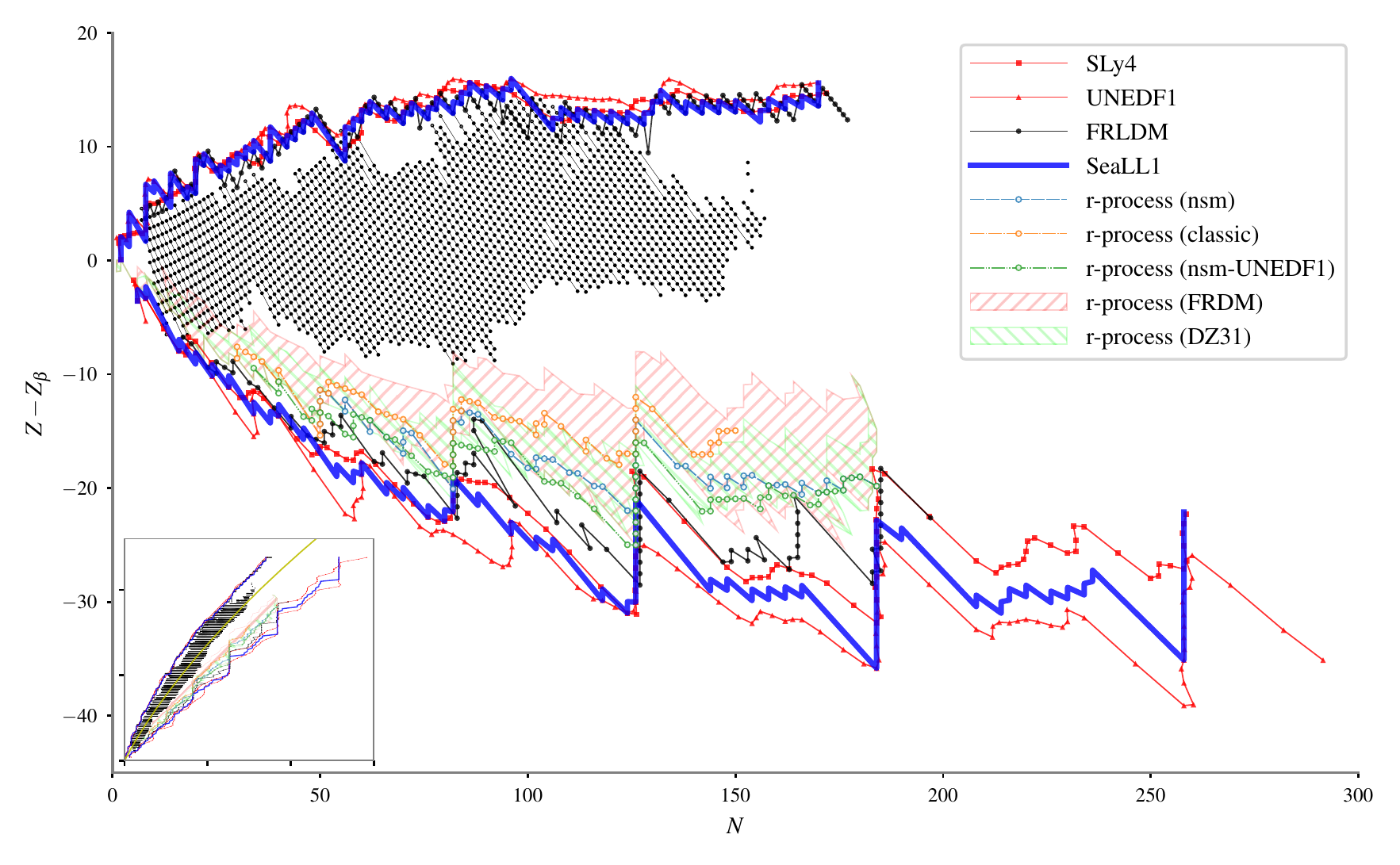}
  \caption{(Color online) Fully self-consistent calculations of the
    proton and neutron driplines for the SeaLL1 \gls{NEDF} (thick blue
    line) compared with predictions of the functionals SLy4 and UNEDF1
    extracted from Ref.~\cite{Erler:2012}, and
    FRLDM~\cite{Moller:1995}.  The vertical axis is shifted by the
    approximate $\beta$-stability line $Z_\beta(N)$ which minimizes
    \cref{eq:Bethe} at constant $A$ with parameters from
    \cref{table:liquid_drop}: $Z_\beta = A/(2+a_CA^{2/3}/2a_I)$,
    $\partial_Z E(A-Z, Z)|_{Z=Z_\beta}=0$.  The inset shows the usual
    $Z$ vs\@. $N$ plot, with the $Z = Z_\beta(N)$ curve as a solid
    (yellow) line.  The 2375 nuclear masses from~\cite{Audi:2012,
      Wang:2012} are displayed as dots.  We have plotted possible
    r-process trajectories predicted to be realized in the case of two
    neutron star mergers~\cite{Mumpower:2016, Mumpower:2016a} (red
    circles), in a classical hot
    $(n, \gamma)\leftrightarrow (\gamma, n)$ in equilibrium
    r-process~\cite{Surman:2017} (green circles) with the FRDM
    model~\cite{Moller:1995} and neutron star merger with the UNEDF1
    functional~\cite{Kortelainen:2012} (blue circles).  With pink
    and green bands we display the r-process paths obtained by
    \textcite{Mendoza-Temis:2015} under various conditions using the
    FRDM model~\cite{Moller:1995} and the Duflo-Zuker model
    \cite{Duflo:1995}.}
  \label{fig:driplines}  
\end{figure*}
%--

This result deserves a few comments.  First, we note that both SkM* and
UNEDF1 were constrained specifically on the height of the first
fission barrier (SkM*) or excitation energy of the fission isomer
(UNEDF1).  By contrast, we did not include any specific information
for nuclei at large deformation in the fit protocol of SeaLL1.  It is,
therefore very encouraging that, without any such constraint, the
resulting \gls{NEDF} is still in reasonable agreement with experimental results,
especially the height of the two barriers.  Our results are definitely
better than predictions with, e.g., SLy4 \cite{Chabanat:1998}, another
popular \gls{NEDF} without constraints on large deformations, which
predicts the second fission barrier much higher than the first one
\cite{Bender:2004}.  Second, the error in fission barriers of
\glspl{NEDF} designed for fission can reach \SI{2.5}{MeV}, as can be
seen in Ref.~\cite{Kortelainen:2014_2} where fission barriers and the
energy of the second isomer in chains of \ce{Ra}, \ce{Th}, \ce{U},
\ce{Pu}, \ce{Cm}, and \ce{Cf}, are compared to the UNEDF1-2, Gogny
D1S~\cite{Berger:1989}, and FRLDM~\cite{Moller:2009} functionals.  We
also point to a recent study of the surface energy coefficient $a_s$
(see \cref{eq:Bethe}) for 76 parameterizations of the Skyrme
\gls{NEDF}~\cite{Jodon:2016} and the rather complex interplay between
the roles of the shell-effects and of the surface energy on the values
of the fission barriers in \ce{^{240}Pu}. The energy of the fission
isomer and the height of the outer fission barrier, are shown to vary
by several \si{MeV}s with respect to the ground state energy.  Third,
we should repeat here the usual warnings about taking at face value
calculations of fission barrier heights: these quantities are not
physical observables, but are extracted from data in a (very)
model-dependent manner.

Ultimately, the predictive power of SeaLL1 (or any other \gls{NEDF}
for that matter) should be judged on their ability to reproduce
fission half-lives, or fission fragment distributions. As recently
shown~\cite{AB_2017}, within a real-time formulation of \gls{DFT}
extended to the \gls{TDSLDA}~\cite{Bulgac:2016}, the SeaLL1 \gls{NEDF}
provides a very accurate description of the features of the dynamics
for the induced fission in \ce{^{240}Pu}, comparable to that of
SkM$^*$, whose fission properties are similar to UNEDF1-HFB.

\subsection{Neutron and Proton Drip Lines}

In \cref{fig:driplines} we compare the proton and neutron drip lines
obtained with SeaLL1 against the predictions of UNEDF1, as well as
those obtained with other Skyrme parametrizations extracted from the
supplemental data of \textcite{Erler:2012} and using
FRLDM~\cite{Moller:1995}.  SeaLL1 predicts that there are 7716 stable
nuclei with $Z\le 120$, as compared with 8450 in case of UNEDF1, and 7212
for SLy4.  The position of the neutron drip line may dramatically
impact the astrophysical $r$-process, which is predicted to follow
lines of constant separation energy in close proximity to the neutron
dripline~\cite{Meyer:1989, Langake:2001}.  \textcite{Meyer:1989}
considered neutron star ejecta as the site of r-process
nucleosynthesis, and determined that the reaction flow is very close
to the dripline.  One should keep in mind also that the precise
position of the drip lines is difficult to pinpoint, since the
fluctuations, comparable to the theoretical errors, in the separation
energies have large fluctuations in their vicinity.  Even though his
simulations were performed for relatively cold matter (recent
simulations seem to indicate that the star material is somewhat
heated~\cite{Goriely:2011, Rosswog:2014}), it will be interesting to
simulate the r-process using SeaLL1.  The predicted position of the
neutron dripline will likely affect the structure of the neutron star
crust inferred from older studies~\cite{Baym:1971yq, Negele:1973,
  Ravenhall:1983uh, Lorenz:1993, Bulgac:2001y, Bulgac:2002x,
  Magierski:2004x, Magierski:2004a, Magierski:2003, Magierski:2002a}.
The corresponding increase in the neutron skin thickness will also
affect the profile and the pinning energy of quantized vortices in the
neutron star crust~\cite{Avogadro:2007, Avogadro:2008,
  Pizzochero:1997, Pizzochero:2008, Pizzochero:2011, Bulgac:2013a,
  Yu:2003a}.

Fusion cross sections~\cite{Gasques:2005, Adelberger:2011} will also
be significantly altered, particularly in stellar environments where
neutron rich nuclei fuse via pycnonuclear reactions~\cite{Schram:1990,
  Afanasjev:2012}, and where the neutron gas surrounding nuclei leads
to their swelling~\cite{Umar:2015}.  A thicker neutron skin with
further enhance this effect.

\subsection{Neutron star crust}

The baryon matter in the Universe organizes itself based on the
short-range nuclear attraction and the long-range Couloumb repulsion.
At densities much lower than the nuclear saturation density,
$\n \approx \SI{0.16}{fm}^{-3}$, the nuclear and atomic length scales
are well separated, and nuclei in matter are expected to form the Coulomb
lattice embedded in the neutron-electron seas that minimizes the
Coulomb interaction energy.  At subsaturation baryon densities,
$0.1\n_0 < \n < 0.8 \n_0$, conditions expected in the bottom layers of
the inner crust of neutron star, there is a strong competition between
the Coulomb and strong interactions, which leads to the emergence of
various complex structures with similar energies that are collectively
referred to as ``nuclear pasta''~\cite{Ravenhall:1983uh, Lorenz:1993,
  Hashimoto:1984}.  Pasta nuclei are eventually dissolved into uniform
matter at a certain nucleon density below $\n_0$.  Existence of pasta
phases would modify some important processes by changing the
hydrodynamic properties and the neutrino opacity in core-collapse
supernovae~\cite{Bethe:1990, Janka:2012} and proto-neutron
stars~\cite{Horowitz:2004, Alloy:2011}.  Also, the pasta phases may
influence neutron star quakes and pulsar glitches via the change of
mechanical properties of the crust matter~\cite{Gearheart:2011,
  Pons:2013, Levin:2001}.

\begin{figure}[tb]
  \includegraphics[width=\columnwidth]{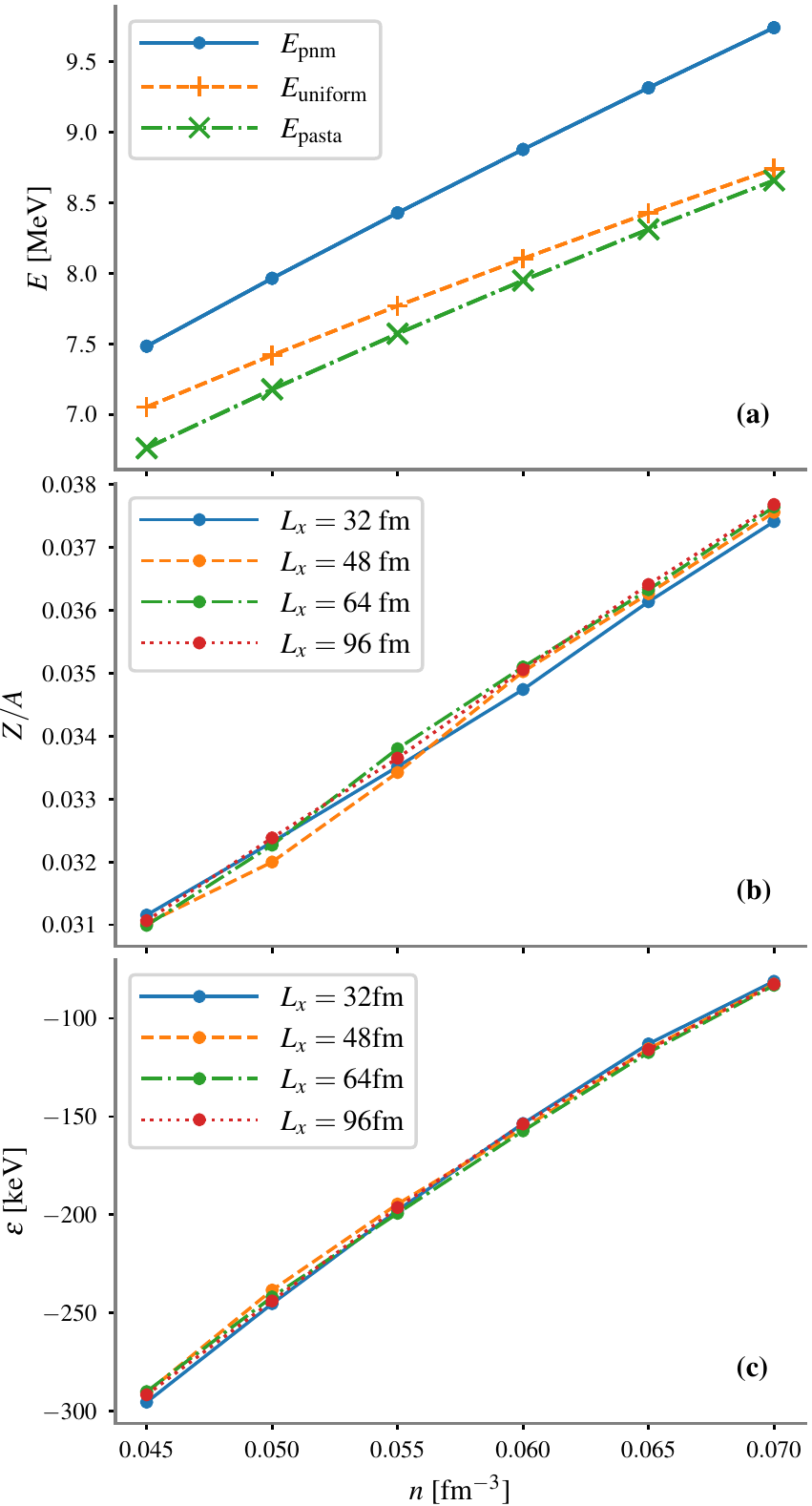}
  \caption{(Color online) \textbf{(a)}: energy per baryon in the pasta
    phase ($E_{\text{past}}$), energy per neutron in pure neutron
    matter ($E_{\text{pnm}}$), and energy per baryon in uniform
    nuclear matter ($E_{\text{uni}}$) as a function of average baryon
    density.  \textbf{(b)}: Charge ratio of the nuclear pasta as a
    function of average baryon density.  \textbf{(c)}: the energy per
    nucleon difference between the uniform and the inhomogeneous
    matter configurations in $\beta$-equilibrium as a function of the
    average baryon density.}
  \label{fig:crust}
\end{figure}

Since its prediction, significant progress has been made in simulating
the pasta phases \cite{Okamoto:2013, Fattoyev:2017, Schuetrumpf:2014}.
In this section, we use the hydrodynamics model to simulate the pasta
phases at average baryon densities
$0.045 \leq \n \leq \SI{0.07}{fm}^{-3}$.  In the nuclear-pasta system,
the chemical potentials of baryons and electrons satisfy the
$\beta$-equilibrium condition
\begin{gather}\label{eq:betaequi}
  \mu_n = \mu_p + \mu_e
\end{gather}
where $\mu_q$ is the chemical potential of species $q=n, p, e$ for
neutrons, protons, and electrons, respectively, and
$\Delta m = m_n - m_p$ is the neutron-proton mass difference.  The
total energy is the sum of the baryon energy $E_{\text{baryon}}$, the
electron density $E_{\text{elec}}$, and the proton-neutron mass
difference
\begin{gather}\label{eq:pasta}
  E_{\text{pasta}} = E_{\text{baryon}} + E_{\text{elec}} - \Delta m c^2 Z .
\end{gather}
For the baryon energy, we use the hydrodynamics model defined in
\cref{sec:hydrodynamic-model} with the SeaLL1 parametrization.  The
electron energy is the Thomas-Fermi energy for relativistic electrons
\begin{gather}\label{eq:electronenergy}
  E_{\text{elec}} = \int \d^3\vect{r}\; (3\pi^2 \n_e)^{4/3} \frac{\hbar c}{4\pi^2}
\end{gather}
where the electron density is determined from \cref{eq:pasta} as
\begin{multline}\label{eq:rhoe}
  \n_e(\vect{r}) = \Theta (\mu_n - \mu_p + V_c(\vect{r}) + \Delta m c^2)\\
  \times \frac{1}{3\pi^2}\left(\frac{\mu_n - \mu_p + V_c(\vect{r}) + \Delta m c^2}{\hbar c}\right)^3.
\end{multline}
where $V_c(\vect{r})$ is the Coulomb potential experienced by
electrons, which includes both the direct and the relativistic
exchange parts~\cite{Dreizler:1990lr} (notice the positive sign,
opposite from the non-relativistic Slater approximation)
\begin{gather}\label{eq:coulomb_electron}
  V_c(\vect{r}) = e^2 \int \d^3 \vect{r}'\; \frac{\n_c (\vect{r}')}{\norm{\vect{r} - \vect{r}'}}
  + \frac{1}{2}e^2 \left( \frac{3}{\pi} \n_e(\vect{r}) \right)^{1/3}
\end{gather}
where $\n_c(\vect{r}) = \n_p(\vect{r}) - \n_e(\vect{r})$ is the charge
density.  Through solving the hydrodynamics equation similar to
\cref{eq:psi} for baryons and \cref{eq:rhoe} for electrons, the charge
number $Z = \int \d^3\vect{r}\; \n_e(\vect{r})$ is determined
self-consistently for a given baryon number $A = N_n + N_p$ where
$N_p = Z$ is satisfied for charge neutrality.  Numerically, we perform
this calculation in a three-dimensional (3D) cubic lattice with periodic boundary
conditions at average baryon densities $\n =$
\SIlist{0.045;0.05;0.055;0.06;0.065;0.07}{fm^{-3}}.  To explore the
role of finite-size effects, the size of cubic lattice is chosen as
$L_x =$ \SIlist{32;48;64;96}{fm} respectively for all $\n$s.  The
lattice constant is fixed as $dx = \SI{1.00}{fm}$.  In
\cref{fig:crust} we compare the energy of uniform pure neutron matter,
with uniform matter in $\beta$-equilibrium, and allowing for the
formation of inhomogeneities.  Even though for various size cubic
boxes the spatial distribution of the matter at a given average
density is not identical, the gain in energy and the proton/neutron
ratios are practically the same and at an average density slightly
above \SI{0.07}{fm^{-3}} the matter distribution becomes homogeneous.

\subsection{Comparison with other \glspl{NEDF}}\label{sec:comp-with-other}

The accuracy of the ground state nuclear properties obtained using
SeaLL1 \gls{NEDF} compares extremely well with other approaches.  The
UNEDF1 nuclear energy functional introduced by
\textcite{Kortelainen:2012} has a residual of
$\chi_E = \SI{1.91}{MeV}$ per nucleus for \num{555} even-even nuclei
from AME2013~\cite{Audi:2003} and an \gls{rms} of $\SI{0.75}{MeV}$
(for $S_{2n}$) and $\SI{0.79}{MeV}$ (for $S_{2p}$) compared to
$\chi_E = \SI{1.74}{MeV}$, and \gls{rms} $\SI{0.69}{MeV}$ (for
$S_{2n}$), and $\SI{0.59}{MeV}$ (for $S_{2p}$) in the case of \gls{SeaLL1}.
\Gls{SeaLL1} delivers better quality single-particle spectra as well,
without introducing them into the fit, unlike UNEDF2. UNEDF2 reports
an \gls{rms} $\chi_r=\SI{0.018}{fm}$ for 49 nuclei only, and we cannot
compare that with that obtained by us, a $\chi_r=\SI{0.034}{fm}$ for
345 measured even-even nuclei.  The UNEDF2 functional of
\textcite{Kortelainen:2014_2} depends on 14 strongly-correlated
parameters.

The BCPM energy density functional introduced by
\textcite{Baldo:2008x, Baldo:2013x}
is based on information extracted from Brueckner-Hartree-Fock
calculations of neutron and symmetric nuclear
matter~\cite{Baldo:2004}, and four additional parameters to describe
pairing correlations in the $T=1$ channel~\cite{Garrido:1999}, 
one for the spin-orbit
interaction and two for the surface properties, in total seven parameters,
not counting the fine-tuning of nuclear saturation properties.  This approach is similar in spirit to
the one suggested by Fayans~\cite{Fayans:1998, Fayans:2000fk}, in the
spirit of the Kohh-Sham \gls{DFT}~\cite{Kohn:1965fk}.  These
authors have also included the beyond the mean-field rotational energy
correction~\cite{Egido:2004}, and the center-of-mass energy
correction~\cite{Butler:1984}, and they find a
$\chi_E=\SI{1.58}{MeV}$ for 579 even-even nuclei in
AME2003~\cite{Audi:2003} and a $\chi_r=\SI{0.027}{fm}$ for 313 nuclei.

\textcite{Goriely:2009, Goriely:2013, Goriely:2015, Goriely:2016} have
produced over the years a series of high-accuracy mass models based on
Skyrme \glspl{NEDF}. Their best model gives an average \gls{rms} around
\SI{0.5}{MeV} for the entire mass table, and a very close value 
$\chi_E=\SI{0.549}{MeV}$ for even-even nuclei.  In the case of
BSk24~\cite{Goriely:2013} the charge radius \gls{rms} is
$\chi_r=\SI{0.005}{fm}$.  However, in contrast with the UNEDF and SeaLL1 \glspl{NEDF}, 
the mass tables evaluated by Goriely \textit{et al.} 
were obtained by adding various phenomenological corrections in order to 
account effectively for beyond mean-field effects. These include corrections
for the center-of-mass motion, the rotational energy correction, and the Wigner 
energy.  These beyond mean-field corrections are hard still to incorporate in 
dynamical calculations, as in the case of fission~\cite{Bulgac:2016} or 
nucleus-nucleus collisions. 

As an exercise, we performed a refit of SeaLL1 after including the 
phenomenological center-of-mass correction due to \textcite{Butler:1984}. For 
spherical even-even nuclei, this term alone reduces the energy \gls{rms} from 
$\SI{1.54}{MeV}$ to $\SI{0.97}{MeV}$. It is thus expected that by adding further
beyond mean-field corrections to SeaLL the value of $\chi_E$ can be
reduced significantly.

We also mention work with the \gls{RMFT} of nuclei.
State-of-the-art parametrizations of the relativistic \gls{NEDF}
yields a $\chi_E$ between \SIrange{2}{3}{MeV} for even nuclei
using the AME2012 data set~\cite{Niksic:2011, Agbemava:2014}.

\begin{figure}[tb]
  \includegraphics[width=\columnwidth]{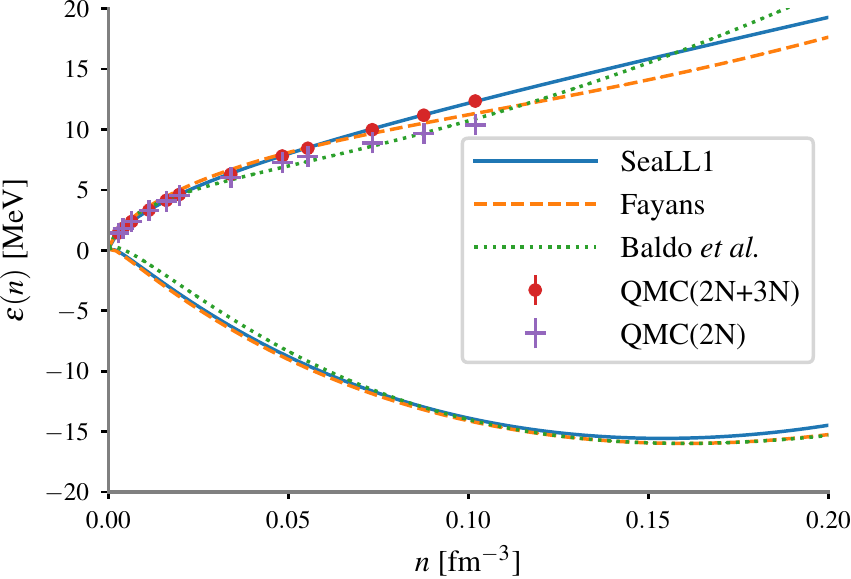}
  \caption{(Color online) The energy per nucleon for pure neutron matter and
    symmetric neutron matter used in \gls{SeaLL1}, compared to the
    corresponding energies used by \textcite{Fayans:1998} and
    \textcite{Baldo:2004, Baldo:2008x, Baldo:2013x}.  For comparison
    we have shown with a dashed line the results of the \gls{QMC}
    calculation of \textcite{Wlazlowski:2014a}, with 2N and 3N
    interactions as well the result with the 2N interactions alone.}
  \label{fig:eos}  
\end{figure}

Finally, we note that phenomenological Skyrme-like \glspl{NEDF}
``predict an inert point'' of the neutron matter \gls{EoS} at
$n\approx \SI{0.12}{fm^{-3}}$, with an energy per
particle~\cite{Goriely:2013, Goriely:2005, PGR_2015} noticeably lower
than the \gls{QMC} calculations and unrealistic low-density behavior,
see Eq.~\eqref{eq:NM} and Fig.~\ref{fig:eos}.  The BCPM \gls{NEDF}
assumes that no quartic terms in isospin $\beta^4$ are present in the
\gls{NEDF}, as their \gls{EoS} for neutron matter is softer than the
\gls{EoS} determined in \gls{QMC} calculations of
\cite{Wlazlowski:2014a}, see discussion in \cref{sec:satur-symm-prop}.
Adding the quartic $\beta^4$ $(j=2)$ terms does not significantly
impact the quality of the fits, see~\cref{sec:homogeneous-terms}.
However, the best fit functional with only quadratic $\beta^2$ $(j=1)$
terms, does not reproduce the neutron matter \gls{EoS}, especially
near $\n \approx \SI{0.12}{fm^{-3}}$ and the low density behavior.
These results demonstrate two important points: 1) quartic terms
$\propto\beta^4$ $(j=2)$ appear to be needed to reproduce the accurate
neutron matter \gls{EoS} only, and 2) known nuclear masses do not
constrain these quartic terms.

\glsreset{NEDF}
\section{Perspectives}\label{sec:perspectives}

\subsection{Static Properties and Correlation Energies}
\label{sec:stat-prop-corr}
  
Additional control may be obtained by introducing generalizations of
the terms included in SeaLL1.  These may be used to refine other
nuclear properties, including the static electric dipole
polarizability, nucleon effective masses, single-particle spectra,
proton and neutron pairing gaps, fission barriers and the second
fission isomer energies.
For example,
\begin{gather}
  \Egrad = \eta_0 \frac{\hbar^2}{2m} \norm{\grad\n_n+\grad\n_p}^2
  + \eta_1 \frac{\hbar^2}{2m} \norm{\grad\n_n-\grad\n_p}^2 
  %\tag{Eq.~\ref{eq:NEDF_grad1}}
  \label{eq:NEDF_grad1_ext}
\end{gather}
with $\eta_0\neq \eta_1$ would allow one to adjust the neutron skin
thickness somewhat independently from the symmetry properties of the
functional and one can also control the static electric
polarizability in the same manner.

The single-particle spectra for \ce{^{48}Ca} and \ce{^{208}Pb}
obtained with SealLL1 have a larger neutron gaps and smaller proton
gaps than measured experimentally (see \cref{fig:spe}).  This could be
remedied by tuning independently the parameters $W_0 \neq W_1$ in a
more general form of the spin-orbit coupling,
\begin{gather}
  \Eso = W_0\vect{J}\cdot\grad\n + W_1(\vect{J}_n-\vect{J}_p)\cdot(\grad\n_n-\grad\n_p).
  \label{eq:NEDF_so1}
\end{gather}
which could be used to independently fine-tune proton and neutron
single particle spectra near the Fermi level.
One can add as well a density dependence of the
spin-orbit coupling, which can lead to fine changes of the
single-particle spectra, see also Ref.~\cite{Goriely:2016} for a
related study.

One could further tune the single-particle spectra, and adjust the
nucleon effective masses, by introducing more generalized
density-dependent terms of the type arising in Eqs.~\eqref{eq:tau_n},
\begin{gather}
  \label{eq:tau_n1}
  \Etau \propto
  \tau n^\sigma - \vect{j}^2n^{\sigma-1}-\frac{3}{5}(3\pi^2)^{2/3}n^{5/3+\sigma}
  \propto \frac{\norm{\grad\n}^2}{n^{1-\sigma}}.
\end{gather}
(The obvious isospin structure has been suppressed.)  The
presence of the current density here is required in order to restore
Galilean covariance~\cite{Engel:1975}.  Since the density gradients
are peaked at the nuclear surface, the dependence of these coupling
constants on density are not expected to lead to a significant
changes in the quality of nuclear mass fits. The corresponding
coupling constants would thus play a subdominant role as discussed in
\cref{subsec:seall1-param}.  This shows that terms like $\tau
\n^\sigma$ in Skyrme-like functionals can be used in the combination
\eqref{eq:tau_n1} where they would play a subdominant role in mass fits.

In connection with gradient corrections, a remark is in order.  Since
the density gradients peak at the surface, allowing the
corresponding coupling constants to acquire a density dependence
could be useful, but such a density dependence of these coupling
constants likely is not going to be very sensitive to different
powers of the density or even a linear combination of different
powers of the densities, though it might be capable of discriminating
between various isospin structures.  This behavior was observed for
example by \textcite{Goriely:2015}, when they introduced various
density dependence of the spin-orbit terms and observed that the energy \gls{rms}
changed only by $\SI{20}{keV}$.

Similarly, a long standing feature of standard \glspl{NEDF} requires
breaking the isospin symmetry of the pairing contribution, even
needing stronger proton pairing than neutron
pairing~\cite{Bertsch:2009, Kortelainen:2012, Kortelainen:2014_2}
despite the Coulomb repulsion.  This can easily be remedied by
using instead a modified form of pairing which conserves the charge
symmetry:
\begin{subequations}
  \begin{multline}\label{eq:NEDF_pairing1}
    \Epair = \int \d^3 \vect{r}\; g_{\mathrm{eff}}(\vect{r})\left(\abs{\nu_n(\vect{r})}^2 + \abs{\nu_p(\vect{r})}^2\right) \\
    +\int \d^3 \vect{r}\; h_{\mathrm{eff}}(\vect{r}) \left(\abs{\nu_n(\vect{r})}^2 - \abs{\nu_p(\vect{r})}^2\right)\beta, 
  \end{multline}
  where $\beta =(\n_n - \n_p)/(\n_n + \n_p)$.  The dependence on
  neutron and proton densities of the bare coupling constants should
  satisfy isospin symmetry:
  \begin{align}
    g\bigl(\n_n(\vect{r}\bigr), \n_p(\vect{r})) &= g\bigl(\n_p(\vect{r}), \n_n(\vect{r})\bigr), \\
    h\bigl(\n_n(\vect{r}\bigr), \n_p(\vect{r})) &= h\bigl(\n_p(\vect{r}), \n_n(\vect{r})\bigr).
  \end{align}
\end{subequations}
Since in measured nuclei one has predominantly $N\ge Z$, see
\cref{fig:driplines}, a phenomenological analysis that leads to a
larger apparent coupling for protons than for neutrons can be
reconciled with renormalized coupling constants
$g_{\mathrm{eff}}(\vect{r}) < 0$ and $h_{\mathrm{eff}}(\vect{r}) >0$.

An additional subdominant term of the type
\begin{gather} \label{eq:spin}
  \tilde{\mathcal{E}}_\text{spin} = 
  \alpha_1 \left ( \vect{s}_n^2 + \vect{s}_p^2 \right ) + \alpha_2 \vect{s}_n\cdot \vect{s}_p , 
\end{gather}
should be considered as well for odd nuclei. The contribution of spin densities 
is typically much smaller than the contributions of the densities in nuclei,
$\int\d^3\vect{r}\;\n_{n, p}(\vect{r}) \gg \Abs{\int \d^3\vect{r}\;
  \vect{s}_{n, p}(\vect{r})}$, as in even-even nuclei
$\vect{s}_{n, p}(\vect{r}) \equiv 0$, and thus these terms will play
a noticeable role in odd $A$ and odd $N$-odd $Z$ nuclei
mainly~\cite{Schunck:2010}.  The term proportional to $\alpha_2$
will be important mostly in odd-odd nuclei.  These type of
contributions will affect in particular $\beta$-decay matrix
elements.

The structure of the double-humped fission barriers also depends
critically on the character of shell-corrections (see
\cref{fig:fissionpath}), and is thus sensitive to the single-particle
spectrum structure.  Hence, fission properties may be tuned by
adjusting all of the subdominant terms discussed above without
degrading the ability of the functional to fit masses and charge
radii.  

We now have a clear path to refine the structure of the SeaLL1
\gls{NEDF}, by systematically adding physically motivated parameters
in order to better describe nuclear physics observables.  While the properties 
of the simple SeaLL1
functional as presented here are quite reasonable without any fine tuning,
there is room for substantial improvement.  For example, one can consider spin-orbit terms
\eqref{eq:NEDF_so1} with $W_0\neq W_1$, gradient terms
\eqref{eq:NEDF_grad1_ext} with $\eta_0\neq\eta_1$, gradient terms
modifying the nucleon effective masses \eqref{eq:tau_n1}, and density
dependent pairing terms \eqref{eq:NEDF_pairing1} with both couplings
$g_{\mathrm{eff}}$ and $h_{\mathrm{eff}}$ non-vanishing.  Subdominant
corrections can be made to the symmetry energy \eqref{eq:SL2}
with $a_1-b_1n_0^{1/3}\neq 0$ and $c_1\neq0$.  Even the
incompressibility $K_0=\tfrac{6}{5}\varepsilon_F-12\varepsilon_0+2a_0n_0^{2/3}$ 
(if $\varepsilon_0$ and $n_0$ are fixed) can be changed 
by $\approx \pm \SI{20}{MeV}$ with the
parameter $a_0$,
see Eq~ \eqref{eq:K0} and Fig.~\ref{fig:hfbfit_a0}.

The next
step is to account for correlation energies; the center-of-mass
corrections, which, in the case of self-bound systems, present some
challenges~\cite{Engel:2007, Barnea:2007, Giraud:2008b, Giraud:2008,
  Giraud:2008a, Messud:2009, Messud:2011, Messud:2013}.  Accounting
for the center of mass correction~\cite{Butler:1984, Goriely:2003}, the
correction due to particle number projection~\cite{Samyn:2004}, the
vibration correlation energy
correction~\cite{Bender:2004a, Reinhard:2016}, the angular momentum
projection~\cite{Delaroche:2010, Egido:2004, Bender:2004,
  Goriely:2005, Reinhard:2016, Bender:2005, Bender:2006}, and Wigner
energy~\cite{Goriely:2015, Goriely:2016} should reduce the \gls{rms} energy
from about \SI{1.7}{MeV} to about \SI{0.5}{MeV}.  Further improvement
may require a proper accounting for quantum chaos like
effects~\cite{Bohigas:2002, *Bohigas:2002E, Aberg:2002, Olofsson:2006,
  Olofsson:2008, Molinari:2004, Molinari:2006, Hirsch:2004,
  Hirsch:2005, Barea:2005}.

\subsection{Nuclear Dynamics and Time-Dependent DFT}
  
One of the main advantages of DFT is the ability to also describe nuclear 
dynamics with the same \gls{NEDF} as for static properties. In time-dependent 
phenomena, additional terms of the \gls{NEDF} become active. We could especially 
consider two types of entrainment terms.
Such terms are never discussed in any standard theory of large amplitude 
collective motion in nuclear physics~\cite{Baranger:1978, Brink:1976, Ring:2004}, 
despite being allowed by symmetry. They are as natural to consider in the 
presence of mixed proton and neutron superfluids in neutron stars as they are 
in mixtures of \ce{^3He} and \ce{^4He} superfluids~\cite{Andreev:1975,
Volovik:1975a, Vardanyan:1981}. 

Entrainment (the Andreev-Bashkin effect) was predicted by
\textcite{Andreev:1975} to occur in superfluid mixtures of $^3$He and
$^4$He, and is rather surprising at first sight, since superfluids are
expected to flow without resistance.  In particular, one might have
expected that if somehow one would bring into motion only one
superfluid component, superfluidity will have the consequence that the
other component remains at rest. The entrainment
term~\eqref{eq:NEDF_entrain} is indeed dissipationless, and thus it
does not violate superfluidity, but allows the motion of one
superfluid to influence (entrain) the other. It is natural 
to expect a similar phenomenon to arise in nuclei, where proton and neutron 
(super)fluids can coexist. 
The entrainment term is
Galilean invariant and in nuclear systems has the form
\begin{gather}
  \mathcal{E}_\text{entrain} = 
  g_{\text{ent}}\left (\frac{n_nn_p}{n^2}\right ) \frac{n}{2m} \Abs{
    \frac{\vect{j}_n}{\n_n} - \frac{\vect{j}_p}{\n_p}}^2,
  \label{eq:NEDF_entrain}
\end{gather}
where $\vect{j}_{n, p}$ are the density currents~\eqref{eq:curr}. 
Since this type of coupling between neutron and proton fluids is
absent when either density vanishes we require that
$g_{\text{ent}}(0)=0$.  The requirement that the total kinetic energy
is always positive leads to the condition $x+g_{\text{ent}}(x)>0$.
Entrainment should also plays a role in neutron 
stars and has been studied intermittently since 1975~\cite{Volovik:1975a, 
Vardanyan:1981, Alpar:1984x, Borumand:1996, Babaev:2004, Gusakov:2005, 
Chamel:2006, Chamel:2013a,  Kobyakov:2017}.
The significant effect of this term is seen in the dynamics only, when
the motion of one fluid will drag along the other, and therefore the
presence of such an additional term will affect strongly the
excitation energies of isovector modes such as the \glspl{GDR} and the
Thomas-Reiche-Kuhn sum rule.  The simplest choice for this coupling is
$g_{\text{ent}}(x) = \alpha x$ with $1+\alpha >0$, which allows for negative 
values of $g(x)$. \textcite{Borumand:1996} recommend 
$g_{\text{ent}}(x)\propto x^{2/3}$, which would restrict $g(x)\ge 0$ for small
values of $x$.

A second type of entrainment contribution can be introduced as
well, with which one can control the Gamow-Teller transitions and
$\beta$-transition matrix elements.
\begin{gather}
  \tilde{\mathcal{E}}_\text{spin entrain} = 
  \tilde{g}_{\text{ent}}\left (\frac{n_nn_p}{n^2}\right ) \frac{n}{2m} \Abs{
    \frac{\vect{J}_n}{\n_n} - \frac{\vect{J}_p}{\n_p}}^2, 
\end{gather}
where $\vect{J}_{n, p}$ are the spin-density currents~\eqref{eq:spin_curr}.

\glsreset{NEDF}
\section{Conclusions}\label{sec:conclusions}

The \gls{NEDF} presented here is physically intuitive, and provides a
clear strategy for further improving the quality of mass fits by
separating contributions of various energy scales in the $\chi_E$ of
nuclear masses.  In this respect, the approach outlined here and
similar ideas used
before by \textcite{Bertsch:2005a}, is similar in spirit to an
effective field theory.  Our starting point was a generalization of
the liquid drop model as suggested by \textcite{Weizsacker:1935},
which aligns with the Hohenberg and Kohn~\cite{HK:1964} formulation of
\gls{DFT} in terms of neutron and proton densities only.  This
formulation allows us to evaluate proton and neutron densities, and
thus the charge radii as well, and the binding energies of 2375 nuclei
with an accuracy superior to the Bethe-Weizsäcker mass formula, but
with the same number of parameters.  Using this as a starting point, three
additional parameters were identified to produce a minimal \gls{NEDF},
in the spirit of the Kohn-Sham \gls{LDA}
formulation~\cite{Kohn:1965fk} of the \gls{DFT}, which is extended to
account for the presence of pairing correlations, shell effects, and
the density dependence of the symmetry energy.  The \gls{NEDF}
developed in this work, which we call SeaLL1, contains thus seven
significant parameters, each clearly related to specific properties of
nuclei.

The SeaLL1 \gls{NEDF} describes the nuclear masses of \num{606}
even-nuclei from the AME2012 evaluation~\cite{Audi:2012, Wang:2012}
with a mean energy error of \SI{0.93}{MeV} and a standard deviation
\SI{1.46}{MeV}, two-neutron and two-proton separation energies with
rms errors of \SI{0.69}{MeV} and \SI{0.59}{MeV} respectively, and the
charge radii of \num{345} even-even nuclei~\cite{Angeli:2013} with a
mean of \SI{0.022}{fm} and a standard deviation of $\SI{0.025}{fm}$.
 
Since in SeaLL1 the effective nucleon mass is equal to the bare mass one
can naturally expect that nuclear level densities~\cite{Egidy:2005}
will be described rather accurately, along with the single-particle
spectra around the Fermi level, unlike many phenomenological
\glspl{NEDF}.
The quality of the single-particle spectra are typically better than
in the case of previous \glspl{NEDF}, even though we did not include
them in the fit.

Nuclear and neutron matter properties are also well reproduced in SeaLL1.
One needs only two parameters to reproduce the symmetric
nuclear binding energy and saturation density. We find a reasonable value for 
the isoscalar nuclear incompressibility, $K_0=\SI{230}{MeV}$, 
although the saturation density is a bit lower than
the canonical value $\SI{0.16}{fm^{-3}}$.  The saturation density is
not well constrained by the mass fits alone, but can be constrained by
also considering the charge radii as discussed in
\cref{fig:hfbfit_rho}.
Two additional parameters control the symmetry properties of
nuclear matter.  
The symmetry energy $S=\SI{31.7}{MeV}$, its density dependence,
the neutron skin thickness $\SI{0.131}{fm}$ of $^{208}$Pb, the
compressibility of nuclear matter all have reasonable values.
SeaLL1 also incorporates information about the \gls{EoS} of pure neutron matter from
quantum Monte Carlo calculations with chiral effective field theory NN
interactions at \gls{N3LO} level and NNN interactions at the \gls{N2LO} level.
The addition
of quartic isovector terms $\propto \beta^4$ permit the \gls{NEDF} to
match the neutron matter \gls{EoS} without significantly affecting the
global mass fit.  We thus find that nuclear masses and the neutron
matter \gls{EoS} are largely uncorrelated, a conclusion somewhat at odds with
previous analyses.

A gradient term with a single parameter controls the diffuseness of
the nuclear surface and the nuclear surface tension.
Two additional parameters are required to describe the
spin-orbit interaction and the pairing correlations.

We have identified the respective role of the parameters of the SeaLL1 
\gls{NEDF} by using a principal component analysis. We have established that a 
number of parameters play an insignificant role in the mass fit. Their values 
can be varied significantly without affecting the quality of the $\chi_E$.  We 
refer to these as insignificant or subdominant parameters, and identify how 
they can be used to fine-tune the values of other observables.

Looking ahead, we note that a number of important nuclear observables such as 
the position of the \gls{GDR}, the Gamow-Teller resonances, the
Thomas-Kuhn-Reiche sum-rule, the nuclear compressibility and correspondingly 
the position of the giant monopole resonances, the dipole electric 
polarizability, the neutron skin thickness and the density dependence of the 
symmetry energy, depend on parameters which can be either freely adjusted 
(spin-orbit splittings and/or effective masses) without affecting the accuracy 
of the ground state binding energies, or which affect very little the ground 
state properties.  In this respect SeaLL1 stands apart from previous 
\glspl{NEDF}, in which many of these properties where often included in the 
fits.

\begin{acknowledgments}
  We are grateful to George F.~Bertsch for numerous discussions and
  suggestions, to Rebecca Surman, Gabriel Martinez-Pinedo and Meng-Ru
  Wu for providing the data for the r-process trajectories, and to Jeremy
  W.~Holt, David~B.~Kaplan, J\'er\^ome Margueron, Piotr Magierski, and
  Sanjay Reddy for comments.  This work was supported in part by US
  DOE Grant No.~DE-FG02-97ER-41014, a WSU Seed Grant, and the
  Scientific Discovery through Advanced Computing (SciDAC) program
  funded by U.S.~Department of Energy, Office of Science, Advanced Scientific
  Computing Research and Nuclear Physics. It was partly performed
  under the auspices of the US Department of Energy by the Lawrence
  Livermore National Laboratory under Contract DE-AC52-07NA27344.
  Some calculations reported here have been performed at the University of
  Washington Hyak cluster funded by the NSF MRI Grant No.~PHY-0922770,
  and with computing support from the Lawrence Livermore  National
  Laboratory (LLNL) Institutional Computing Grand Challenge program.
\end{acknowledgments}

\appendix
\section{Orbital-Free Functional}\label{app:orbit-free-funct}

Here we discuss some details of the orbital-free theory described in
\cref{sec:orbit-free-functional}.

As mentioned there, the main challenge in formulating an orbital-free
theory is to express terms with the auxiliary densities $\tau_{n, p}$,
$\nu$, $\vect{J}_{n, p}$, and $\vect{j}_{n, p}$ by an appropriate
functional of the number densities $\n_{n, p}$.  One approach is to
start with a semiclassical expansion.  Neglecting the spin-orbit
interaction~\eqref{eq:NEDF_so}, the kinetic density $\tau$ admits the
following semiclassical expansion~\cite{Jones:1989, Dreizler:1990lr,
  Brack:1997}:
\begin{gather}
  \label{eq:kin}
  \tau \approx
    \overbrace{\tfrac{3}{5}(3\pi^2)^{2/3} \n^{5/3}}^{\tau_{TF}[\n]}
    + \overbrace{\frac{1}{9} \Norm{\grad\sqrt{\n}}^2}^{\tau_{2}[\n]}
    + \tau_4[\n]
    + \cdots
\end{gather}
The factor of $1/9$ can be derived rigorously for smoothly varying
densities, along with higher order terms discussed in \cref{eq:grad4}
below.  This should be compared with the factor of unity originally
suggested by \textcite{Weizsacker:1935}, later shown to be valid only
if the density has small amplitude rapid
oscillations~\cite{Brack:1997, Dreizler:1990lr, Jones:1989}.  For
nuclei, the semiclassical result is relevant for the bulk, but gives
incorrect asymptotic behavior, while Weizsäcker's result reproduces
the correct asymptotic behavior, but is a poor approximation in the
bulk, see~\cite{Brack:1985x} for a discussion.  Resolving this tension
is an active area of research in \gls{DFT}, and many suggestions have
been compared~\cite{Garcia-Aldea:2007}.

The simplest option is to treat the coefficient $1/9 = \eta$ as a
phenomenological parameter, since gradient terms can also be generated
by interactions~\cite{Negele:1972, Negele:1975, Gebremariam:2010x}.
Fitting the nuclear masses yields values of $\eta$ close to \num{0.5},
roughly half-way between the semiclassical and Weizsäcker values.
\textcite{Stocker:1988} used a similar approach in order to discuss
the anomaly in the nuclear curvature energy -- the term in the nuclear
mass formula $\propto A^{1/3}$.

Another appealing approach suggested by~\textcite{DePristo:1987} and
advocated in~\cite{Dreizler:1990lr} is to use a Padé approximant
$F(X)$ to interpolate between the semiclassical and asymptotic
results:
\begin{align}\label{eq:tau1}
  \tau &\approx \tau_{TF}[\n]F(X), 
  &
    X &= \frac{\tau_2[\n]}{\tau_{TF}[\n]}.
\end{align}
\textcite{DePristo:1987} motivate a rather complicated form $F(X)$,
but for nuclei, we find little improvement over the following
single-parameter form:
\begin{gather}
  \label{eq:F_GGA1}
  F(X) = \frac{1+(1+\bGGA) + 9\bGGA X^2}{1+\bGGA X}
  = \begin{cases}
    1+X & X \ll 1\\
    9X & X \gg 1.
  \end{cases}
\end{gather}
Note: the approximation $\eta \approx 1/9$ mentioned above is
implemented with $F(X) = 1 + 9\eta X$.

The next order in the semiclassical expansion of non-interacting
fermions~\cite{Dreizler:1990lr, Brack:1997} is:
\begin{gather}
  \tau_{4}[\n] = \frac{1}{810(3\pi^{2})^{2/3}}f(\n), 
  \label{eq:grad4}\\
  f(\n) = \n^{1/3}\left[
    \left(
      \frac{\grad\n}{\n}
    \right)^4
    - \frac{27}{8}\left(
      \frac{\grad\n}{\n}
    \right)^2
    \frac{\laplacian\n}{\n}
    + 3\left(
      \frac{\laplacian\n}{\n}
    \right)^2
  \right]. \nonumber
\end{gather}
This type of correction has been studied in nuclear physics and shown
to lead to quite accurate estimates of the kinetic energy density
within the extended Thomas-Fermi approximation~\cite{Brack:1976,
  Brack:1985x, Brack:1997}.  Within a \gls{DFT}, such terms can also
arise due to the finite range of the interactions in a matter similar
to some Skyrme interactions~\cite{Negele:1972, Negele:1975,
  Gebremariam:2010x}.  However, these terms -- even with adjustable
parameters -- do not significantly change the quality of the mass
fits, so we do not consider them in our main analysis.  Including them
perturbatively in the fit, however, does improve the fit of the charge
radii.  For example, fitting the overall coefficient reduce the charge
radii residual $\chi_r$ (see details in \cref{sec:hydrodynamic-model})
from $\chi_r\approx \SI{0.14}{fm}$ to $\chi_r\approx \SI{0.09}{fm}$.
Fitting each of the three terms independently further reduces the
residuals to $\chi_r\approx \SI{0.06}{fm}$.  Fourth-order terms are
neglected as they can lead to a complex behavior of the emerging
equation for the densities, which can be difficult to rationalize.
(See, for example, the analysis of fourth order differential equations
arising in case of non-local potentials by \textcite{Bulgac:1988x}.)
Higher order gradient corrections than \cref{eq:grad4} lead to an
unphysical behavior of the densities in the classically forbidden
regions.  Furthermore, the semiclassical expansion has an asymptotic
character~\cite{Jones:1989}, and corrections beyond second order do
not always improve the functional.  Finally, when using a properly fit
Padé approximant \cref{eq:F_GGA}, we find that
$\int \tau_{TF}[\n]F(X) - \tau_{TF}[\n] - \tau_2[\n] \d^{3}\vect{x}
\approx \int\tau_{4}[\n] \d^{3}\vect{x}$ for many nuclei.  Thus, the
Padé approximant \cref{eq:F_GGA} seems to incorporate the qualitative
effects of the $\tau_4[\n]$ term.  For these reasons, we do not
include fourth-order corrections $\tau_4[\n]$ in our orbital-free
theory.

When spin-orbit interactions are included, they modify the
semiclassical expansion.  Thus, to properly express the orbital-free
theory, we must consider both terms together.  The correct
semiclassical expansion of this combined energy density to second
order is~\cite{Brack:1976, Brack:1985x}:
\begin{multline}
  \Ekin + \Eso = \frac{\hbar^2}{2m}(\tau_n + \tau_p) + W_0\vect{J}\cdot\grad\n\\
  \approx \frac{\hbar^2}{2m}\left(
    \tau_{TF}[\n_n] + \tau_{TF}[\n_p] + \tau_2[\n_n] + \tau_2[\n_p]
  \right)  \\
  -\frac{W_0^2}{2}\frac{2m}{\hbar^2} \n (\grad\n)^2.
\end{multline}
Note that the sign of the last term differs from the expression (7)
in~\cite{Brack:1976} which contains only the kinetic component.  The
result here combines both the kinetic and spin-orbit contributions,
altering the sign.
(The remaining terms in the functional only alter the mean-field
potential, and so they do not affect this result.)

This expansion suffers the same problems as the pure semiclassical
expansion of the kinetic energy \cref{eq:kin}.  Thus, for the reasons
discussed above, we replace $\tau_{TF} + \tau_2$ with the Padé
approximant \cref{eq:F_GGA}.  In principle, a similar correction could
be used with the spin-orbit term, however, this term has the form
$\n(\grad\n)^2$ instead of $\tau_2 \propto (\grad\n)^2/\n$.  It is
therefore suppressed in the tails and does not effect the asymptotic
behavior of the nuclear density profile.  Note that the scaling is
similar to the gradient correction.  For this reason we keep the
semiclassical form, but refit the coefficient $\eta_s$ to compensate
for any inaccuracies.

The equations that determine the equilibrium densities of a nucleus in
the orbital-free theory are obtained by minimizing the energy of a
given nucleus $E(N, Z) = \int\d^3\vect{r}\; \mathcal{E}[\n_n, \n_p] $
with respect to the densities, while constraining the total numbers of
neutrons $N$ and protons $Z$ with two chemical potentials
$\mu_{n, p}$:
\begin{subequations}
  \begin{gather}
    -\frac{\hbar^2}{2m} \grad\cdot\left(
      \frac{F'(X_q)}{9}\vect\nabla \n_q^{1/2}
    \right)
    + U_q\n^{1/2}_q =
      \mu_q\n^{1/2}_q, \label{eq:psi}\\
      U_q=\frac{ \partial \mathcal{E} [\n_n, \n_p] }{ \partial \n_q}, 
      \text{ for } q \in \{n, p\}.
  \end{gather}
\end{subequations}
We present these here as the inclusion of $F(X)$ acts as a
density-dependent effective mass.  No such complication appears in the
\gls{HFB} formulation, which proceeds as described in~\cite{Jin:2017}.

\section{Orbital-Free \gls{NEDF} parameters}\label{sec:hydrodynamic-model}

\begin{table*}[pt]
  \centering
  \resizebox{\textwidth}{!}{%
  \sisetup{round-mode = places,
           %round-precision = 3,
           group-digits = false,
           table-number-alignment = center,
           scientific-notation = fixed, fixed-exponent = 0,
           omit-uncertainty = true,
           %table-format = 4.4,
           %table-space-text-pre = xx,
           %table-space-text-post = xx,
           } %
  %\begin{ruledtabular}
    \begin{tabular}{cS[table-format=1.4]S[table-format=1.4]SS[table-format=+2.2]S[table-format=3.2]S[table-format=+3.2]S[table-format=+3.3]S[table-format=+3.2]S[table-format=3.3]S[table-format=3.1]S[table-format=+3.2]S[table-format=3.2]S[table-format=2.3]S[table-format=1.2]S[table-format=1.2]}
      NEDF & {$\eta$}& {$\eta_{s}$}& {$W_{0}$}& {$a_{0}$}& {$a_{1}$}& {$a_{2}$}& {$b_{0}$}& {$b_{1}$}& {$b_{2}$}& {$c_{0}$}& {$c_{1}$}& {$c_{2}$}& {$\delta$}& {$\chi_{E}$}& {$\chi_{r}$} \\
           & & [\si{fm^3}]& [\si{MeV.fm^5}]& [\si{MeV.fm^2}]& [\si{MeV.fm^2}]& [\si{MeV.fm^2}]& [\si{MeV.fm^3}]& [\si{MeV.fm^3}]& [\si{MeV.fm^3}]& [\si{MeV.fm^4}]& [\si{MeV.fm^4}]& [\si{MeV.fm^4}]& [\si{MeV}]& [\si{MeV}]& [\si{fm}] \\
      \hline
      0& 4.719+-0.130 e-01& 0& 0& 0& 1.311+-0.180 e+02& 0& -7.41570+-0.35528 e+02& -1.43+-0.38 e+02& 0& 9.4050+-0.7253 e+02& 0& 0& 1.146+-0.983 e+01& 2.589679& 0.144826\\
      1& 4.742+-0.222 e-01& 0& 0& 0& 1.226+-1.283 e+02& 0& -7.38302+-0.35404 e+02& -1.28+-2.75 e+02& 0& 9.3438+-0.7220 e+02& 0& 0& 1.147+-0.732 e+01& 2.581880& 0.134558\\
      2& 4.743+-0.199 e-01& 0& 0& 0& 1.201+-1.025 e+02& 0& -7.40226+-0.36225 e+02& -1.23+-2.21 e+02& 0& 9.3826+-0.7393 e+02& 0& 0& 0& 2.713880& 0.140327\\
      1r& 4.807+-0.088 e-01& 0& 0& 0& 1.359+-0.017 e+02& 0& -7.02003+-0.02463 e+02& -1.57+-0.04 e+02& 0& 8.6133+-0.0481 e+02& 0& 0& 1.175+-0.031 e+01& 2.707050& 0.051065\\
      3& 4.800+-6.826 e-01& 0& 0& -0.01+-1.35 e+03& 1.250+-8.752 e+02& 0& -0.69508+-5.34150 e+03& -0.13+-5.34 e+03& 0& 0.8921+-5.2513 e+03& -0.0+-7.9 e+03& 0& 1.141+-0.983 e+01& 2.582060& 0.138972\\
      3n& 4.739+-0.214 e-01& 0& 0& -7.59+-9.50 e+00& 1.957+-0.675 e+02& -2.207+-0.683 e+02& -7.07006+-0.02673 e+02& -3.22+-7.91 e+02& 9.13194+-7.89586 e+02& 9.0250+-0.4084 e+02& 0.1+-1.5 e+03& -0.8738+-1.5222 e+03& 1.157+-0.981 e+01& 2.569704& 0.133055\\
      3nr& 4.815+-0.096 e-01& 0& 0& -7.63+-0.00 e+00& 1.954+-0.046 e+02& -2.204+-0.046 e+02& -6.74608+-0.08333 e+02& -3.17+-0.12 e+02& 8.76220+-0.13353 e+02& 8.3729+-0.1648 e+02& 7.5+-1.3 e+01& -8.0321+-0.1987 e+02& 1.245+-0.182 e+01& 2.671408& 0.050314\\
      E& 4.885+-0.136 e-01& 0& 0& 0& 3.460+-0.202 e+01& 0& -7.40950+-0.35272 e+02& 6.51+-0.38 e+01& 0& 9.3863+-0.7195 e+02& 0& 0& 1.121+-0.134 e+01& 2.643127& 0.128690\\
      Er& 4.957+-0.118 e-01& 0& 0& 0& 3.298+-0.105 e+01& 0& -7.07394+-0.03698 e+02& 6.21+-0.20 e+01& 0& 8.7091+-0.0714 e+02& 0& 0& 1.271+-0.082 e+01& 2.740375& 0.051105\\
      En& 4.866+-0.136 e-01& 0& 0& 0& 3.401+-0.203 e+01& -6.660+-0.203 e+01& -7.41546+-0.35145 e+02& 6.40+-0.38 e+01& 5.62093+-0.31986 e+02& 9.4002+-0.7175 e+02& 0& -8.3090+-0.7175 e+02& 1.126+-0.093 e+01& 2.619855& 0.132884\\
      Enr& 4.970+-0.118 e-01& 0& 0& 0& 3.254+-0.112 e+01& -6.513+-0.112 e+01& -7.07031+-0.03705 e+02& 6.12+-0.21 e+01& 5.30344+-0.03379 e+02& 8.7015+-0.0715 e+02& 0& -7.6103+-0.0715 e+02& 1.251+-0.098 e+01& 2.736572& 0.051131\\
      En-rho& {$1/9$}& 4.9731+-0.1631 e+00& 0& 0& 2.971+-0.105 e+01& -6.229+-0.105 e+01& -6.72625+-0.01336 e+02& 5.59+-0.20 e+01& 5.01277+-0.01414 e+02& 9.3485+-0.0000 e+02& 0& -8.2573+-0.0000 e+02& 1.178+-0.068 e+01& 2.643758& 0.052903\\
      Enr-rho& {$1/9$}& 5.0397+-0.1572 e+00& 0& 0& 2.952+-0.104 e+01& -6.211+-0.104 e+01& -6.72986+-0.01290 e+02& 5.56+-0.20 e+01& 5.01986+-0.01415 e+02& 9.3485+-0.0000 e+02& 0& -8.2573+-0.0000 e+02& 1.372+-0.624 e+01& 2.677247& 0.051599\\
      En-so& {$1/9$}& 5.4751+-0.0438 e+00& 7.62e+01& 0& 1.368+-0.000 e+02& -1.694+-0.000 e+02& -6.69776+-0.00000 e+02& 5.15+-0.15 e+01& 5.02814+-0.01522 e+02& 9.3485+-0.0000 e+02& 0& -8.2573+-0.0000 e+02& 1.173+-0.027 e+01& 3.184413& 0.048368 \\
      & {$\kappa$}\\
      \hline
      En-pade-1& {\num{6.500e-02}}& 5.0941+-0.0031 e+00& 0& 0& 3.014+-0.000 e+01& -6.273+-0.000 e+01& -6.72785+-0.00150 e+02& 5.67+-0.00 e+01& 5.00620+-0.00151 e+02& 8.0220+-0.0000 e+02& 0& -6.9308+-0.0000 e+02& 1.040+-0.000 e+01& 2.818597& 0.067707\\
      En-pade-2& {\num{1.500e-01}}& 4.6365+-0.1565 e+00& 0& 0& 3.037+-0.110 e+01& -6.296+-0.110 e+01& -6.72213+-0.01386 e+02& 5.72+-0.21 e+01& 4.99610+-0.01474 e+02& 8.0141+-0.0000 e+02& 0& -6.9229+-0.0000 e+02& 1.149+-1.032 e+01& 2.886304& 0.070308\\
      En-pade-3& {\num{2.000e-01}}& 4.4318+-0.1291 e+00& 0& 0& 3.033+-0.109 e+01& -6.291+-0.109 e+01& -6.71889+-0.01195 e+02& 5.71+-0.20 e+01& 4.99374+-0.01456 e+02& 8.0097+-0.0000 e+02& 0& -6.9185+-0.0000 e+02& 1.194+-0.069 e+01& 2.927845& 0.071709\\
      En-pade-4& {\num{3.000e-01}}& 4.2098+-0.0261 e+00& 0& 0& 3.150+-0.020 e+01& -6.409+-0.020 e+01& -6.72625+-0.00000 e+02& 5.93+-0.04 e+01& 4.97894+-0.00374 e+02& 8.0198+-0.0000 e+02& 0& -6.9286+-0.0000 e+02& 1.207+-0.000 e+01& 3.108949& 0.073894\\
      Hydro& {\num{2.000e-01}}& 3.3696+-0.0499 e+00& 0& 0& 5.088+-0.074 e+01& -8.347+-0.074 e+01& -6.85597+-0.00192 e+02& 9.49+-0.14 e+01& 4.75237+-0.01482 e+02& 8.2876+-0.0000 e+02& -160& -5.5964+-0.0000 e+02& 0& 2.861315& 0.041483 \\
%      & & & & & & & & & & & & & {$g_0$}\\
%      & & & & & & & & & & & & & {\clap{[\si{MeV.fm^3}]}} \\
      & & & & & & & & & & & & & {\clap{${g_0}$ [\si{MeV.fm^3}]}} \\

      \hline
      SeaLL1& {N/A}& 3.93& 73.5& 0& 64.3& -96.8& -684.5& 119.9& 449.2& 827.26& -256& -461.7& {\num{-200}}& 1.74& 0.034
    \end{tabular}
  }%\resizebox  
  %\end{ruledtabular}
  \caption{\label{table:NEDF_A_supp} 
Fit parameters and residuals for the various NEDFs.
The top set of functionals uses the simplified form $F(X) = 1+9\eta X$ while the second set uses the form in Eq.~\eqref{eq:F_GGA1} with the parameter $\kappa$ instead.
The SeaLL1 parameters are shown in the last row for comparison.
}
\end{table*}

\begin{table}[htbp]
  \sisetup{round-mode = places,
           round-precision = 3,
           } %
  \begin{ruledtabular}
    \begin{tabular}{ccccccccc}
           & &&&&&& \multicolumn{2}{c}{Neutron skin}\\
      NEDF & {$\n_{0}$}& {$-\varepsilon_0$}& {$K$}& {$S$}& {$L$}& {$L_{2}$}& {\ce{^{208}Pb}}& {\ce{^{48}Ca}} \\
           & [\si{fm^{-3}}]& & & & & & [\si{fm}]& [\si{fm}] \\
      \hline
0& 0.136& 15.24& 222.5& 26.8& 34.1& 32.8& 0.082& 0.118\\
1& 0.136& 15.22& 222.4& 26.7& 35.9& 34.7& 0.087& 0.123\\
2& 0.136& 15.21& 222.2& 26.7& 36.8& 35.6& 0.089& 0.125\\
1r& 0.148& 15.48& 227.7& 27.1& 30.9& 29.6& 0.078& 0.116\\
3& 0.136& 15.21& 216.5& 26.7& 34.7& 33.4& 0.088& 0.124\\
3n& 0.137& 15.20& 218.2& 30.0& 29.3& 16.7& 0.068& 0.107\\
3nr& 0.147& 15.44& 222.9& 31.0& 31.2& 15.5& 0.068& 0.107\\
E& 0.136& 15.28& 223.1& 29.7& 68.2& 66.9& 0.159& 0.174\\
Er& 0.147& 15.53& 228.1& 30.6& 70.2& 68.9& 0.161& 0.176\\
En& 0.136& 15.27& 222.9& 30.1& 29.1& 66.1& 0.152& 0.172\\
Enr& 0.147& 15.53& 228.2& 31.1& 31.1& 68.3& 0.156& 0.174\\
En-rho& 0.160& 15.85& 234.4& 32.3& 33.5& 68.9& 0.138& 0.149\\
Enr-rho& 0.160& 15.87& 234.6& 32.4& 33.5& 68.6& 0.138& 0.149\\
En-so& 0.160& 15.74& 233.1& 32.2& 33.5& 65.4& 0.120& 0.139\\
En-pade-1& 0.160& 15.86& 234.5& 32.4& 33.5& 69.6& 0.157& 0.176\\
En-pade-2& 0.160& 15.83& 234.2& 32.3& 33.5& 69.9& 0.166& 0.189\\
En-pade-3& 0.160& 15.82& 234.1& 32.3& 33.5& 69.8& 0.170& 0.194\\
En-pade-4& 0.160& 15.85& 234.4& 32.3& 33.5& 71.6& 0.181& 0.206
    \end{tabular}
  \end{ruledtabular}
  \caption{\label{table:NEDF_B} Saturation, symmetry, and neutron skin
    properties for the various NEDFs.  All values in \si{MeV} unless
    otherwise specified.}
\end{table}

We start by considering the functional with the simplified kinetic
energy
\begin{gather}
  \Ekin[\n_n, \n_p] = \frac{\hbar^2}{2m}\sum_{q=n, p}\tau_{TF}[\n_q]F(X_q), 
\end{gather}
where $\tau_{TF}$, $X_q$, and $F(X)$ are given in Eqs. \eqref{eq:tau1}
and \eqref{eq:F_GGA1}.

As discussed above, when using the simplified form
$F(X) = 1 + 9\eta X$, the best fit value of $\eta \approx 0.5$.  One
might naïvely think that this corresponds to a dynamical theory of
superfluid neutron and proton pairs with an effective nucleon pair
mass $m_{\text{eff}} \approx 2 m$ (see i.e.~\cite{Forbes:2012b} and
references therein).  Such a theory with $\eta=0.5$, however, leaves
the potentials $U_q$ wrong by a factor of 2.  To correctly describe a
dynamical theory of superfluid neutron and proton pairs, one would
need a value of $\eta = 1/4$.  Thus, in this approximation, the
parameter $\eta$ must simply be interpreted as an approximate way to
control the falloff of the densities in the surface region where the
interaction effects are still strong.

We now consider our \gls{NEDF} as an hydrodynamic model for nuclei and
fit the parameters to the same $N_E = 2375$ measured nuclear masses
with $A\geq 16$ from~\cite{Audi:2012, Wang:2012} used to fit the
liquid drop models in \cref{table:liquid_drop}.  However, unlike the
liquid drop model, our hydrodynamic model allows us also to consider
properties of the density distribution.  Thus, we also fit the
$N_r = 883$ nuclear charge radii from~\cite{Angeli:2013} with
$\chi_r^2 = \sum\abs{\delta r}^2/N_r$.  When we include the charge
radii in the fit, we minimize the following quantity
$\chi^2_E/(\SI{3}{MeV})^2 + \chi^2_r/(\SI{0.05}{fm})^2$ which roughly
equalizes the weight of the mass and radii contributions in the fit.

At this point, we have 7 parameters in our \gls{NEDF}: $\eta$,
$a_{0, 1}$, $b_{0, 1}$, and $c_{0, 1}$ (the $j=2$ parameters are fixed
by the neutron matter \gls{EoS}).  In addition, we include by hand the
conventional even-odd staggering \cref{eq:odd-even} with a coefficient
$\delta$ to describe pairing correlations, even though this has very
little significance in the fits.  The results of various fits
scenarios we have considered are summarized below in
\cref{table:NEDF_A_supp} where we present sets of parameters for
various fit strategies, and in \cref{table:NEDF_B} where we present
the saturation, symmetry, and neutron skin properties.

We have considered the following type of fits:
\begin{description}
\item[NEDF-0] A six parameter least-squares fit of the $N_E = 2375$
  nuclear masses~\cite{Audi:2012, Wang:2012} including $\eta$, $b_0$,
  $c_0$, $a_1$, $b_1$, and $\delta$ but setting the nucleon charge
  form factors \cref{eq:FF} ${G}^{p}_{E} \equiv 1$ and
  ${G}^{n}_{E} \equiv 0$.
\item[NEDF-1] The same as NEDF-0, but including the measured charge
  form factors.  Comparing with NEDF-0 we see that the electric form
  factors are not significant for the overall mass fits, but slightly
  impact the charge radii at the \SI{0.01}{fm} level (for the reduced
  $\chi_r$).
\item[NEDF-2] The same as NEDF-1, but without the pairing parameter
  $\delta=0$.  Comparing with NEDF-1 we see that odd-event staggering
  is also relatively unconstrained at the level of \SI{0.1}{MeV} per
  nucleus.  This is consistent with the results from the mass formulas
  in \cref{table:liquid_drop}.
\item[NEDF-1r] The same as NEDF-1, but including the $N_r = 883$
  charge radii into the fit.  We see that there is significant room to
  improve the description of the charge radii without significantly
  degrading the mass fits.
\item[NEDF-3] The same as NEDF-1, but with all 8 parameters, including
  $a_0$ and $c_1$ that we omitted from the previous fits.  In
  conjunction with the principal component analysis shown in
  \cref{fig:PCA}, this fit demonstrates that the terms with parameters
  $a_0$ and $c_1$ are unconstrained.
\item[NEDF-3n] The same as NEDF-1, but with all 8 parameters,
  including $a_0$ and $c_1$ that we omitted from the previous fits,
  and the $\beta^4$ parameters for the terms quartic in isospin,
  constrained by the \gls{QMC} neutron matter
  \gls{EoS}~\cite{Wlazlowski:2014a} using Eqs.~\eqref{eq:nm}.  That
  the quality of the fit, isoscalar, and isovector parameters change
  very little, demonstrates that the neutron matter \gls{EoS} is
  essentially independent of the nuclear masses.
\item[NEDF-3nr] The same as NEDF-3n but including the charge radii as
  in fit NEDF-1r.  That the $a_0$ and $c_1$ terms are unconstrained
  for both masses and radii is also emphasized by this fit.
\end{description}
\begin{description}
\item[NEDF-E] Following the principal component analysis of NEDF-3n
  (discussed below) we find the combination $a_1 - b_1\n_0^{1/3}$ to
  be only weakly constrained by the mass fit.  To test this, we set
  $a_1 = b_1\n_0^{1/3}$ where $\n_0 = \SI{0.154}{fm^{-3}}$ is a
  constant.  The combination $a_1 - b_1\n_0^{1/3}$, to which the
  masses are insensitive, allows independent control the slope $L_2$
  of the symmetry energy (see \cref{eq:L}).  From the fits we see that
  this same combination also controls the neutron skin thicknesses.
\item[NEDF-Er] The same as NEDF-E but including the charge radii as in
  fit NEDF-1r.
\item[NEDF-En] This is our main fit.  It is the same as NEDF-E but
  includes the $\beta^4$ parameters adjusted to reproduce the neutron
  matter \gls{EoS} as in fit NEDF-3n.
\item[NEDF-Enr] The same as NEDF-En but including the charge radii as
  in fit NEDF-1r.
\end{description}

In all fits above, the parameter $\eta$ is around $1/2$, which
deviates from the Weisz\"{a}cker value $1/9$.  In our latest fits, we
fix $\eta = 1/9$ and introduce a new gradient term $\eta_s$.

From the equilibrium condition of symmetric nuclear matter we get a
relationship between $\tilde{a}_0$, $\tilde{b}_0$, and $\tilde{c}_0$
\begin{align}
  \label{eq:satrho}
  0 = \frac{3}{5} + \tilde{a}_0 + \frac{3}{2}\tilde{b}_0 + 2\tilde{c}_0
\end{align}
or by using the original parameters:
\begin{align} \label{eq:satrhoa}
  a_0 = -\frac{3\varepsilon_F}{5k_0^2} - \frac{3}{2}b_0 k_0 - 2c_0k_0^2
\end{align}
where $k_0 = \n_0^{1/3}, \quad \n_0 = 0.16$.  If $a_0$ is set to be 0,
there is a relationship between $b_0$ and $c_0$:
\begin{align}
  \label{eq:satrhoc}
  c_0 = -\frac{3\varepsilon_F}{10 k_0^4}- \frac{3b_0}{4k_0}
\end{align}

Using this relationship, the saturation density derived from the NEDF
will be fixed to be $\n_0 = 0.16$.
\begin{description}
\item[NEDF-En-rho] We fix $\eta = 1/9$ and add $\Egrad$ into the
  \gls{NEDF}.  The saturation density $\n_0$ is fixed to be $0.16$ by
  adding a constraint between $b_0$ and $c_0$.  Then the number of
  significant parameters in this \gls{NEDF} is reduced to 3.
\item[NEDF-Enr-rho] The same as NEDF-En-rho but including the charge
  radii as in fit NEDF-1r.
\end{description}

In our earlier fits, we do not include the contribution of spin-orbit
interaction, which is crucial for the proper description of nuclear
static properties.

\begin{description}
\item[NEDF-En-so] Following NEDF-En-rho, we add $\Eso$ into the
  \gls{NEDF}.  The spin-orbit strength $W_0$ is fixed to be the value
  suggested in \cite{Fayans:1998}.  The significant fitting parameters
  are the same with NEDF-En-rho.
\end{description}

When we fix $\eta = 1/9$ and neglect higher order \gls{ETF} expansion
in the kinetic energy, the asymptotic form of density can be proved to
be
\begin{gather}
  \n(r) \underset{r \to \infty}{ \longrightarrow} \frac{1}{r^2} e^{-r/a}, \quad a = \sqrt{-\frac{1}{36} \frac{\hbar^2}{2m} \frac{1}{\mu}}.
\end{gather}
where $\mu$ is the chemical potential (which is negative).
Unfortunately, the diffuseness $a$ is too small by a factor of 3
compared with the realistic nuclear surfaces, which corresponds to
$\eta = 1$ in the asymptotic region.  In order to obtain a nucleus
density with correct asymptotic behavior, we suggest using the
following Padé approximation in the representation of extended
Thomas-Fermi approximation for the kinetic density, see
Eqs.~\eqref{eq:tau1} and \eqref{eq:F_GGA1}:
\begin{gather}
  \tau_q = \tau_{TF, q} F(X)
\end{gather}
where the function $F(x)$ has the asymptotic behavior:
\begin{gather}
  F(X) = \begin{cases}
    1+X, & X \ll 1 \\
    9X, & X \gg 1
  \end{cases}
\end{gather}
In this approximation, we can get both correct behavior for the
nucleus density in the near and asymptotic region.  Through varying
the parameter $\kappa$ we obtain the following fits.
\begin{description}
\item[NEDF-En-pade-1] Following NEDF-En-so, we use the Padé
  approximation for the kinetic energy, and the parameter $b=0.065$
\item[NEDF-En-pade-2] Same with NEDF-En-pade-1, but $\kappa=0.15$
\item[NEDF-En-pade-3] Same with NEDF-En-pade-1, but $\kappa=0.2$
\item[NEDF-En-pade-4] Same with NEDF-En-pade-1, but $\kappa=0.3$
\end{description}

\begin{figure}[tb]
  \includegraphics[width=\columnwidth]{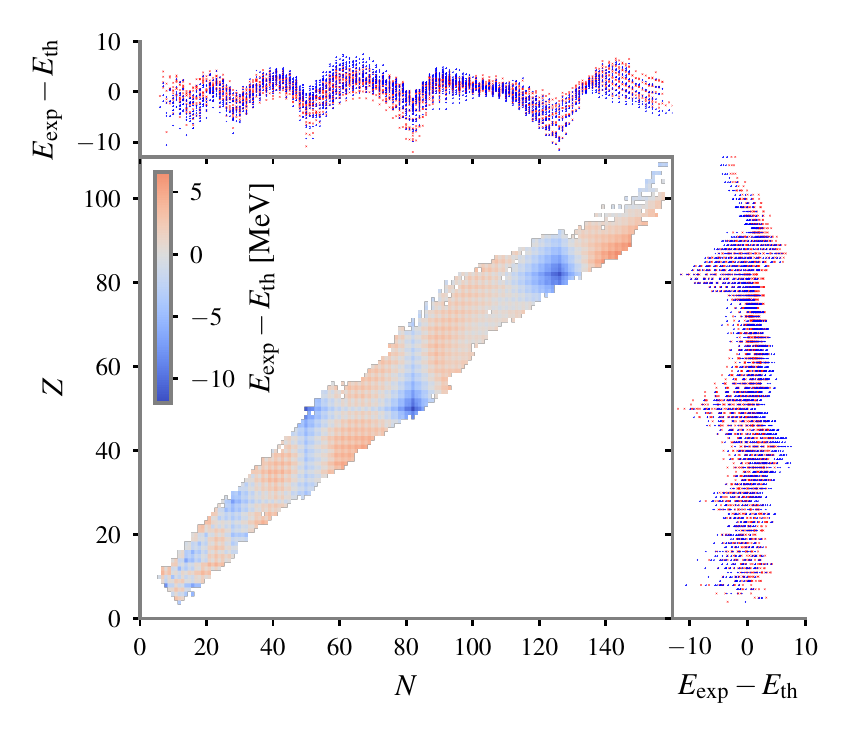}
  \caption{(Color online) The blue pluses show the results obtained using the
    orbital-free approximation with $\chi_E = \SI{2.86}{MeV}$, while
    the red crosses are the results of the fits using nuclear mass
    formula \cref{eq:masses}, with $\chi_E = \SI{2.64}{MeV}$.  When
    compared against each other, the \gls{rms} energy deviation
    between the two fits is $\Delta \chi_E = \SI{1.10}{MeV}$.  Thus,
    the orbital-free theory essentially reproduces the nuclear mass
    formula \cref{eq:masses}.  The main plot is the same as in
    \cref{fig:masses} in which one can see clearly the magic numbers
    separately for neutrons and protons.}
  \label{fig:be_zn}
\end{figure}

These fits are summarized in \cref{table:NEDF_A_supp}, with the
saturation and symmetry properties in \cref{table:NEDF_B}.  The
residuals for fit NEDF-1 are shown in \cref{fig:be_zn} and compared
with a fit to the nuclear with mass formula \cref{eq:masses}.

The reduced $\chi_E$ for these fits is comparable to that obtained
using the nuclear mass formulas \cref{eq:Bethe} with four parameters
(plus $\delta$) and \cref{eq:masses} with five parameters (plus
$\delta$).  This is consistent with our hypothesis that a \gls{NEDF}
for masses should contain no more than five significant parameters.
Note, however, that unlike the mass formulas, the \gls{NEDF} also
gives a good description of charge radii -- for which the mass formula
says nothing -- and provides access to nuclear dynamics.

\section{Principal Component Analysis}\label{sec:PCA1}

\newcommand{\myincludegraphics}[4][]{
  \begin{overpic}[#1]{#2}%
    \put #3 {\textbf{(#4)}}
  \end{overpic}
}

\begin{figure}[tp]
  %\subfloat[]{\includegraphics[width=\columnwidth]{PCA_1}}\\
  %\subfloat[]{\includegraphics[width=\columnwidth]{PCA_3}}
  \myincludegraphics[width=\columnwidth]{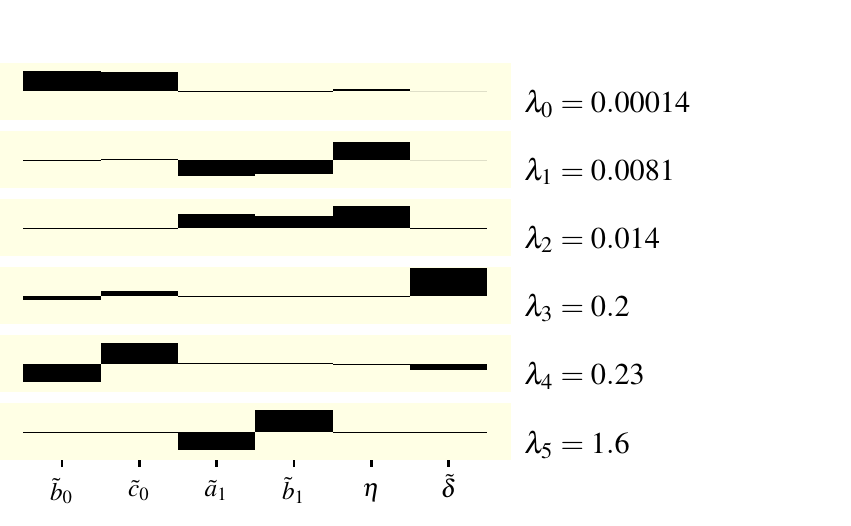}{(90,10)}{a}\\
  \myincludegraphics[width=\columnwidth]{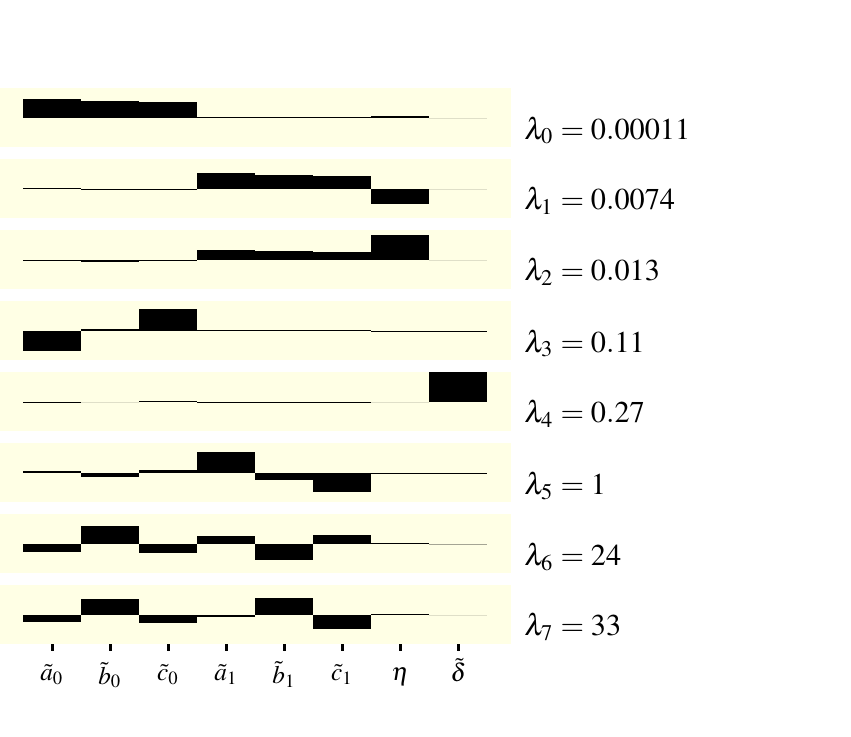}{(90,10)}{b}
  \caption{Principal component analysis for the NEDF-1 fit \textbf{(a)} and
    the NEDF-3 fit \textbf{(b)}.  Plotted are the components of the
    eigenvectors $\vect{v}_n$ defining the principal component
    \cref{eq:pca_p} as linear combinations of the various
    dimensionless parameters.  From this we see that for NEDF1 the
    most-significant component $p_0 \approx \tilde{b}_0 + \tilde{c}_0$
    which fixes the saturation energy to high precision.  At the same
    time the component $p_4\approx \tilde{b}_0-\tilde{c}_0$ in NEDF-1
    (and similarly in NEDF-3n) is not well constrained.  We also see
    that the least-significant component
    $p_{5} \approx \tilde{a}_1 - \tilde{b}_1$ is essentially
    unconstrained.  For NEDF-3, we find three insensitive components,
    two of which can be used to set the smallest parameters
    $a_0=c_1=0$.  After removing these, one obtains a similar analysis
    as for NEDF-1 above.}
  \label{fig:PCA}
\end{figure}

The principal components for fits NEDF-1 and NEDF-3 are shown in
\cref{fig:PCA}.

In the case of NEDF-3, we see that two parameters are completely
unconstrained.  These include $\tilde{a}_0 \approx -0.088$ and
$\tilde{c}_1 = -0.017$.  These values are an order of magnitude
smaller than the other coefficients: hence, the unconstrained
components can be easily removed by setting $a_0 = c_1 = 0$ which we
do in most of our fits.

Finally, both plots indicate that a combination of the $j=1$
parameters is highly unconstrained.  Thus, in NEDF-1, the combination
$\tilde{b}_1 - \tilde{a}_{1}$ can be given almost any value of order
unity without changing $\chi_E$ more than \SI{0.1}{MeV}.  This is
directly tested in the changes from NEDF1 to NEDF-E, NEDF-Er, NEDF-En,
and NEDF-Enr, where we change the sign of $b_1$ and set
$a_1 = b_1 \n_0^{2/3}$.  Indeed, we see that $\chi_E$ changed by about
\SI{0.1}{MeV}.  Notice from \cref{table:NEDF_B} that the slope of the
symmetry energy $L_2$ changes from about $\SI{30}{MeV}$ to
$\SI{70}{MeV}$ while the other parameters remain about the same.  This
also significantly changes the neutron skin thickness, demonstrating a
correlation between $L_2$ and the skin thickness, similar to that seen
in other mean-field models~\cite{Vinas:2014}.  This is consistent with
\cref{eq:L} where we see that $\tilde{b}_1$ gives us a direct handle
on $L_2$.  Finally, we have some unconstrained parameters, including
$\tilde{\delta}$.

\begin{figure}[tbp]
  \includegraphics[width=\columnwidth]{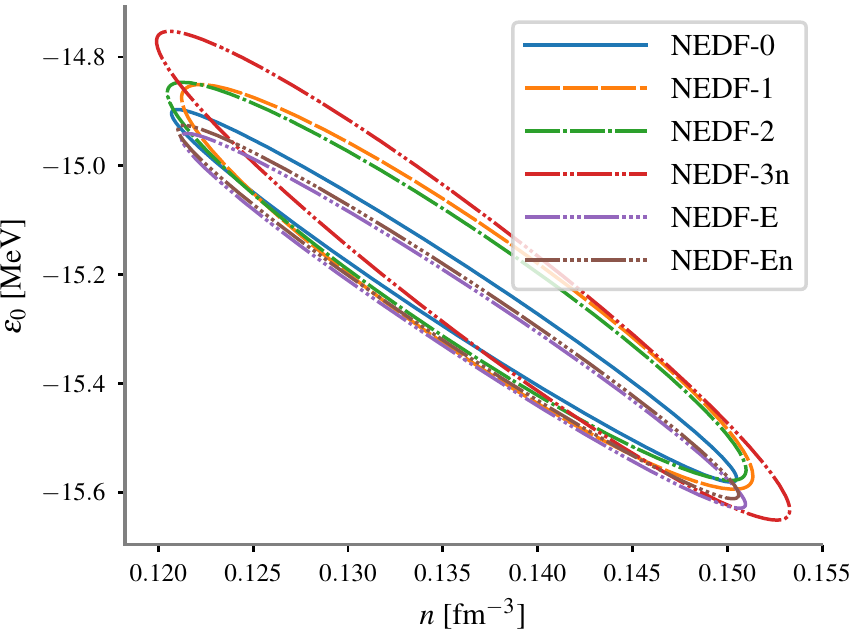}
  \caption{(Color online) The various ellipses show the region in the
    $(\varepsilon_0, \n_0)$ plane, in which the \gls{NEDF} parameters
    can be changed and to lead to changes in the residual
    $\delta \chi_E< \SI{0.2}{MeV}$.  While the equilibrium energy
    $\varepsilon_0$ and density $\n_0$ are controlled mainly by the
    combination $\tilde{b}_0+\tilde{c}_0$, which is constrained with
    very high precision, the combination $\tilde{b}_0-\tilde{c}_0$
    is significantly less constrained, see \cref{sec:PCA}.  This
    aspect allows us to manipulate to a certain degree the saturation
    properties, while affecting the overall fit only slightly.}
  \label{fig:ellipses}
\end{figure}

\section{Saturation, Symmetry Properties, and Neutron Matter}\label{sec:satur-symm-prop}

\begin{figure}[tb]
  \includegraphics[width=\columnwidth]{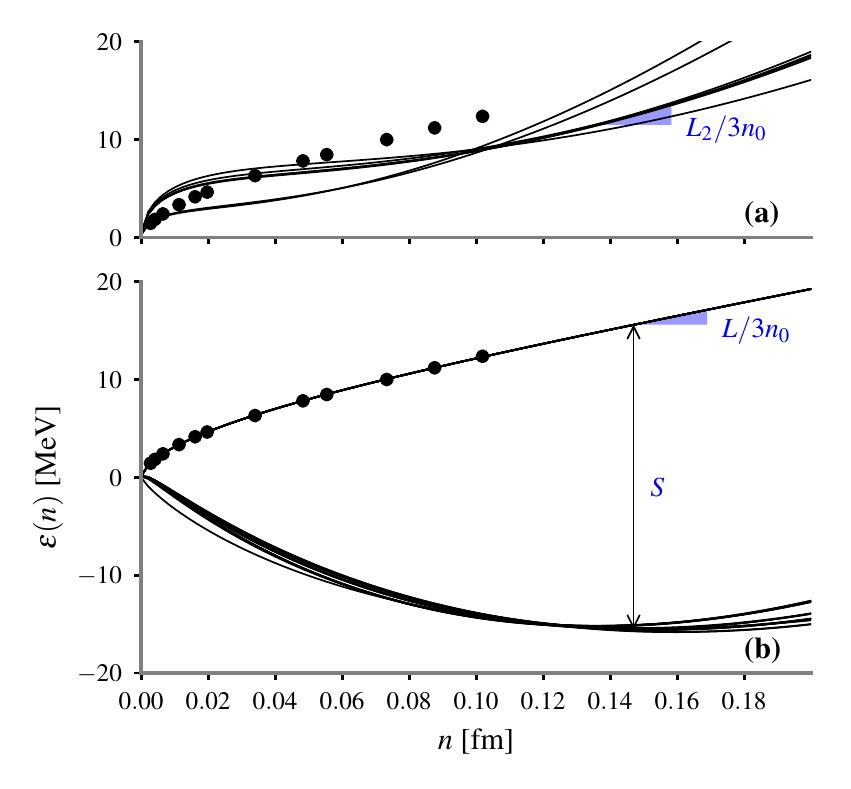}
  \caption{(Color online) The energy density per nucleon for:
    \textbf{(a)}, pure neutron matter for NEDF-0, 1, 1r, 2, 3, E, and
    Er, which do not constrain the neutron \gls{EoS} and have only
    $\beta^2$ contributions; and \textbf{(b)}, symmetric nuclear matter
    for all orbital-free functionals, and neutron matter for NEDF-3n,
    3nr, En, and Enr which collapse to the single curve fitting the
    \gls{QMC} results~\cite{Wlazlowski:2014a} (dots).}
  \label{fig:INM}
\end{figure}

When only $\beta^2$ isospin contributions are included in the
functional, our fits to the nuclear binding energies display a feature
reported in other \glspl{NEDF} discussed in literature: The energy per
neutron in pure neutron matter appears to be well constrained at a
density of $\n_n\approx \SI{0.1}{fm}^{-3}$ where all functionals
cross; see \cref{fig:INM}.  The symmetry energy $S$ is indicated for
the functionals NEDF-En and Enr.  The slope $L\approx \SI{30}{MeV}$ is
fixed by the neutron matter \gls{EoS} alone (if used as a constraint,
see \cref{eq:LL}).  In this case the slope $L_2/3\n_0$ may be tuned
without significantly affecting the mass fit by adjusting the
insensitive combination $a_1 - b_1\n_0^{1/3}$ or $c_1$, see
\cref{sec:PCA}.  Functionals with only quadratic isospin contributions
($\beta^2$) appear to cross near $\n \approx \SI{0.1}{fm^{-3}}$, see
also Ref.~\cite{Horowitz:2014a} and references therein.  However, the
value for the energy per neutron $\approx \SI{9}{MeV}$ at this point
in our fits is significantly smaller than the value
$\approx \SI{12.19}{MeV}$ obtained in \gls{QMC} calculations of
\textcite{Wlazlowski:2014a} or the equations of state for neutron
matter used by \textcite{Fayans:1998} and \textcite{Baldo:2004,
  Baldo:2008x, Baldo:2013x}, see \cref{fig:eos}.  This feature is not
present when the $\beta^4$ terms are included (NEDF-3n, NEDF-3nr,
NEDF-En and NEDF-Enr) and the \gls{QMC} results are thus automatically
reproduced.

\begin{figure}[tbp]
  \includegraphics[width=\columnwidth]{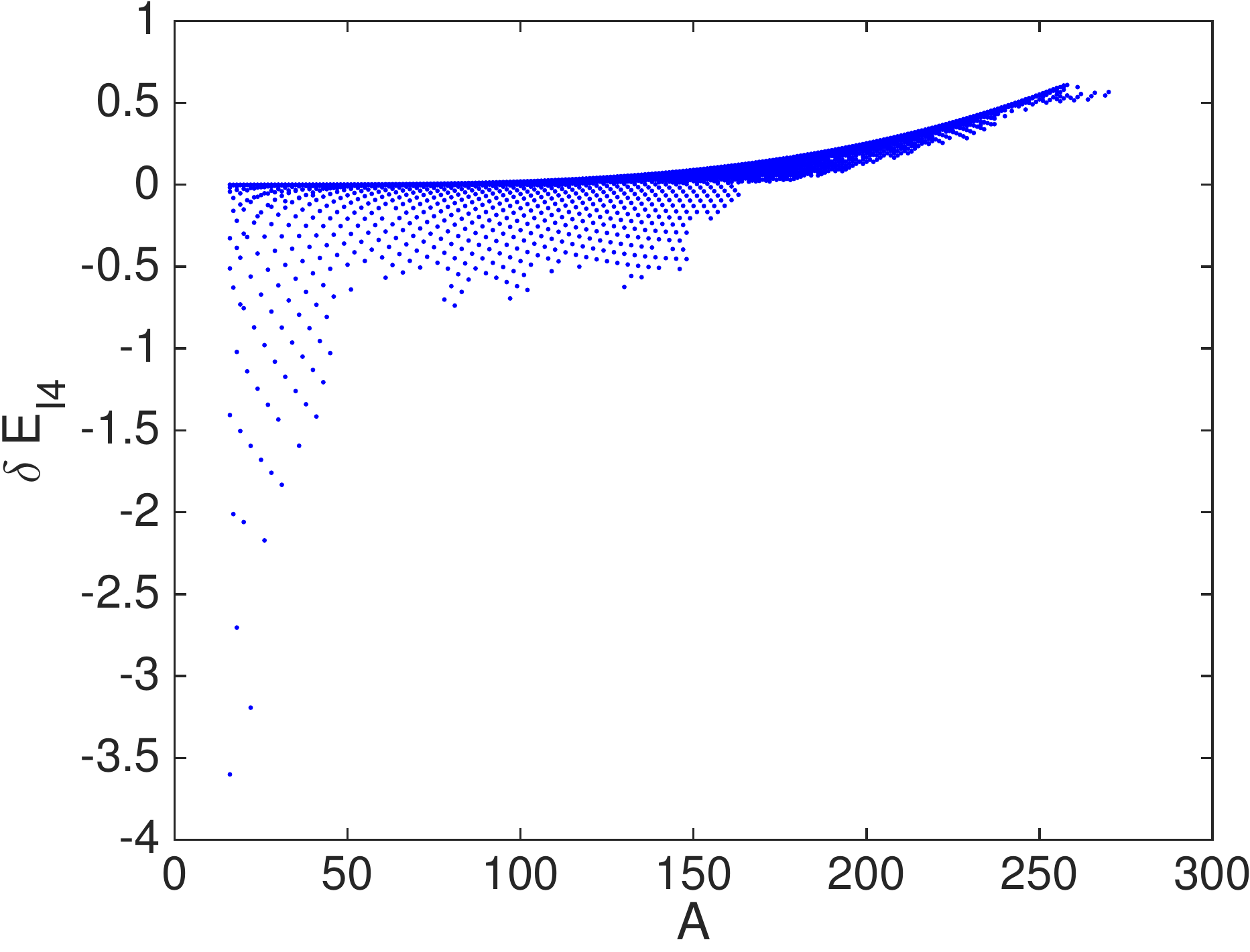}
  \caption{(Color online) The contribution to the ground state energies of
    the terms quartic in isospin density
    $\delta E_{I4}= \int \d^3 \vect{r}\; \mathcal{E}_2(\n)\beta^4$,
    evaluated perturbatively with NEDF-1, see
    \cref{table:NEDF_A_supp}.}
  \label{fig:E4}
\end{figure}

The inclusion of the $j=2$ terms quartic in $\beta^4$ have very little
significance on mass fits.  This demonstrates an important point: the
\gls{EoS} of pure neutron matter has very little impact on the form of
the \gls{NEDF}, if only nuclei are considered.  In measured nuclei,
the ratio $\beta = (\n_n-\n_p)/\n\approx (N-Z)/A$ is
$\abs{\beta} <1/4$ (with a very small number of exceptions), hence
nuclear masses are essentially insensitive to the presence of the
$\beta^4$ terms, as $\abs{\beta}^4 < 1/256$.  To assess the magnitude
of these effects, we have evaluated the $\beta^4$ contributions to the
nuclear binding energies perturbatively, see \cref{fig:E4}.  This
contribution is quite small and can be easily overlooked when
discussing known nuclei, but is crucial in order to correctly
reproduce the energy of neutron matter.  By evaluating \cref{eq:e024} at
$\n=\SI{0.1}{fm}^{-3}$ one obtains
\begin{gather}
  \left. \frac{\mathcal{E}}{\n}\right|_{\n=0.1}= [-4.399 + 13.961 \beta^2 + 2.635 \beta^4] \; \si{MeV}.
\end{gather}
When one averages $\beta^2$ and $\beta^4$ over all nuclei one obtains
the values \num{0.028} and \num{0.001} respectively, which are
noticeably lower than the ``maximum'' values of $1/16\approx 0.062$
and $1/256\approx \num{0.004}$ and thus the contribution of the terms
in $\beta^4$ to $\chi_E$ and nuclear masses is further reduced.  The
contributions of these terms to the averaged energy density per
nucleon over $\beta$ at $\n=\SI{0.1}{fm}^{-3}$ are
\begin{gather}
  \left. \frac{\mathcal{E}}{\n}\right| _{\n=0.1} = [-4.399 + 0.391 + 0.0026] \; \si{MeV}, 
\end{gather}
and the contribution of the quartic term in $\beta$ to the total
energy is practically invisible in nuclei.

Thus, using properties of the neutron matter to constrain the form of
the \gls{NEDF} and/or arguing against the inclusion of higher powers
of $(\n_n-\n_p)$~\cite{Fayans:1998, Baldo:2008x, Baldo:2013x,
  Brown:2013, Brown:2014, Goriely:2013, Fantina:2014, Reinhard:2010,
  Erler:2013} is an ill-advised procedure, and the applications of
functionals constructed in this manner, in particular to star
environments, should be regarded with suspicion.  The statement often
made in the literature (see e.g\@. \textcite{Horowitz:2014a} and
references therein) that the value of the symmetry energy at
$\n \approx \SI{0.1}{fm}^{-3}$ is well constrained by nuclear masses
must only be applied to the local expansion $S_2$ at this density, but
not to the symmetry energy difference $S$ between symmetric and pure
neutron matter.

\section{Charge Form Factors}\label{app:charge-form-factors}

The charge form factors are determined experimentally, and we
approximate the Fourier transforms of the form factors with the dipole
term for the proton,
${G}^{p}_E(Q) \approx (1 +
Q^2/\SI{0.71}{GeV^2})^{-2}$~\cite{Perdrisat:2007}, and
${G}_E^{n}(Q) \approx a(1 + Q^2r_+^2/12)^{-2} - a(1 +
Q^2r_-^2/12)^{-2}$ with
$r_\pm^2 = r_{\text{avg}}^2 \pm \braket{r_n^2}/2a$,
$\braket{r_n^2} = -\SI{0.1147(35)}{fm^2}$,
$r_{\text{avg}} = \SI{0.856(32)}{fm}$, and
$a =\num{0.115(20)}$~\cite{Gentile:2011}.

%%%%%%%%%%%%%%%%%%%%%%%%%%%%%%%%%%%%%%%%%%%%%

% These are needed to avoid a babel error.
\providecommand{\selectlanguage}[1]{}
\renewcommand{\selectlanguage}[1]{}

\pagebreak

\bibliography{master,local}

\end{document}